\documentclass[prb,aps,twocolumn,superscriptaddress]{revtex4-1}
\usepackage{graphicx}
\usepackage{dcolumn}
\usepackage{bm}
\usepackage{bbm}
\usepackage{amsfonts}
\usepackage{amsmath}
\usepackage{natbib}

\usepackage[mathlines]{lineno}
\usepackage[colorinlistoftodos]{todonotes}
\usepackage[colorlinks=true, allcolors=blue]{hyperref}

\begin{document}

\newcommand{\unamone}{Departamento de Sistemas Complejos, Instituto de Fisica,
Universidad Nacional Aut\'onoma de M\'exico, Apartado Postal 20-364,01000,
Ciudad de M\'exico, M\'exico.}
\newcommand{\unamtwo}{Instituto de F\'isica,
Universidad Nacional Aut\'onoma de M\'exico, Apartado Postal 20-364, 01000,
Ciudad de  M\'{e}xico, M\'{e}xico}
\newcommand{\uama}{\'Area de F\'isica Te\'orica y Materia Condensada,
Universidad Aut\'onoma Metropolitana Azcapotzalco, Av. San Pablo 180,
Col. Reynosa-Tamaulipas, 02200 Cuidad de M\'exico, M\'exico}

\title[Sample title]{ Floquet spectrum and electronic transitions of tilted anisotropic Dirac materials under electromagnetic radiation: monodromy matrix approach.}

\author{A. Kunold}
\email{akb@azc.uam.mx}
\affiliation{\uama}
\author{J. C. Sandoval-Santana}
\email{jcarlosss@fisica.unam.mx}
\affiliation{\unamtwo}
\author{V. G. Ibarra-Sierra}
\email{vickkun@fisica.unam.mx}
\affiliation{\unamone}
\author{Gerardo G. Naumis}
\email{naumis@fisica.unam.mx}
\affiliation{\unamone}

\date{\today}

\begin{abstract}
We analyze the quasienergy-spectrum and the
valence to conduction-band transition probabilities
of a tilted anisotropic Dirac material subject
to linearly and circularly polarized electromagnetic fields.
The quasienergy-spectrum is numerically calculated from the
monodromy matrix of the Schr\"odinger equation
via the Floquet theorem for arbitrarily intense
electromagnetic fields. 
To asses the valence to conduction-band transition times
we deduced
a Rabi-like formula in the rotating wave approximation.
In the strong-field regime the spectrum
as a function of the momentum components
divides into two very distinctive regions.
In the first, located around the Dirac point,
the quasi-spectrum
is significantly distorted by the field as
the electronic parameters are
renormalized by electronic-dressing.
In the second, all the characteristics
of the free carrier spectrum are retained.
Linearly polarized light anisotropically
deforms the spectrum according to the field
polarization direction. Dirac-like points
form around the original Dirac point.
The quasi-spectrum of
circularly polarized
light, instead, exhibits a gap formation
in the Dirac point and has elliptical symmetry.
We show that,
in contrast to the single-photon resonant transitions that
characterize the weak-field regime,
the strong-field regime is dominated
by multiphoton resonances.
\end{abstract}
\maketitle

\section{Introduction}

The extraordinary electronic and optical properties of graphene
make it an ideal  platform for the development of
diverse optoelectronic devices and applications
\cite{bonaccorso2010graphene,bao2017graphene,ponraj2016photonics}
such as THz generators \cite{SALEN20191},
plasmonic devices
\cite{grigorenko2012graphene,fan2019graphene},
polarization-sensitive, broad band photodectectors
\cite{doi:10.1021/acsnano.9b03994f,SCAGLIOTTI2019643},
broad band optical modulators
\cite{liu2011graphene,sorianello2018graphene,hao2019experimental},
infrared photodetectors
\cite{safaei2019dirac} and
solar cells
\cite{yin2014graphene,o2019graphene}.

Because of its broadband and ultrafast optical response
and weak screening\cite{neto2009electronic},
graphene is a particularly attractive material
for the implementation of attosecond 
science applications
\cite{higuchi2017light,heide2018coherent,
doi:10.1063/1.5086773,SCAGLIOTTI2019643}.
This field has been rapidly growing\cite{krausz2014attosecond}
since the first demonstration of sub-femtosecond pulse
generation \cite{hentschel2001attosecond}
with the prospect of new time-resolved spectroscopic techniques
and overcoming the speed limitations of electronics \cite{krausz2014attosecond}.
These applications rely on the high light-matter coupling
between carriers and strong
optical fields.

The limit of strong electromagnetic fields is of great interest
also because it may give rise to exotic and novel quantum phases
coherently induced by light.
Two striking examples of Floquet-engineered topological
phases in graphene are the Photovoltaic Hall effect
\cite{PhysRevB.79.081406} and
the light-induced anomalous Hall effect
\cite{mciver2020light}. Astonishingly
the Hall effect,
produced in the absence of a magnetic field,
arises from a light-induced Berry curvature
absent in the static case
\cite{mciver2020light}.
The dressing of electrons, i.e.  electrons bounded to a strong
electromagnetic field
\cite{lopez2008analytic,lopez2010graphene,bonaccorso2010graphene,
kibis2010metal,
Foa2011Tuning,
sun2012ultrafast,
kibis2015,
kristinsson2016control,
kibis2017all,
kibis2018electromagnetic,safaei2019dirac},
has become a key concept
in understanding the interaction of electromagnetic
radiation with charge carriers in graphene
\cite{lopez2008analytic,doi:10.1063/1.3597412,kibis2015}.
Electromagnetic dressing
substantially renormalizes the energy spectrum as well
as other electronic parameters
of graphene\cite{kibis2015,kibis2017all,
kibis2018electromagnetic,sandoval2020floquet}
and could, therefore, be exploited to adjust its
optoelectronic
properties.

Unlike metals, semiconductors and other conventional materials used
in electronics, Dirac materials have a linear dispersion
relation near the band edge that can be characterized
by the Dirac Hamiltonian. This hinders the application of
some of the standard solid-state theoretical tools.  
Such is the case of carriers interacting with a strong
electromagnetic field.
The purpose of this paper is to analyze
the dynamics of electrons in a Dirac material
coupled to an intense electromagnetic field, and at the
same time, to introduce a numerical technique to find the temporal
evolution of quantum, systems.

The research presented here is a follow up of our previous work
concerning electromagnetic waves in the strong field regime
acting on borophene
\cite{Champo2019,ibarra2019dynamical}.
The methods and results introduced
in this paper
are of a far more general character.
In former investigations we were only able to tackle
the strong-field regime in an approximate manner,
due to the lack of a
method capable
to bridge the gap between the low and high
electromagnetic field intensity.
In this work we have developed  methods that allow us to
approach the calculations of the quasienergy-spectrum,
time-dependent wave function and transition
probabilities without any restrictions in
field parameters such as polarization, intensity
and time duration. These methods allow us
to continuously go from the weak to the
strong-field regime without any approximations.
This is a fundamental requirement in the comparison
of spectra and transition probabilities in both regimes.
Moreover, the deviced methods open the door
to the study of many other time-driven quantum systems\cite{Goldman,sandovalannderphysik,Calvo}.

We center our discussion in the effects
on the quasienergy spectrum
and the valence to conduction band transition probability.
In order for our study to be general enough
we utilize the low energy Hamiltonian of an anisotropic tilted Dirac
material. Specifically, we present results for borophene
\cite{peng2016electronic,verma2017effect,Villanova2016Spin,
Champo2019,Villanova2016Spin}, the paradigmatic
example of tilted Dirac materials.
For the sake of comparison, we calculated the quasi-spectra
in the weak and strong-field regimes for linearly
and circularly polarized light.

We show that, regardless of the field polarization,
under strong illumination
the quasi-spectrum can be separated in two
very well defined regions in $\boldsymbol{k}$-space.
The first, where the electromagnetic field is perturbative,
resembles that of the
free carrier spectrum. In the second,
where the field effects are more dramatic,
new Dirac-like points as well as gaps are generated.
The boundary where the sharp transition from one regime
to the other takes place is accurately determined.
Under circularly polarized light we clearly demonstrate
the emergence of a gap in the Dirac point.
In the strong field region, an anisotropic quasi-spectrum
emerges for linearly polarized light. The anisotropy is
oriented in accordance with the filed polarization direction.
We demonstrate that, in stark contrast with the perturbative regime,
where transitions happen 
between energy levels whose energy difference match
the photon energy, in the strong field regime
they take place well outside the single photon resonance condition.
This indicates the presence of multiphoton
excitations.

The paper is organized as follows.
In Sec. \ref{sec:hamiltonian} we introduce
the low energy Hamiltonian of a 2D tilted anisotropic Dirac
material,  while in Sec. \ref{sec:elliptical}  we set up the
basic equations for the model system under an electromagnetic field with arbitrary
polarization. A method to compute the quasienergy-spectrum
via the Floquet theorem and the monodromy matrix
is presented in Sec. \ref{sec:quasienergy}.
Section \ref{sec:rabiformula} is devoted to the derivation
of a Rabi-like formula by adopting the rotating wave
approximation. A comparison between the diverse quasienergy-spectra
and valence to conduction-band transition probabilities
under different illumination conditions is shown
in Sec. \ref{sec:results}, while a discussion of the results is
presented in Sec. \ref{Discussion}.
We conclude and summarize
in \ref{sec:conclusions}.

%
%
\begin{figure*}
\includegraphics[width=0.28\textwidth]{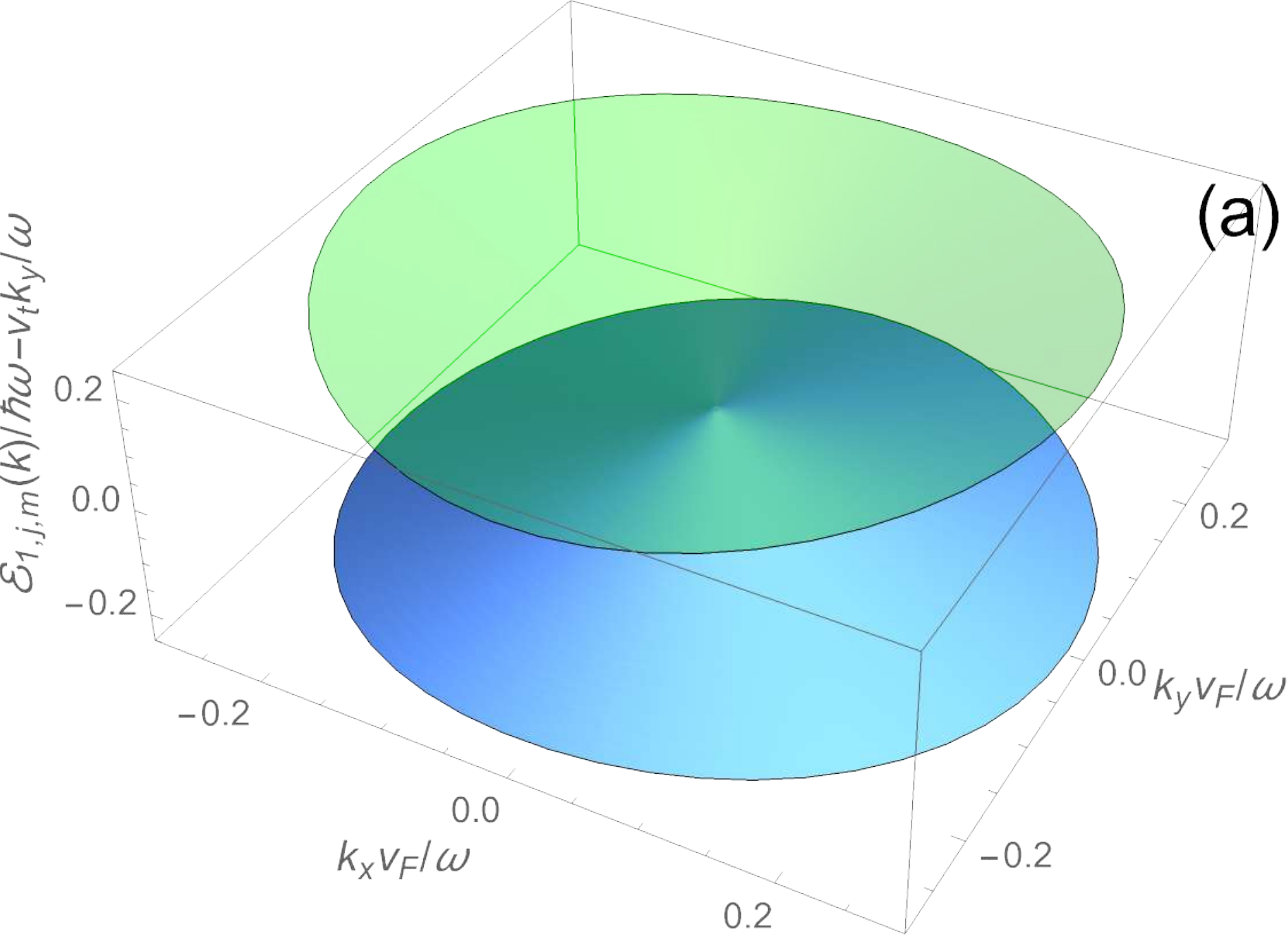}
\includegraphics[width=0.28\textwidth]{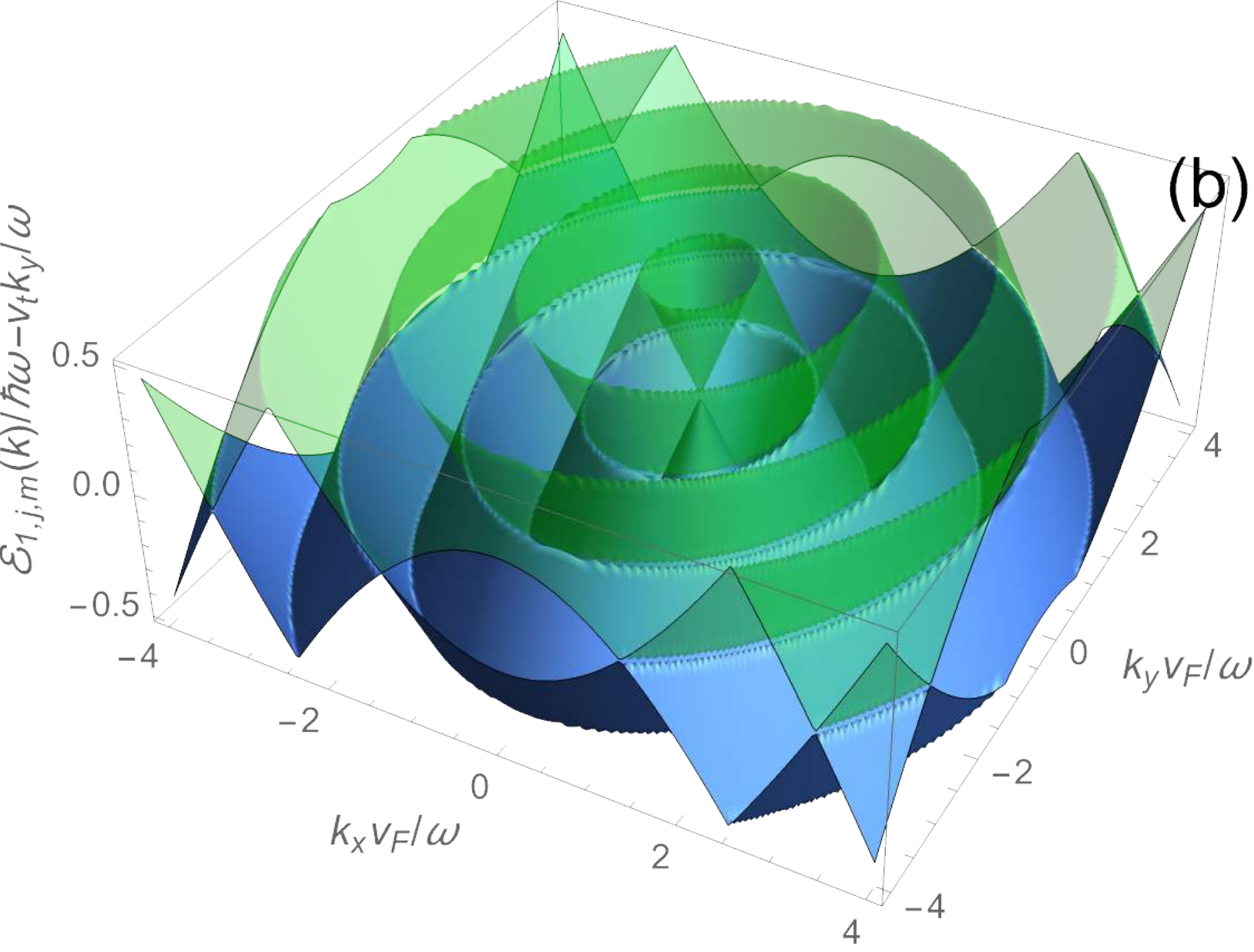}
\includegraphics[width=0.20\textwidth]{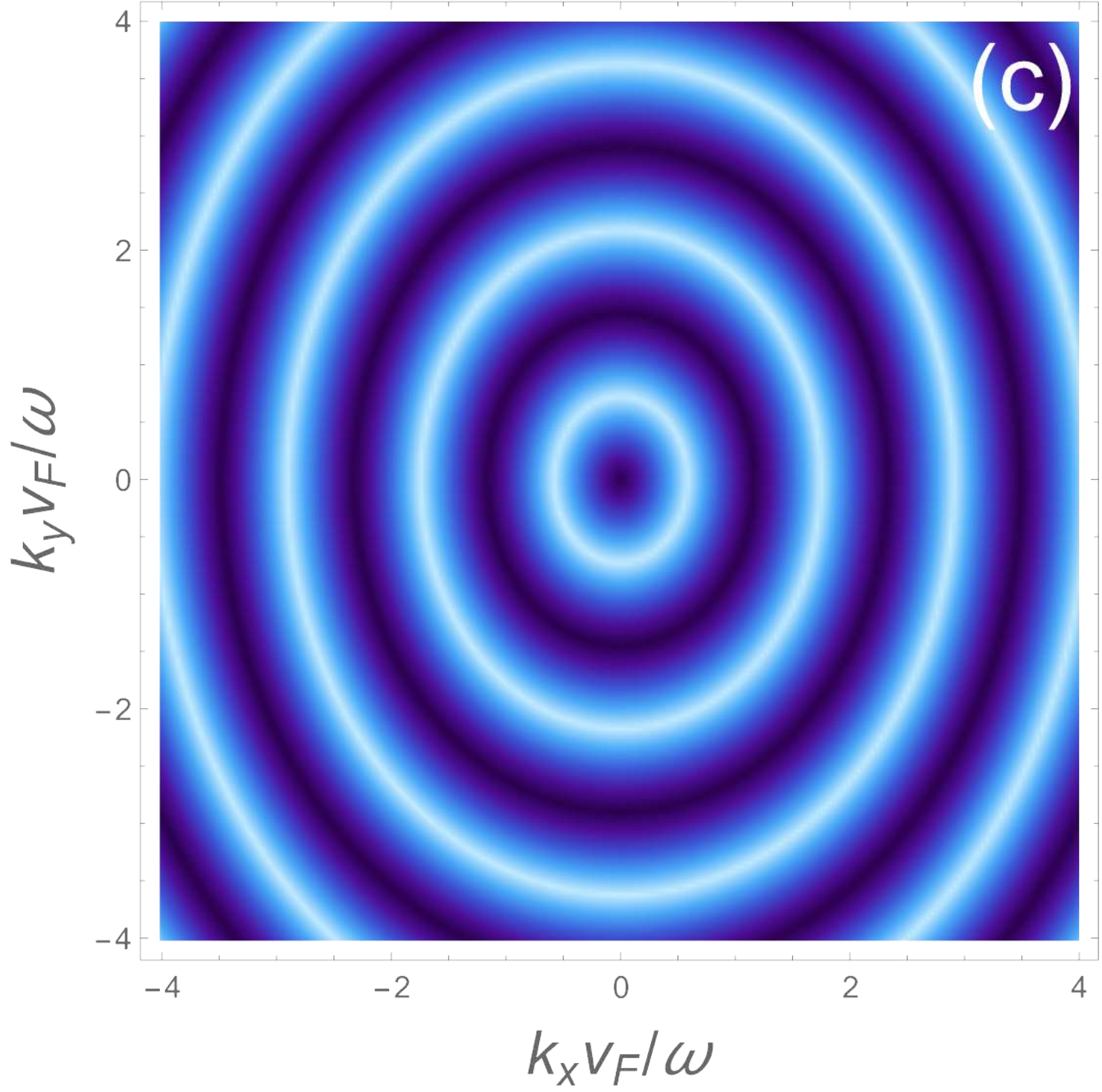}
\includegraphics[width=0.405\textwidth]{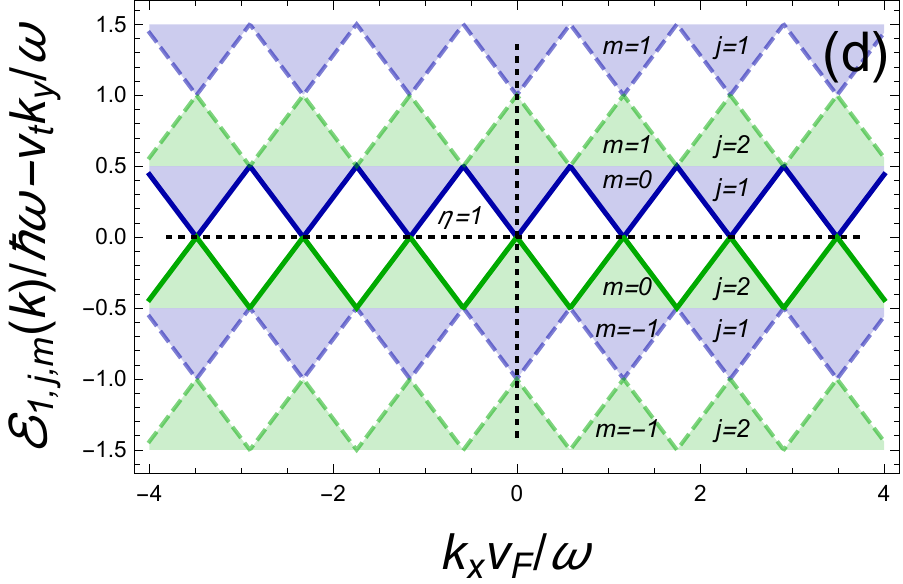}
\includegraphics[width=0.40\textwidth]{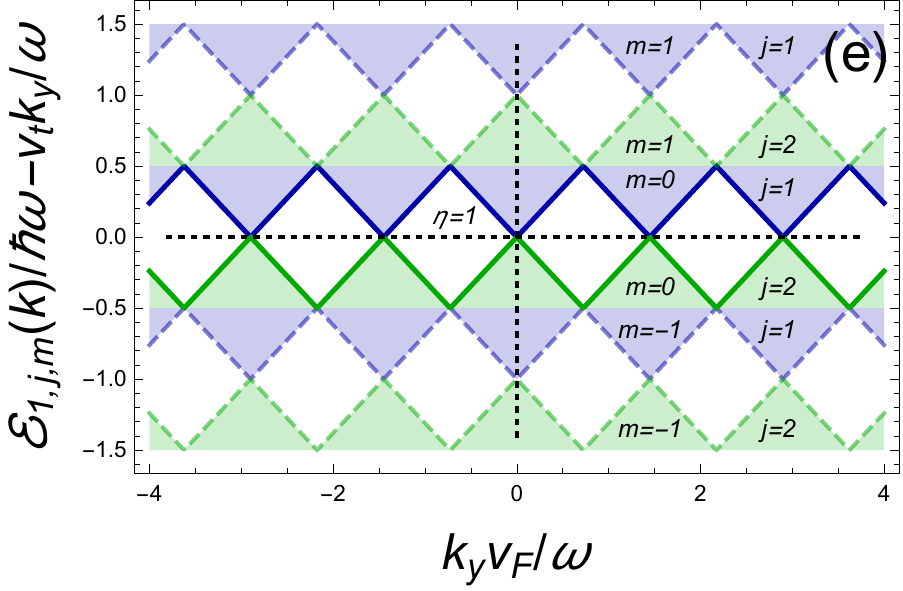}
\caption{Quasienergy-spectrum 
$\mathcal{E}_{\eta,j,m}(\boldsymbol{k})-\hbar v_t k_y$
in the zero field regime
($E_x=E_y=0$)
for the $K$ Dirac point $\eta =1$;
(a) Quasienergy spectrum
as a function of the
momentum components $k_x$ and $k_y$ in the
vicinity of the Dirac point for
the $j=1$ (solid blue surface) and $j=2$ (transparent
green surface) bands. Only the first Floquet zone
$m=0$ is plotted. (b) Quasienergy-spectrum
as a function of the
momentum components $k_x$ and $k_y$
covering many different sections of the Dirac cone for
the $j=1$ (solid blue surface) and $j=2$ (transparent
green surface) bands. Only the first Floquet zone
$m=0$ is plotted.
(c) Density plot of the quasienergy spectrum
as a function of the
momentum components $k_x$ and $k_y$ for the
first band $j=1$.
(d) Quasienergy spectrum as a function of the
momentum component $k_x$ and fixed $k_y=0$
for the $j=1,2$ bands
(blue and green solid lines respectively) and
the first three Floquet zones ($m=-1,0,1$).
(e) Quasienergy spectrum as a function of the
momentum component $k_y$ and fixed $k_x=0$
for the $j=1,2$ bands
(blue and green solid lines respectively) and
the first three Floquet zones ($m=-1,0,1$).
}\label{figure1}
\end{figure*}

%
%
\begin{figure*}
\includegraphics[width=0.28\textwidth]{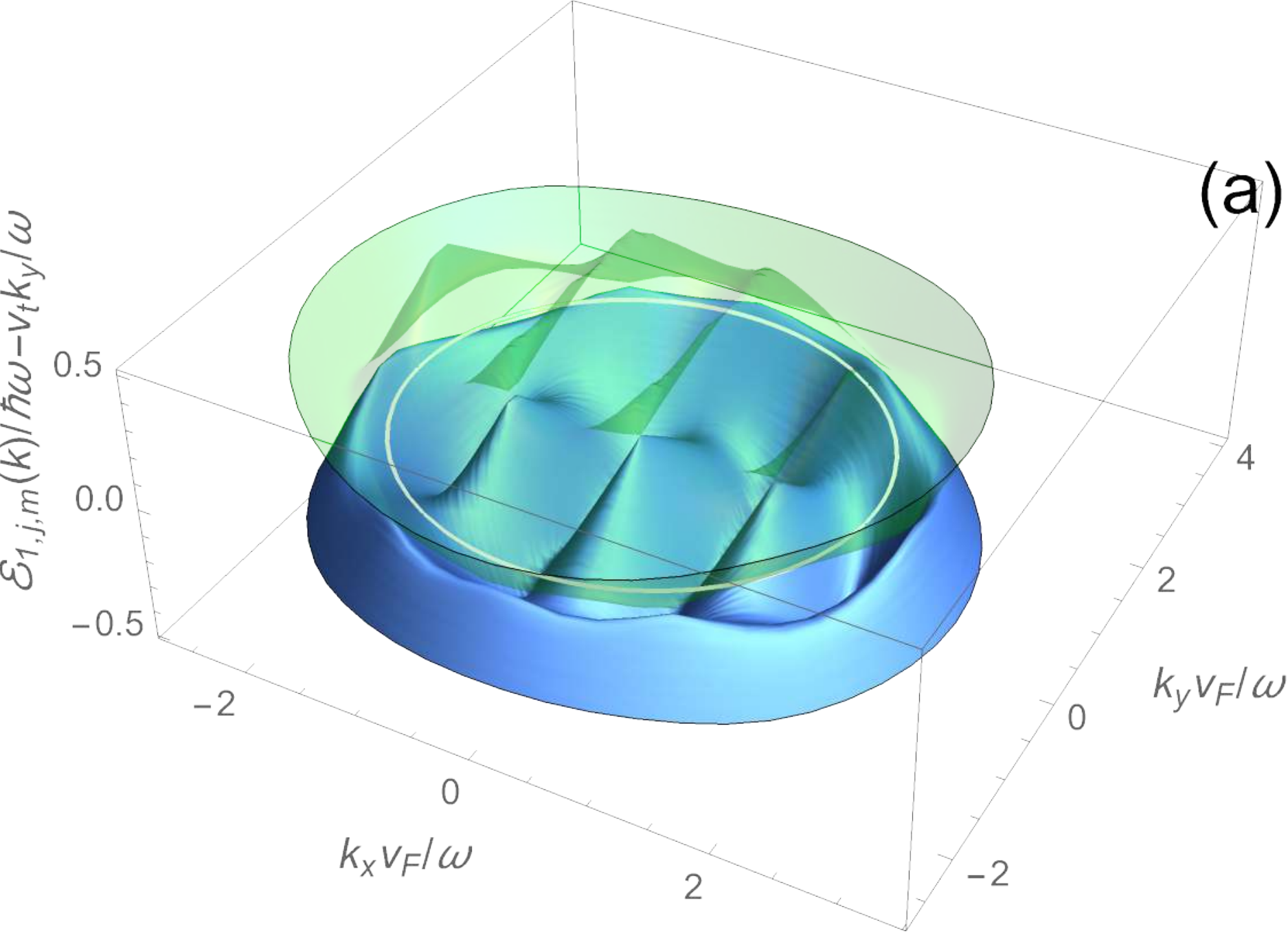}
\includegraphics[width=0.28\textwidth]{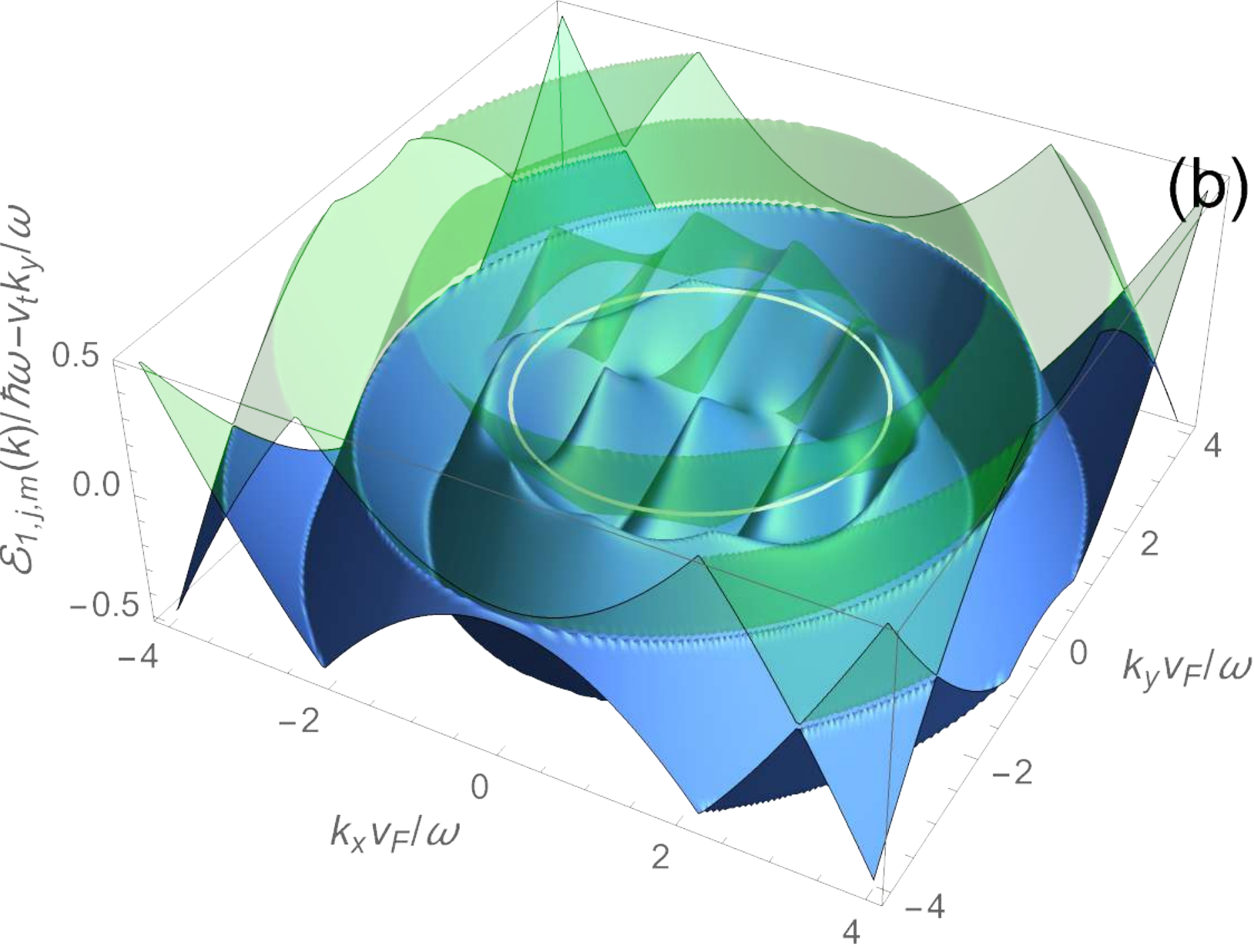}
\includegraphics[width=0.2\textwidth]{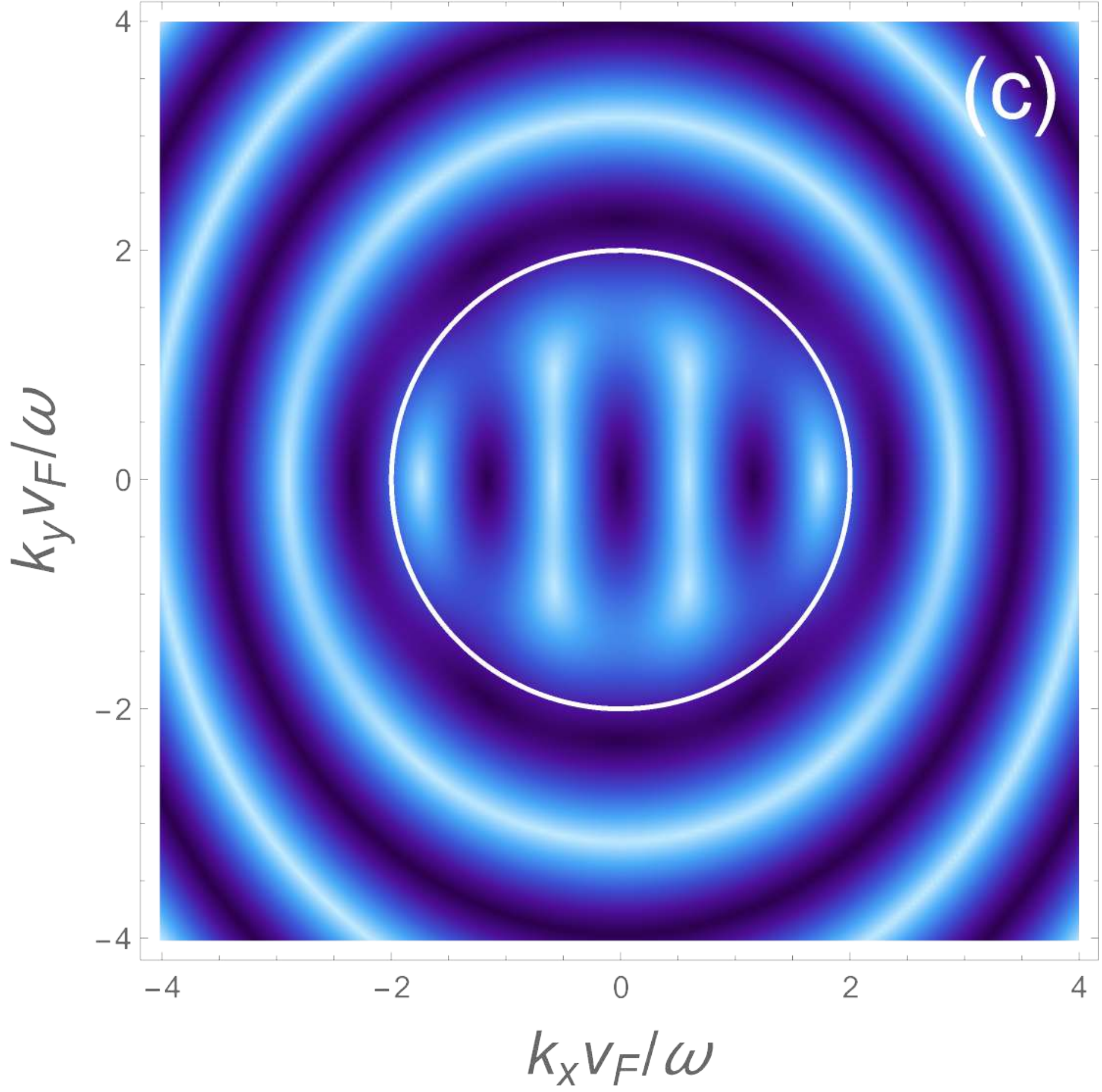}
\includegraphics[width=0.40\textwidth]{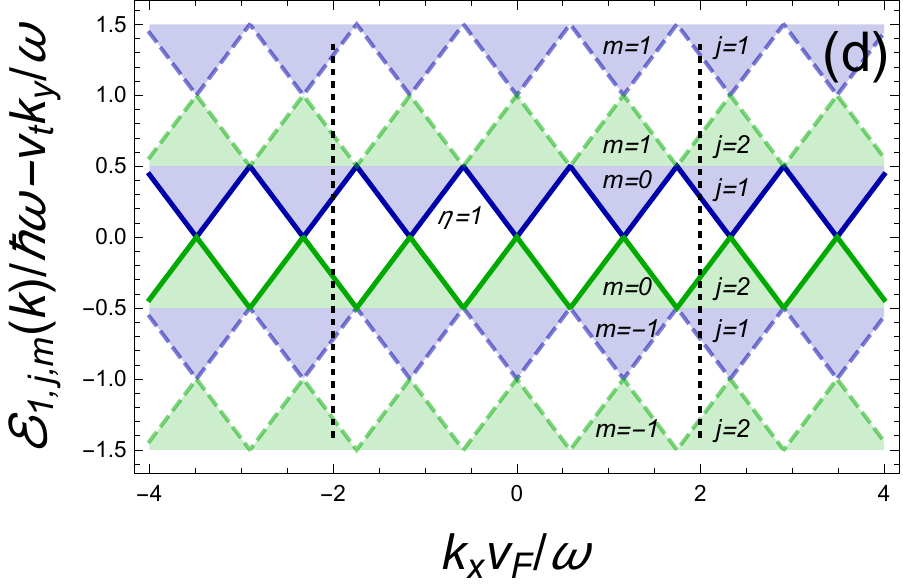}
\includegraphics[width=0.40\textwidth]{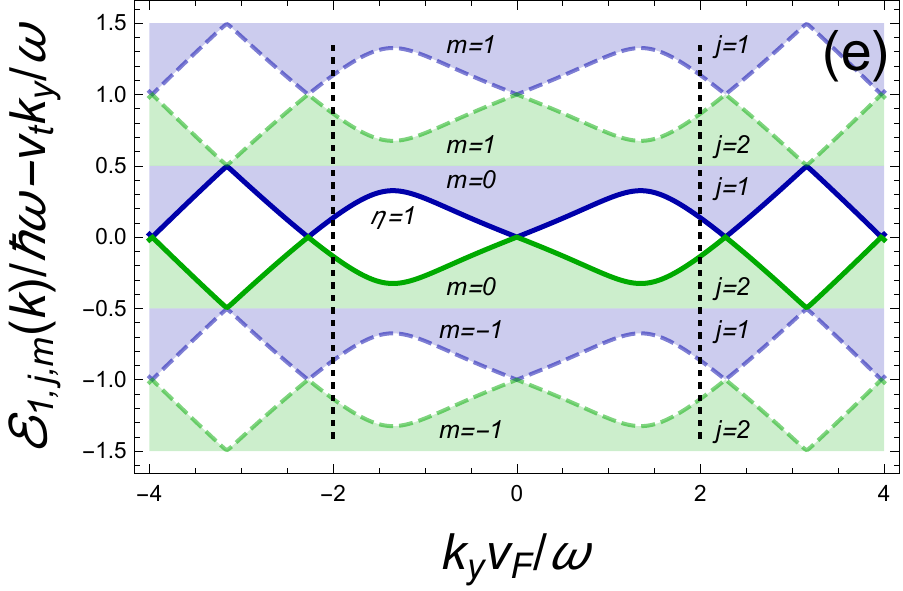}
\caption{Quasienergy-spectrum
$\mathcal{E}_{\eta,j,m}(\boldsymbol{k})-\hbar v_t k_y$
in the strong
field regime for a linearly polarized wave
($E_x e v_F/\hbar \omega^2=2.0$, $E_y=0$, $\phi=0$)
in the $K$ Dirac point $\eta =1$;
(a) Quasienergy-spectrum as a function of the
momentum components $k_x$ and $k_y$ for
the $j=1$ (solid blue surface) and $j=2$ (transparent
green surface) bands. Only the first Floquet zone
$m=0$ is plotted. A zoom of the original and
two new Dirac points is shown.
(b) Quasienergy-spectrum as a function of the
momentum components $k_x$ and $k_y$ for
the $j=1$ (solid blue surface) and $j=2$ (transparent
green surface) bands. Only the first Floquet zone
$m=0$ is plotted. The wider range
of $\boldsymbol{k}$ allows to see the three Dirac points
emerging in the middle of the zero-field quasi-spectrum.
(c) Density plot of the quasienergy-spectrum as
a function of the
momentum components $k_x$ and $k_y$ for
the $j=1$ band.
(d) Quasienergy-spectrum as a function of the
momentum component $k_x$ and fixed $k_y=0$
for the $j=1,2$ bands
(blue and green solid lines respectively) and
the first three Floquet zones ($m=-1,0,1$).
(e) Quasienergy-spectrum as a function of the
momentum component $k_y$ and fixed $k_x=0$
for the $j=1,2$ bands
(blue and green solid lines respectively) and
the first three Floquet zones ($m=-1,0,1$).
The white curve
in panels (a), (b) and (c) 
indicates the circle $k=e E/\hbar\omega=2v_F/\omega$ that
divides the strong ($k < e E/\hbar\omega$)
and weak field regions ($k > e E/\hbar\omega$).
The vertical dotted lines in panels (d) and (e)
correspond to the position of edges of the circle
$k=e E/\hbar\omega=\pm 2v_F/\omega$.
}\label{figure2}
\end{figure*}

%
%
\begin{figure*}
\includegraphics[width=0.28\textwidth]{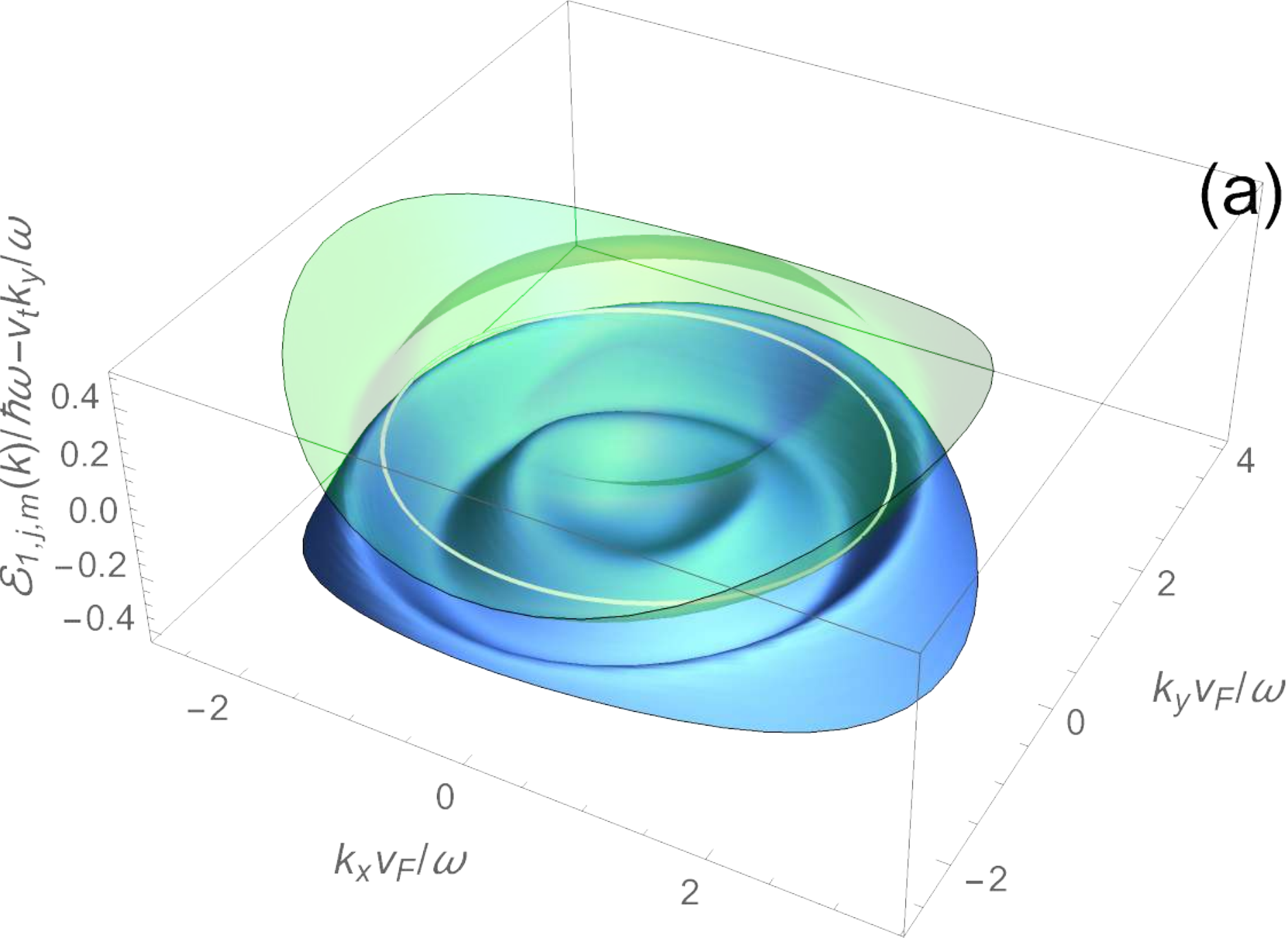}
\includegraphics[width=0.28\textwidth]{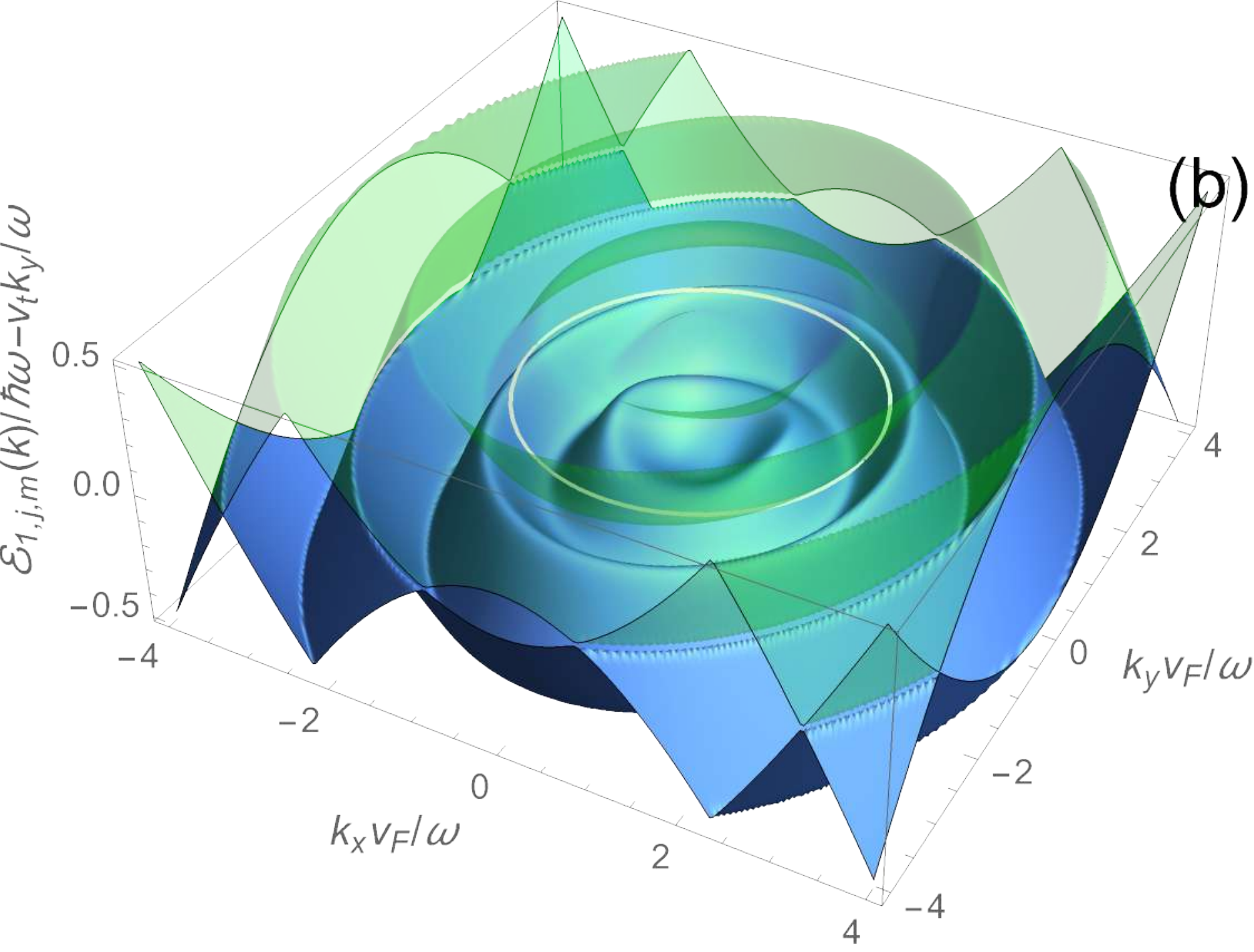}
\includegraphics[width=0.2\textwidth]{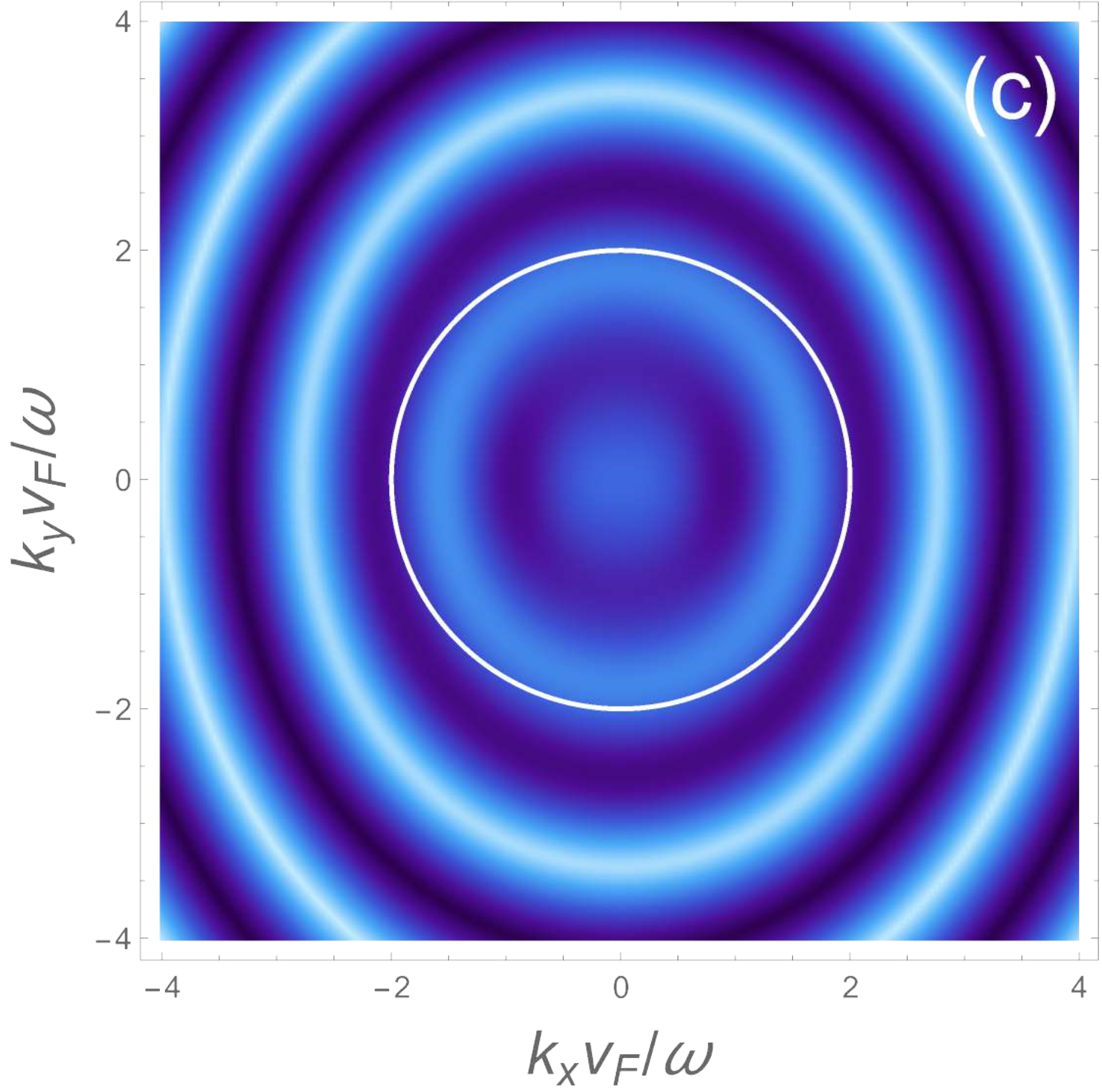}
\includegraphics[width=0.40\textwidth]{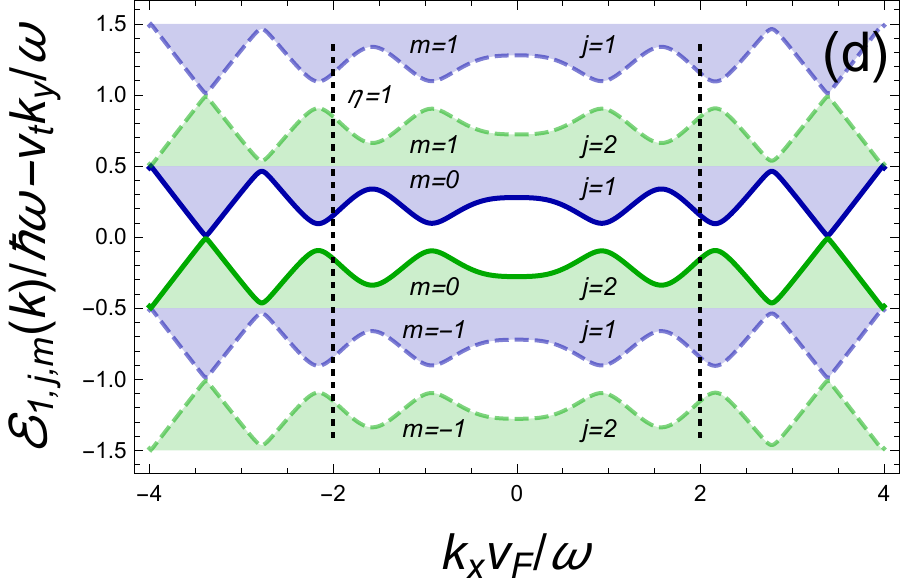}
\includegraphics[width=0.40\textwidth]{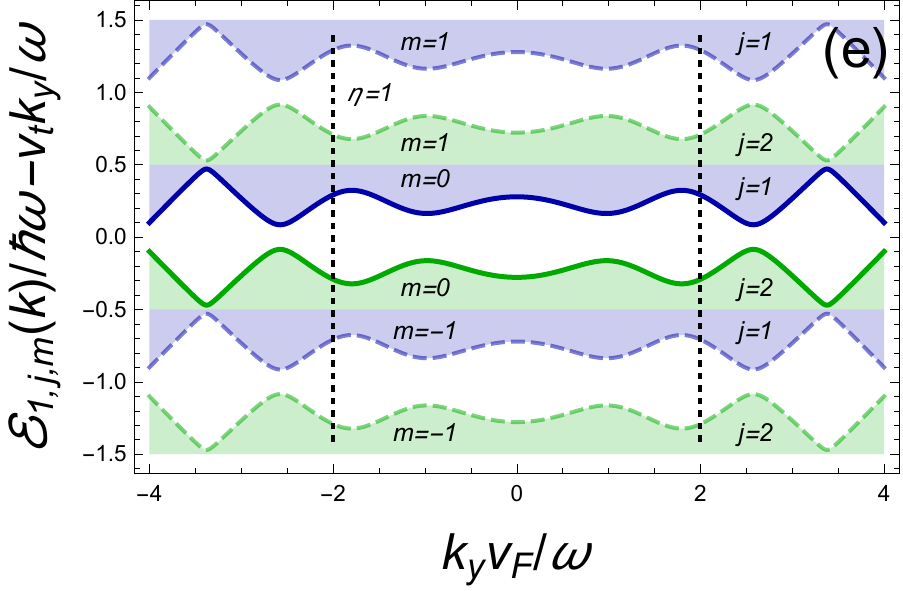}
\caption{Quasienergy-spectrum
$\mathcal{E}_{\eta,j,m}(\boldsymbol{k})-\hbar v_t k_y$
in the strong
field regime for a circularly polarized wave
($E_x e v_F/\hbar \omega^2=E_y e v_F/\hbar \omega^2=\sqrt{2}$, $\phi=-\pi/2$)
in the $K$ Dirac point $\eta =1$;
(a) Quasienergy-spectrum as a function of the
momentum components $k_x$ and $k_y$ for
the $j=1$ (solid blue surface) and $j=2$ (transparent
green surface) bands. Only the first Floquet zone
$m=0$ is plotted. (b) Quasienergy-spectrum as a function of the
momentum components $k_x$ and $k_y$ for
the $j=1$ (solid blue surface) and $j=2$ (transparent
green surface) bands. Only the first Floquet zone
$m=0$ is plotted. The wider range
of $\boldsymbol{k}$ allows to see the three Dirac points
emerging in the middle of the zero-field quasi-spectrum. (c) Density plot
of the quasienergy-spectrum as a function of the
momentum components $k_x$ and $k_y$ for
the $j=1$ band.
(d) Quasienergy-spectrum as a function of the
momentum component $k_x$ and fixed $k_y=0$
for the $j=1,2$ bands
(blue and green solid lines respectively) and
the first three Floquet zones ($m=-1,0,1$).
(e) Quasienergy-spectrum as a function of the
momentum component $k_y$ and fixed $k_x=0$
for the $j=1,2$ bands
(blue and green solid lines respectively) and
the first three Floquet zones ($m=-1,0,1$).
The white curve
in panels (a), (b) and (c) 
indicates the circle $k=e E/\hbar\omega=2v_F/\omega$ that
divides the strong ($k < e E/\hbar\omega$)
and weak field regions ($k > e E/\hbar\omega$).
The vertical dotted lines in panels (d) and (e)
correspond to the position of edges of the circle
$k=e E/\hbar\omega=\pm 2v_F/\omega$.
}\label{figure3}
\end{figure*}

%
%
\begin{figure}
\includegraphics[width=0.238\textwidth]{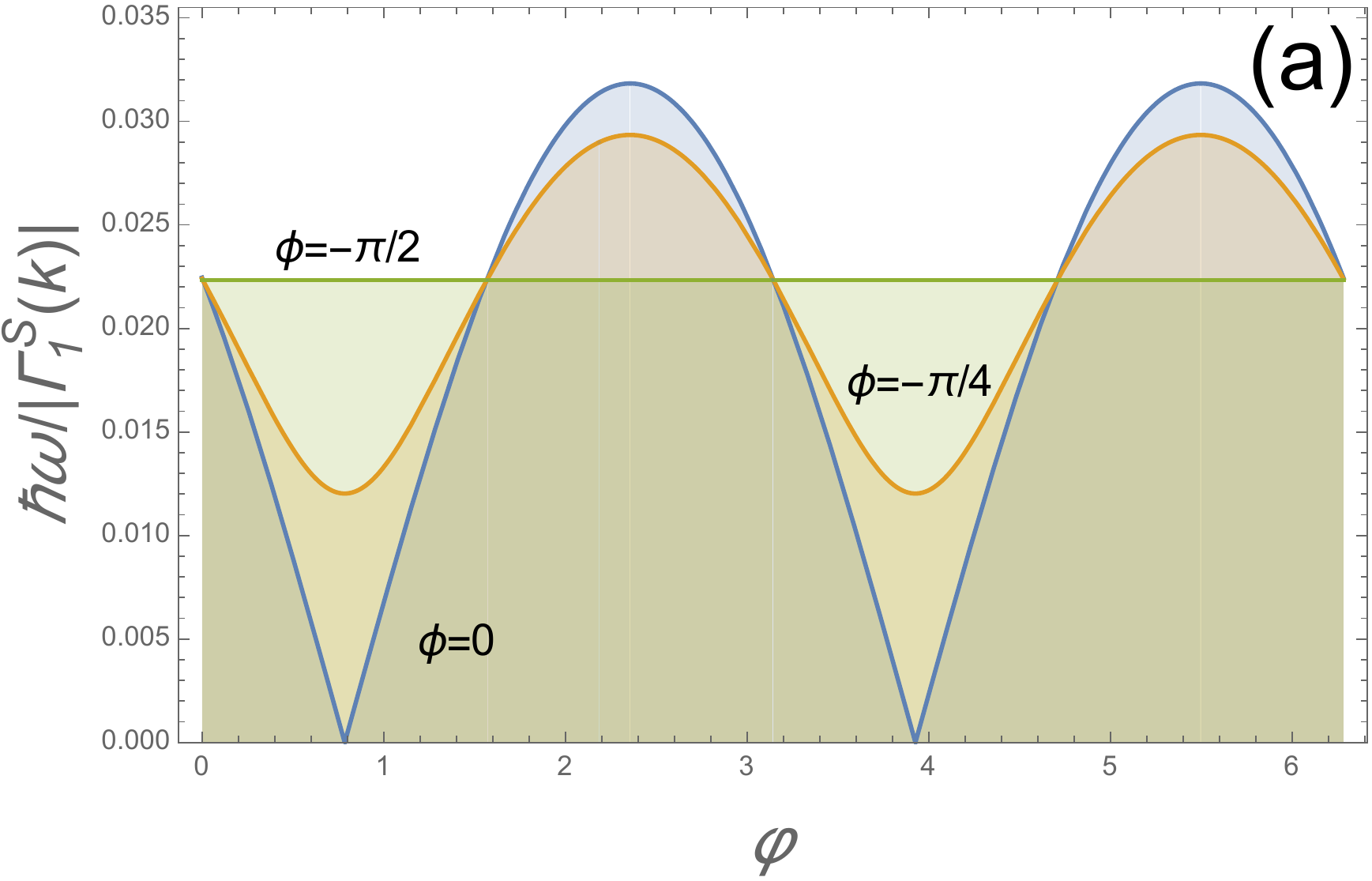}
\includegraphics[width=0.238\textwidth]{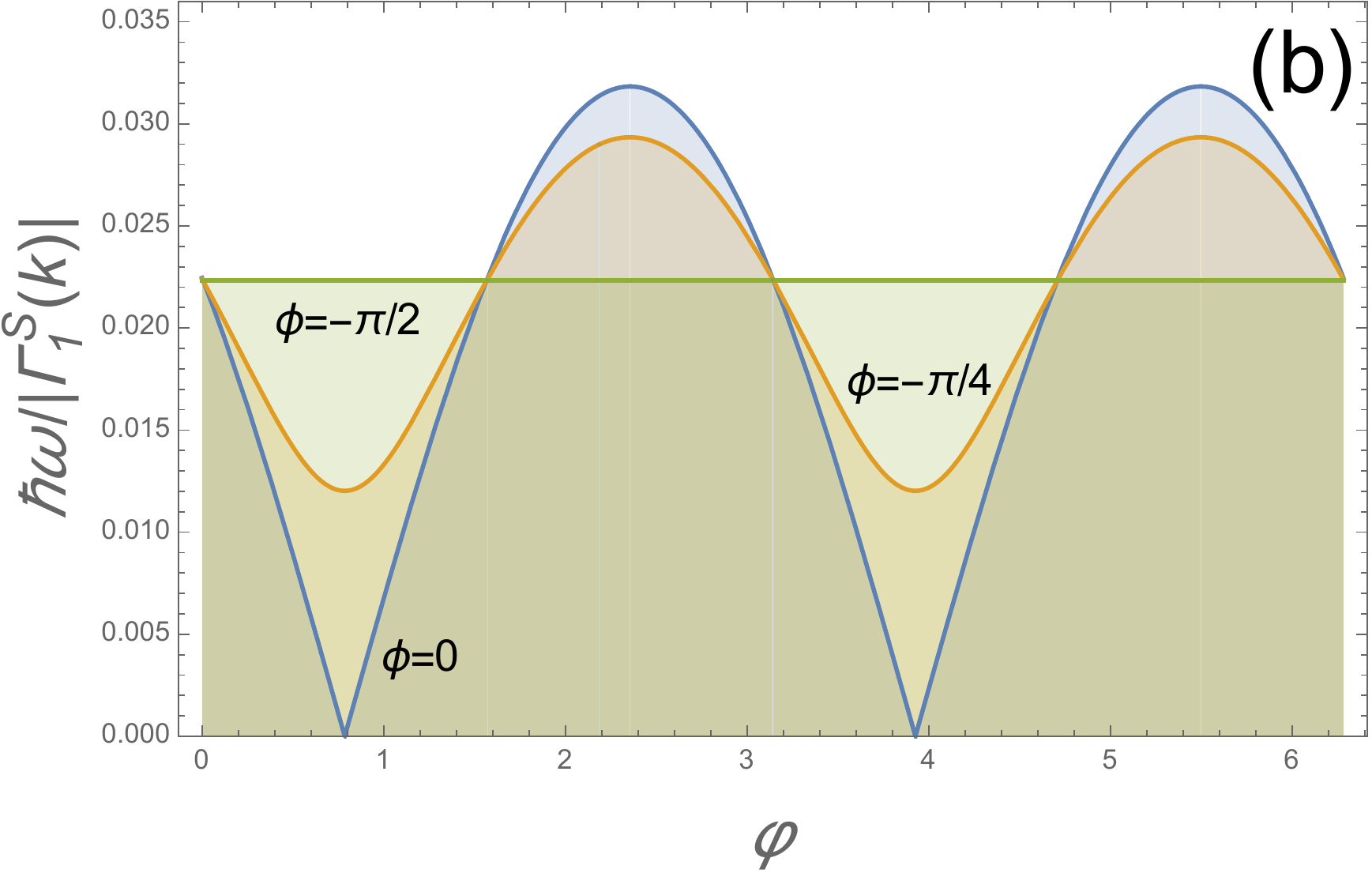}
\includegraphics[width=0.238\textwidth]{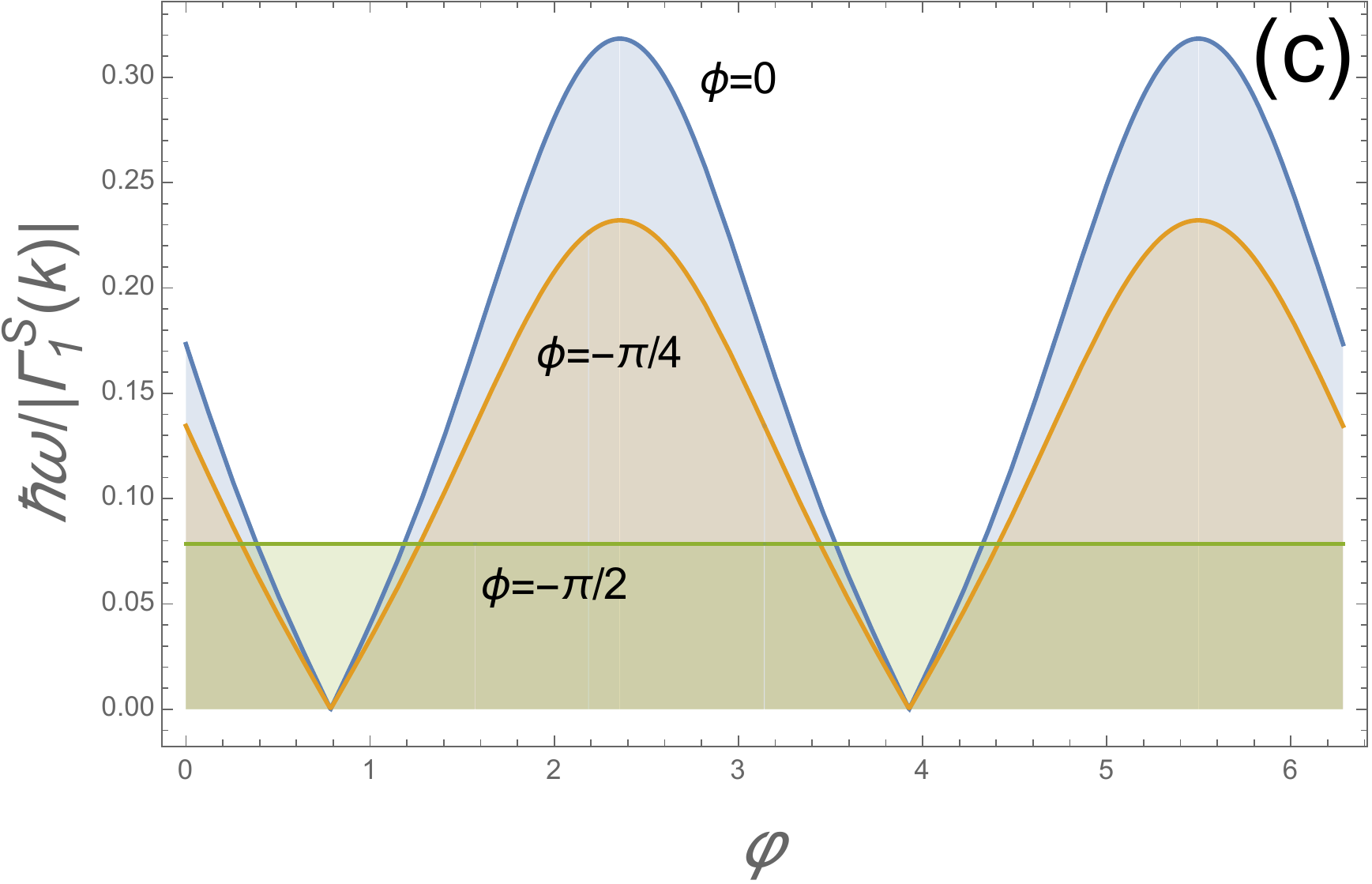}
\includegraphics[width=0.238\textwidth]{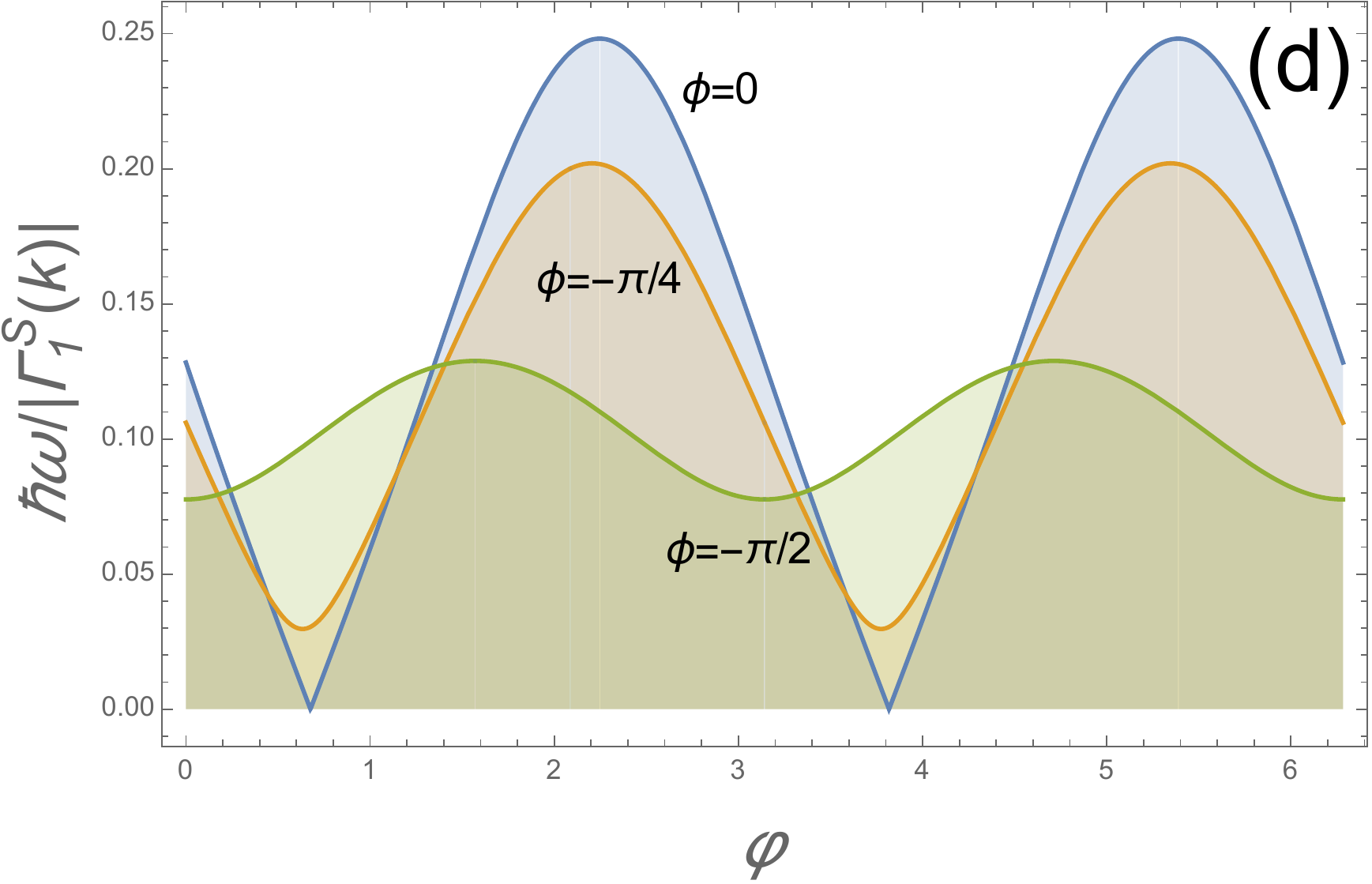}
\caption{
Comparison of the single-photon mode transition time
$\hbar \omega/\left\vert \Gamma_{1}^S(\boldsymbol{k})\right\vert$
as a function of the momentum direction given by the angle $\varphi$
for graphene and borophene-like
materials in the weak and strong-field regimes.
(a) $v_t=0$, $v_x=v_F$, $v_y=v_F$
(graphenel)
and $E_x e v_F/\hbar \omega^2=E_y e v_F/\hbar \omega^2=0.1/\sqrt{2}$
(weak-field regime);
(b) $v_t=0.32 v_F$, $v_x=v_F$, $v_y=v_F$
(borophene) and
$E_x e v_F/\hbar \omega^2=E_y e v_F/\hbar \omega^2=0.1/\sqrt{2}$
(weak-field regime);
(c) $v_t=0$, $v_x=v_F$, $v_y=v_F$
(graphene)
and $E_x e v_F/\hbar \omega^2=E_y e v_F/\hbar \omega^2=1/\sqrt{2}$
(strong-field regime);
(d) $v_t=0.32 v_F$, $v_x=v_F$, $v_y=v_F$
(borophene) and
$E_x e v_F/\hbar \omega^2=E_y e v_F/\hbar \omega^2=1/\sqrt{2}$
(strong-field regime).
Three polarization cases are presented:
$\phi =0$ (linear polarization),
$\phi=-\pi/2$ (circular polarization) and
$\phi=-\pi/4$ (elliptical polarization). 
}\label{figure4}
\end{figure}

%
%
\begin{figure}
  \includegraphics[width=0.28\textwidth]{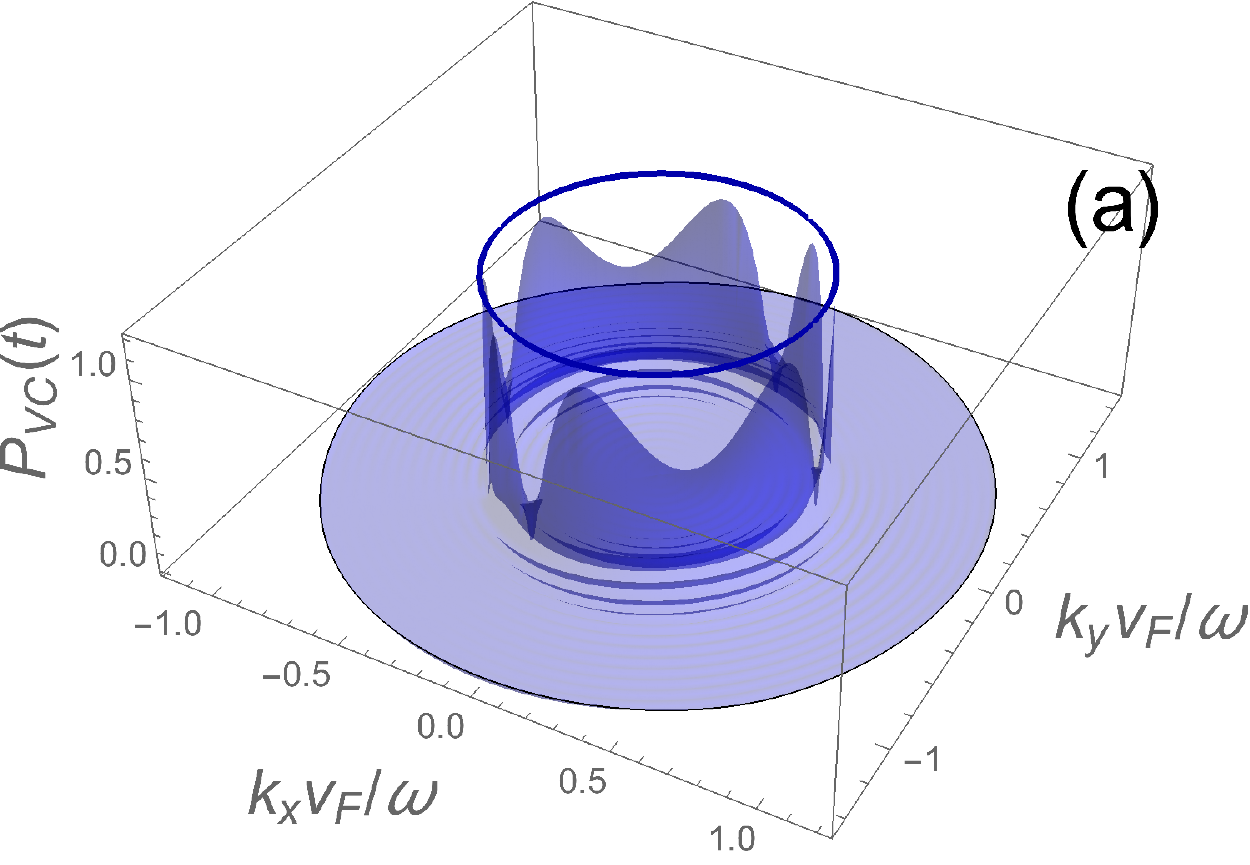}
  \includegraphics[width=0.18\textwidth]{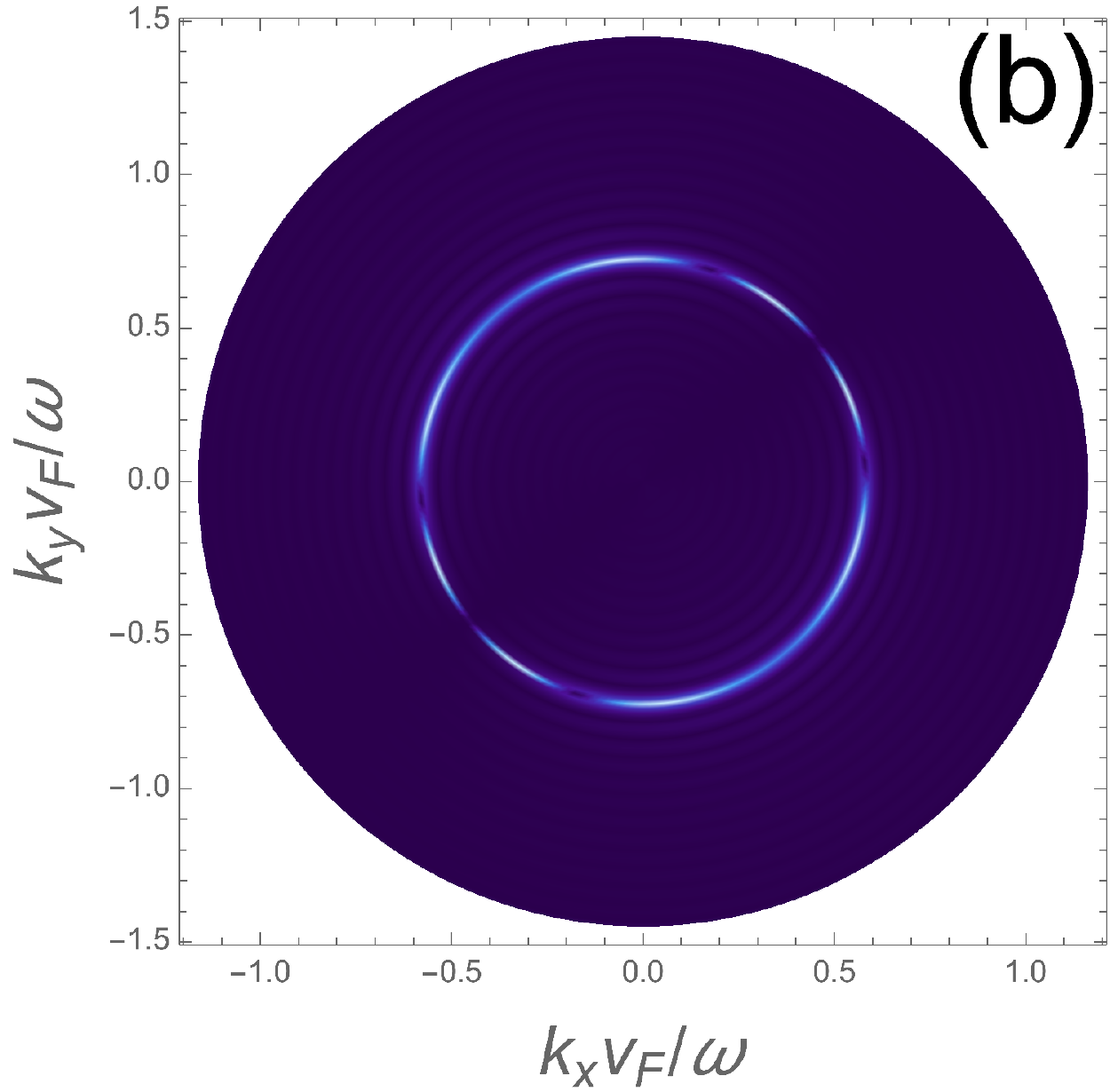}
  \caption{(a) Three-dimensional and (b) density plots  of the transition probability $P_{CV}(t)$ as a function
  of the momentum components $k_x$ and $k_y$ under linearly polarized light
  ($\phi=0$) in the weak-field regime
  ($E_x e v_F/\hbar \omega^2=E_y e v_F/\hbar \omega^2=0.01$).
  In panel (a) the solid dark blue elliptical line over the plot marks the zone
  where the single-photon resonant condition
  $2\epsilon(\boldsymbol{k})=\hbar\omega$ is fulfilled.
  }
    \label{figure5}
\end{figure}
%
%
\begin{figure}
  \includegraphics[width=0.28\textwidth]{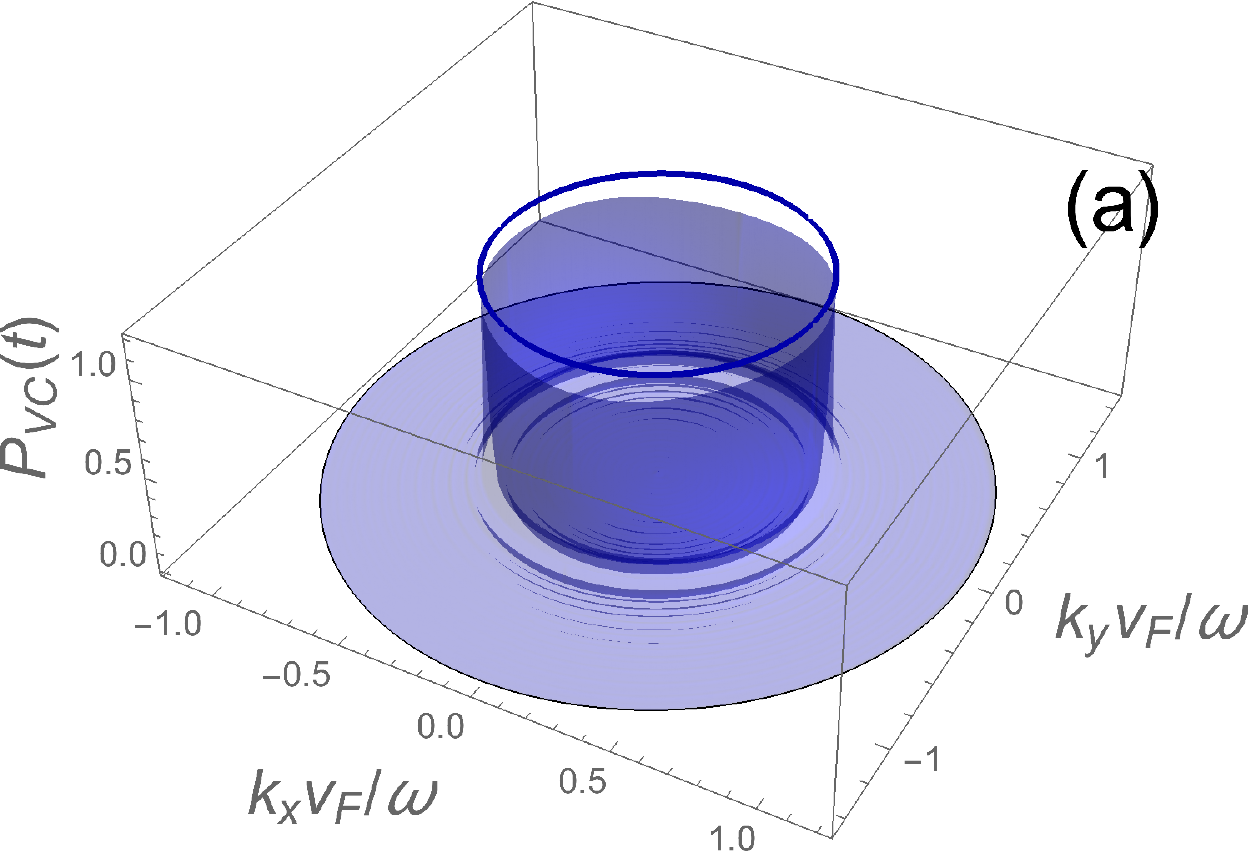}
  \includegraphics[width=0.18\textwidth]{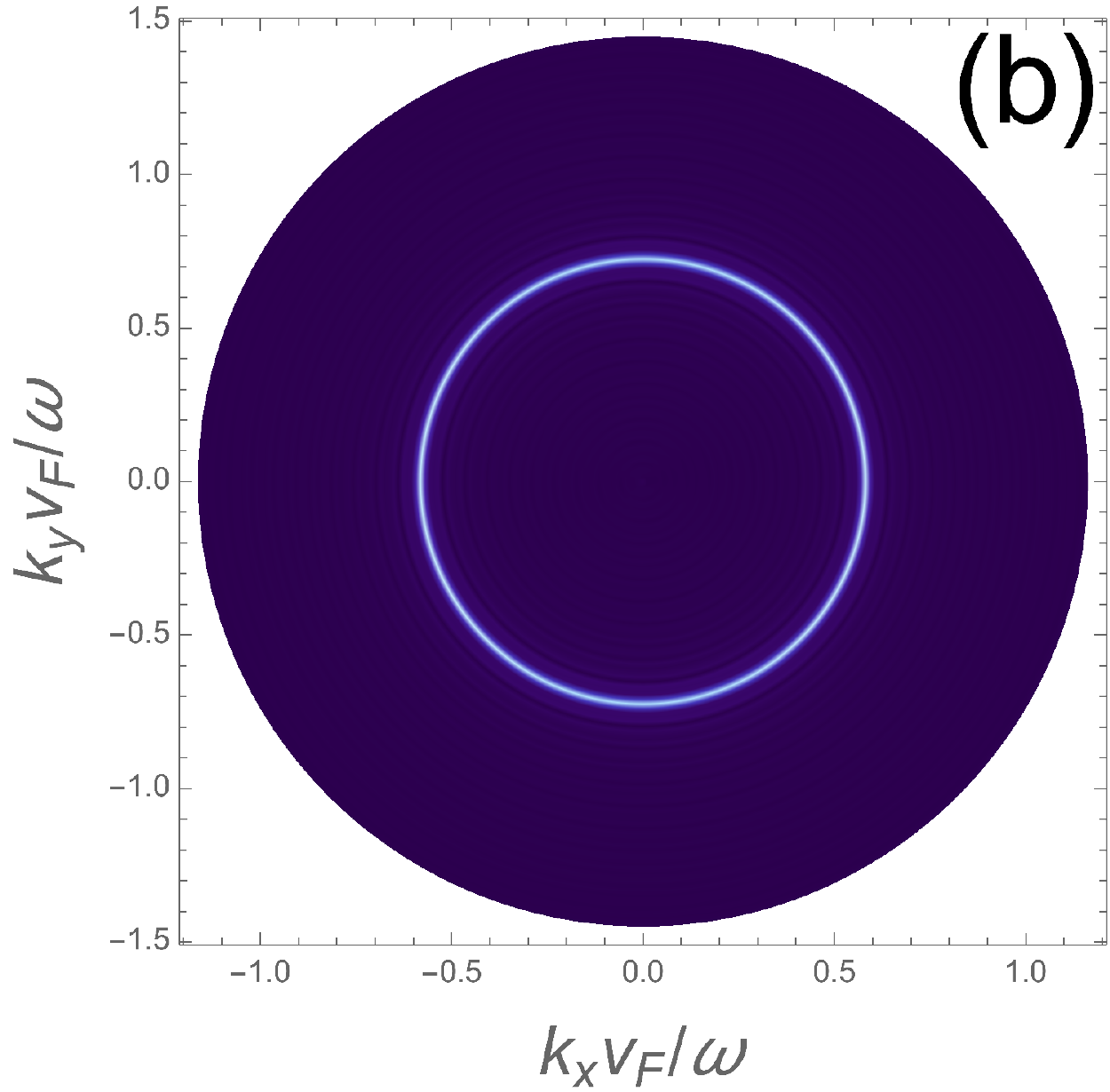}
  \caption{(a) Three-dimensional and (b) density plots
  of the transition probability $P_{CV}(t)$ as a function
  of the momentum components $k_{x,y}$ under circularly polarized light
  ($\phi=-\pi/2$) in the weak-field regime
  ($E_x e v_F/\hbar \omega^2=E_y e v_F/\hbar \omega^2=0.01$).
  In panel (a) the solid dark blue elliptical line over the plot
  marks the zone
  where the single-photon resonant condition
  $2\epsilon(\boldsymbol{k})=\hbar\omega$ is fulfilled.}
    \label{figure6}
\end{figure}
%
%
\begin{figure}
  \includegraphics[width=0.28\textwidth]{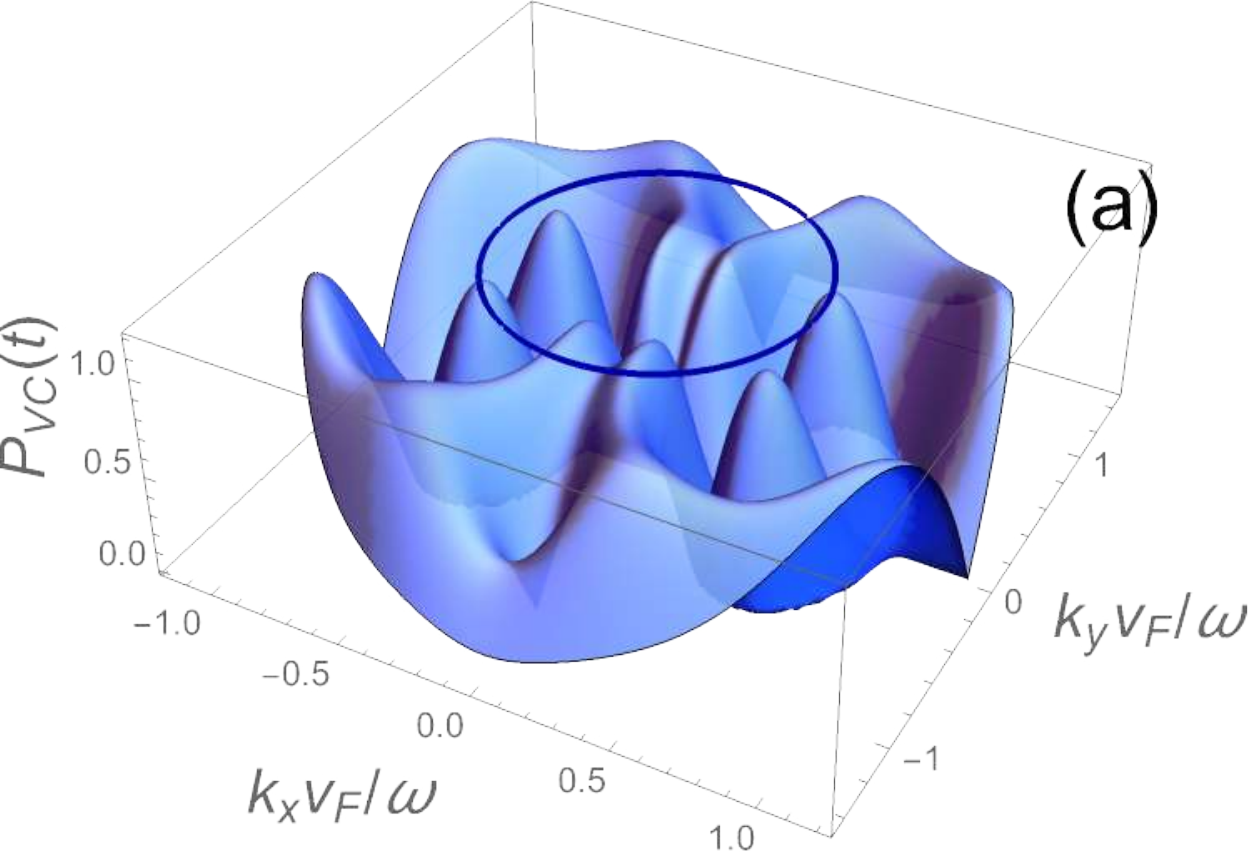}
  \includegraphics[width=0.18\textwidth]{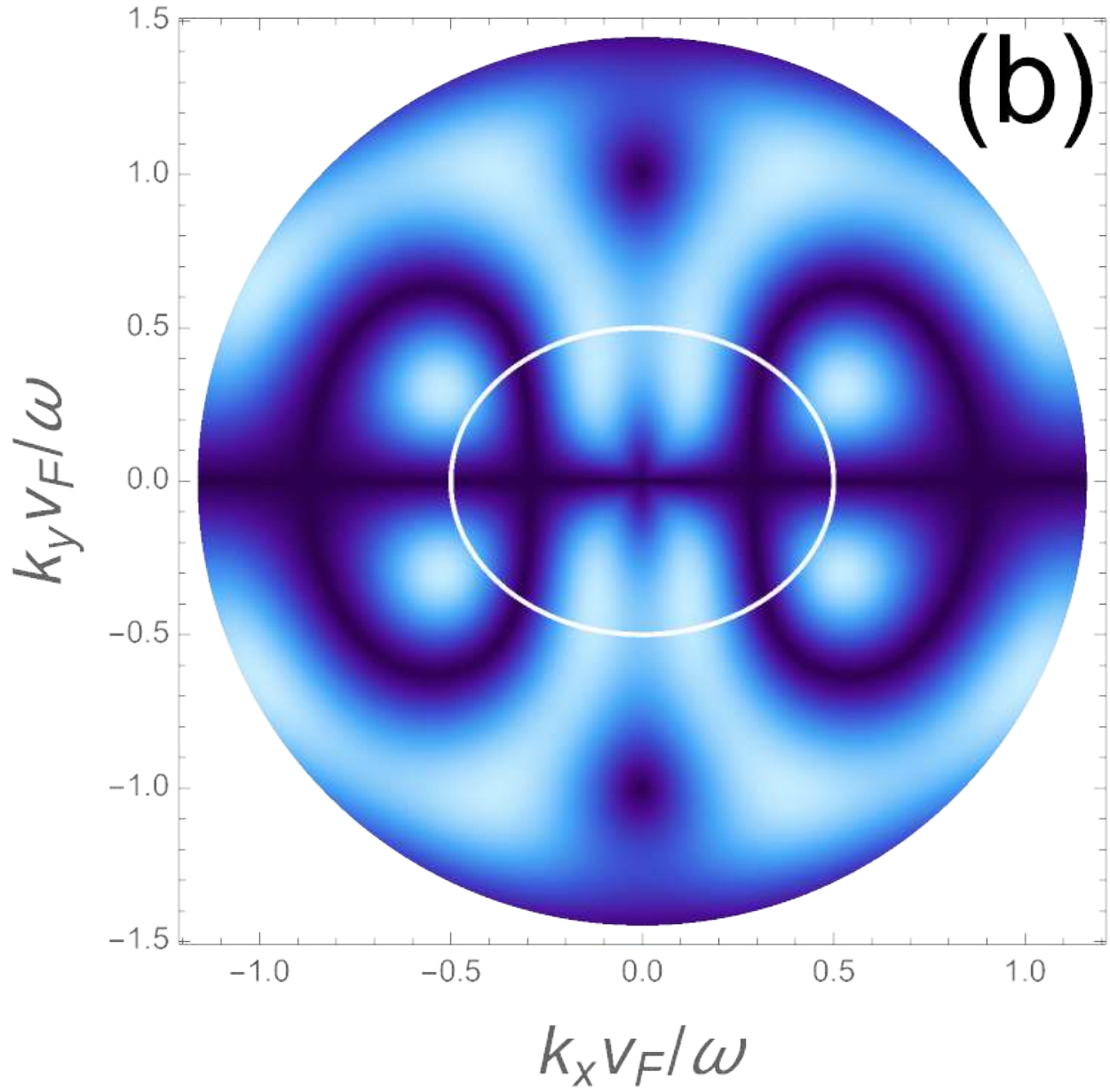}
  \caption{(a) Three-dimensional and (b) density plots
  of the transition probability $P_{CV}(t)$ as a function
  of the momentum components $k_{x,y}$ under linearly polarized light
  ($\phi=0$) in the strong-field regime
  ($E_x e v_F/\hbar \omega^2=1,E_y e v_F/\hbar \omega^2=0$).
 The solid dark blue (a) and white (b) elliptical lines
 mark the zone where the single-photon resonant condition
  $2\epsilon(\boldsymbol{k})=\hbar\omega$ is fulfilled.
  }
    \label{figure7}
\end{figure}

%
%
\begin{figure}
  \includegraphics[width=0.28\textwidth]{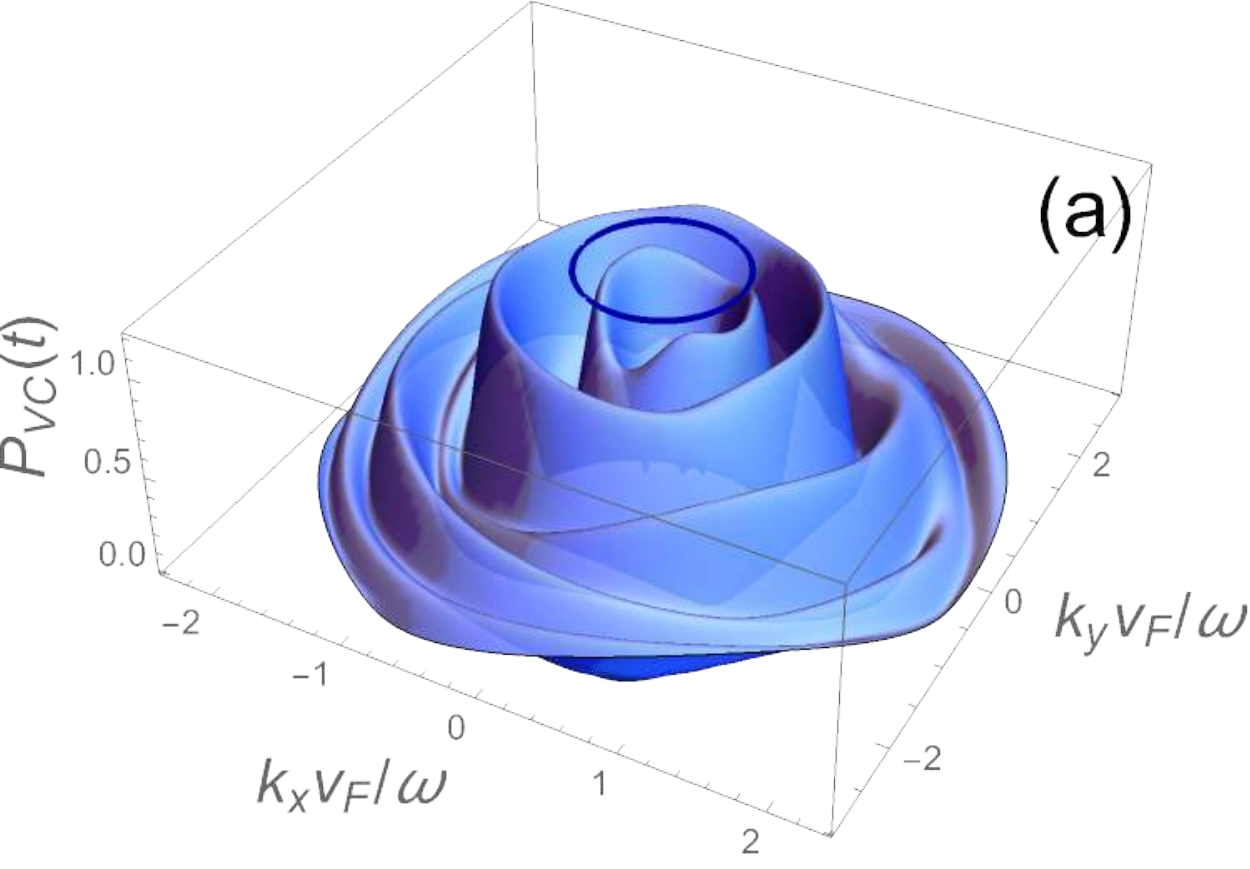}
  \includegraphics[width=0.18\textwidth]{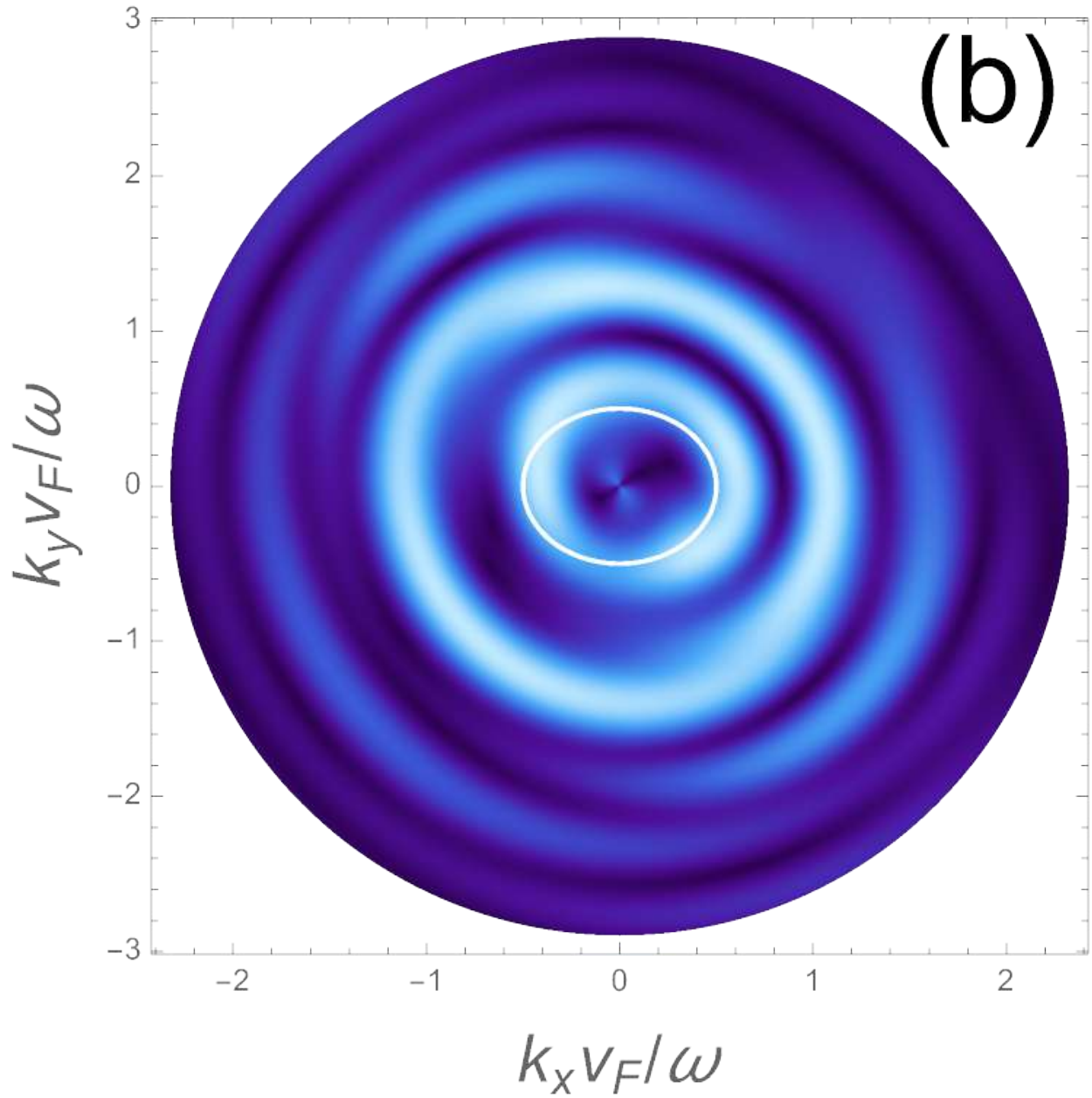}
  \caption{(a) Three-dimensional and (b) density plots
  of the transition probability $P_{CV}(t)$ as a function
  of the momentum components $k_{x,y}$ under circularly polarized light
  ($\phi=-\pi/2$) in the strong-field regime
  ($E_x e v_F/\hbar \omega^2=E_y e v_F/\hbar \omega^2=1/\sqrt{2}$).
 The solid dark blue (a) and white (b) elliptical lines
 mark the zone where the single-photon resonant condition
  $2\epsilon(\boldsymbol{k})=\hbar\omega$ is fulfilled.
  }
    \label{figure8}
\end{figure}

\section{Tilted Anisotropic Dirac Hamiltonian Model}\label{sec:hamiltonian}

The  $\boldsymbol{K}$ valley
low energy Hamiltonian of a 2D tilted anisotropic Dirac
material is given by
\cite{Zabolotskiy2016,verma2017effect,Champo2019,ibarra2019dynamical,HerreraNaumisKubo2019}
\begin{equation}
\mathcal{H}_0= v_t p_y \sigma_0+ v_xp_x\sigma_x+ v_yp_y\sigma_y,
\label{eq:hamp}
\end{equation}
where $\sigma_0$ is the $2\times 2$ identity matrix,
$\sigma_{x,y}$ are the Pauli matrices, $p_{x,y}$ are the components
of the 2D momentum operator $\boldsymbol{p}$ and $v_{x}$, $v_{y}$ are the
anisotropic Fermi velocities. The velocity $v_{t}$ gives the tilting
of the energy dispersion with respect to the energy
axis\cite{Zabolotskiy2016,verma2017effect}.
The well known graphene  
energy dispersion is recovered by setting $v_t=0$ and
$v_x=v_y=v_F\approx c/300\approx10^6\,m/s$
with $v_F$ and $c$ being the Fermi velocity\cite{Zabolotskiy2016}
and the speed of light respectively.
In $8$-$Pmmm$ borophene,
such velocities are $v_t=0.32 v_F$, $v_x=0.86 v_F$ and $v_y=0.69 v_F$.
To simplify the presentation, from here on,
we restrict our calculations to the $\boldsymbol{K}$ valley.
To retrieve the expressions corresponding to
the $\boldsymbol{K}^{\prime}$ valley it suffices to
invert the sing of $v_t$. The sings
of $v_x$ and $v_y$ depend on the chosen basis\cite{verma2017effect}.
The free electron wave function corresponding to \eqref{eq:hamp}
is given by the spinor
\begin{equation}
\Psi_{\eta}(\boldsymbol{r})
=\frac{\exp(i \boldsymbol{k}\cdot\boldsymbol{r})}{\sqrt{2}}
\left(\begin{array}{c}
1 \\ \eta \mathrm{e}^{i\theta(\boldsymbol{k})}\\
\end{array}\right),\label{eq:freepsi}
\end{equation}
where $k_{x}$ and $k_y$ are the wave vector $\boldsymbol{k}$ components,
$\theta(\boldsymbol{k})=\tan^{-1}(v_y k_y/v_x k_x)$
and $\eta=\pm 1$ is the band index.
The eigenenergies associated with this wave function
are
\begin{equation}
    \mathcal{E}_{\eta}(\boldsymbol{k})
=\hbar v_tk_y+\eta\epsilon(\boldsymbol{k}),
\end{equation}
where
\begin{equation}
\epsilon(\boldsymbol{k})=\hbar\sqrt{v_x^2k_x^2+v_y^2k_y^2}.
\label{eq:freepsilon}
\end{equation}
It is important here to underline
that the Hamiltonian matrix \eqref{eq:hamp}, as well as the
wave function vector \eqref{eq:freepsi}, are expanded in
terms of the sublattice state basis.\\

\section{Tilted Anisotropic Dirac Materials under
an elliptically polarized field}\label{sec:elliptical}

Under the action of a
normally incident electromagnetic plane wave,
the Hamiltonian \eqref{eq:hamp} transforms according
to the Peierls substitution
$\boldsymbol{p}\rightarrow \boldsymbol{p}-e\boldsymbol{A}$
as \cite{sakurai1995modern}
\begin{multline}
    \mathcal{H}=v_t\left(p_y-e A_y\right)\sigma_0\\
      +v_x\left(p_x-e A_x\right)\sigma_x
      +v_y\left(p_y-e A_y\right)\sigma_y,
      \label{eq:hamfield}
\end{multline}
where $\boldsymbol{A}=A_x\hat{\boldsymbol{\imath}}+A_y\hat{\boldsymbol{\jmath}}$ is
the radiation vector potential.
It is convenient to adopt a gauge in which
the vector potential only depends on time as
\begin{equation}
    \boldsymbol{A}(t)=-\frac{E_x}{\omega}
    \cos\left(\omega t+\delta\right)\hat{\boldsymbol{\imath}}
       -\frac{E_y}{\omega}
       \cos\left(\omega t+\delta+\phi\right)\hat{\boldsymbol{\jmath}},
       \label{eq:vecpot}
\end{equation}
where $\omega$ is the angular frequency of the radiation.
For $E_x=E_y$ the parameter $\phi$ allows to continuously
vary the field polarization
from linear ($\phi=0$) to circular ($\phi=\pm \pi/2$).
The initial field phase is given by $\delta$.
We drop the initial phase $\delta$ for the moment
and will restore it later by doing $\omega t\rightarrow \omega t+\delta$.
The electric field of the electromagnetic wave is given by
\begin{equation}
\boldsymbol{E}=-\frac{\partial \boldsymbol{A}}{\partial t}
  =-E_x\sin\left(\omega t\right)\hat{\boldsymbol{\imath}}-E_y\sin\left(\omega t+\phi\right)\hat{\boldsymbol{\jmath}}.
\end{equation}

The time-dependent Schr\"odinger equation
that arises from the Hamiltonian \eqref{eq:hamfield},
\begin{equation}
i\hbar\frac{\partial}{\partial t}\Psi(t,\boldsymbol{r})
=\mathcal{H}\Psi(t,\boldsymbol{r}),
\end{equation}
is easily solved by making the ansatz
\begin{equation}
    \Psi(t,\boldsymbol{r})=\exp(i\boldsymbol{k}\cdot\boldsymbol{r})\psi(t).
\end{equation}
This substitution considerably simplifies
the problem by yielding the Schr\"odinger
equation in $\boldsymbol{k}$-space
\begin{equation}
    i\hbar\frac{d}{d t}\psi(t)=H(t)\psi(t),
    \label{eq:schrok}
\end{equation}
where the wave function $\psi(t)$ and the Hamiltonian
\begin{multline}
    H(t)=v_t\left[\hbar k_y-e A_y(t)\right]\sigma_0\\
     +v_x\left[\hbar k_x-e A_x(t)\right]\sigma_x
     +v_y\left[\hbar k_y-e A_y(t)\right]\sigma_y,
     \label{eq:hamk}
\end{multline}
depend entirely on time.

Substituting the particular form
of the vector potential \eqref{eq:vecpot}
into the Hamiltonian \eqref{eq:hamk}
gives
\begin{multline}
    H(t)=v_t\left[\hbar k_y
      +e \frac{E_y}{\omega}\cos(\omega t+\phi)\right]\sigma_0\\
     +v_x\left[\hbar k_x
     +e \frac{E_x}{\omega}\cos(\omega t)\right]\sigma_x\\
     +v_y\left[\hbar k_y
     +e \frac{E_y}{\omega}\cos(\omega t+\phi)\right]\sigma_y.
     \label{eq:hamkexplicit}
\end{multline}
A simple inspection of this Hamiltonian
reveals the limit between the
strong and the weak field regimes.
When the kinetic energy excels the
radio frequency strength, $\hbar k>eE/\omega$
the system is in a weak field regime.
Otherwise, it is in a strong field regime.
In other words, the circle 
with radius
\begin{equation}
    k=eE/\hbar\omega,
\end{equation}
is the boundary between the weak ($k>eE/\hbar\omega$)
and strong field ($k<eE/\hbar\omega$)
regions.\\

\section{Time-evolution, quasienergy-spectrum and the monodromy matrix}
\label{sec:quasienergy}

Only a small number of very restrictive simple cases of Eq. \eqref{eq:schrok}
are exactly solvable. The most general and interesting cases are
only approachable via numerical methods.
The foregoing differential equation
is usually addressed through the Floquet theorem
and the subsequent Fourier time-frequency decomposition
of the periodical part of the solution \cite{Shirley1965Solution,Goldman}.
This approach trades the time-dependent differential
equation \eqref{eq:schrok} by an infinite-dimensional
time-independent Hamiltonian
matrix eigenvalue problem. This method presents two
major drawbacks. First, to determine the quasienergy-spectrum
through the diagonalization of the infinite-dimensional
matrix it has to be chopped up to a required accuracy.
The diagonalization process of such large matrices is
of course numerical.
Second,
recovering the wave function in its spinor form requires
the inverse Fourier transform of a very large dimensional
eigenvector.
Although this last step is not numerical, it relies in
the numerically obtained eigenvectors.
Since the complexity of the problem presses
for the use of numerical methods, instead,
we  compute the monodromy matrix
from the numerical solution
of the system of ordinary differential equations
that stem from the Schr\"odinger equation.
Conveniently, the quasi-states
in the spinor form
are a direct outcome of the previous calculation.
The Schr\"odinger equation for the time evolution operator
is \cite{sakurai1995modern}
\begin{equation}
    i\hbar\frac{d}{d t}\mathcal{U}(t)=H(t)\mathcal{U}(t),
    \label{eq:schroUk}
\end{equation}
where $H(t)$ if given by \eqref{eq:hamk}, and $\mathcal{U}(t)$ is the time
evolution operator. For the sake of simplicity,
we set the initial condition $\mathcal{U}(0)=\boldsymbol{1}$
where $\boldsymbol{1}$ is the $2\times 2$ identity matrix.
Eq. \eqref{eq:schroUk} consists
of four coupled scalar ordinary differential equations
for the elements $U_{ij}(t)$
with initial conditions $U_{ij}(0)=\delta_{ij}$.
Therefore, the solution of \eqref{eq:schroUk}
encodes all the dynamical information of the system.
Moreover, as a result of the Floquet theorem,
the solution in any time interval can be extracted
from the domain $t\in \left[0,T\right]$.
Indeed, due to the periodicity of the Hamiltonian,
$H(t)=H(t+T)$ where $T=2\pi/\omega$,
the evolution operator
must comply with the Floquet theorem.
It states that \cite{verhulst2006nonlinear,Goldman}
\begin{equation}
    \mathcal{U}(t)=\exp\left(-\frac{i}{\hbar}H_e t\right)\mathcal{W}(t),
    \label{eq:evolfloquet}
\end{equation}
where $\mathcal{W}(t+T)=\mathcal{W}(t)$ and $H_e$ is termed the effective Hamiltonian.
The eigenvalues of $H_e$ are precisely the quasienergies of $H(t)$.
Thus, \eqref{eq:evolfloquet} allows us to compute the quasienergies
and the time-dependent wave function $\psi(t)$ for $t\in \left[0,\infty\right)$
provided that the evolution operator is known in the interval
$t\in \left[0,T\right]$.
Let us assume that $U(t)$ is known in the domain $t\in \left[0,T\right]$,
for example,
through the numerical solution of Eq. \eqref{eq:schroUk},
for a given set of $k_{x}$ and $k_y$ values.
Since, by definition
\begin{equation}
\mathcal{U}(T)
=\exp\left(-\frac{i}{\hbar}H_e T\right)\mathcal{W}(T)
=\exp\left(-\frac{i}{\hbar}H_e T\right),
\label{eq:evolopUT}
\end{equation}
the quasienergies are given by
\begin{equation}
    \mathcal{E}_{\eta,j,m}(\boldsymbol{k})
      =-\frac{\hbar\omega}{2\pi} \arg[u_{\eta,j}(\boldsymbol{k})]+m\hbar \omega,
      \label{eq:quasieneries}
\end{equation}
where $u_{\eta,j}(\boldsymbol{k})$ are the two eigenvalues of 
$\mathcal{U}(T)$,
and the subscripts $j=1,2$
and  $m=0,\pm 1, \pm 2,\dots$ tag the band and the Floquet
zone respectively.
In the language of ordinary differential equations theory,
$\mathcal{U}(T)$ is called the monodromy matrix \cite{verhulst2006nonlinear,hale2009ordinary},
its eigenvalues $u_{\eta,j,\boldsymbol{k}}$ are named characteristic multipliers
and $i\arg(u_{\eta,j,\boldsymbol{k}})/T$ are the characteristic exponents.
Meanwhile, the evolution operator at any time is
\begin{equation}
    \mathcal{U}(t)=\mathcal{W}(T)^{\left\lfloor t/T\right\rfloor}
    \mathcal{U}\left(\mathrm{mod}(t,T)\right),
\end{equation}
where $\left\lfloor t/T\right\rfloor=\mathrm{floor}(t/T)$.

\section{Rabi formula}\label{sec:rabiformula}

In many 2D materials, like graphene or borophene,
the conduction and valence bands are symmetric with
respect to the Fermi level. Thus, numerous similarities are shared 
with two-level systems \cite{mojarro2020}. In this section, we 
develop such analogy  for the present model.

In a two-level system,
two important parameters ensue from the Rabi formula
\cite{sakurai1995modern,dittrich1998quantum}:
the detuning parameter $\Delta$
and the characteristic frequency $\Omega$.
The detuning parameter is a measure of how
off is the field frequency from the quantum
two-level system's resonant frequency.
When the field
is tuned to the resonant frequency, i.e. $\Delta=0$,
the system is capable of
transitioning from one-level to the other
with probability equal to 1 after a certain time.
In contrast, when $\Delta \ne 0$, the system
will at best reach a superposition of both states.
The change rate from one state to the other
is given by the characteristic frequency:
a fourth of its period
$\tau=\pi/2\Omega$ is the time elapsed between
the start and the completion of the quantum transition \cite{sakurai1995modern}.
Even though, it is usually proportional to the
quantum system-radiation interaction strength,
we will see further on that
the added complexity of 2D Dirac materials
yields a considerably richer and more intricate behaviour.

Our main goal is now to identify these two Rabi parameters
from the Schr\"odinger equation \eqref{eq:schrok}.
This will allow us to predict the conditions
for producing Rabi cycles and their duration.
We follow a three stage strategy in which we first
move to the conduction-valence band basis, then
remove the time-dependent diagonal elements of the Hamiltonian
and finally adopt
the rotating wave approximation.
This procedure requires a total of four unitary transformations.

In the case studied here, the two-state quantum system
is embodied by the conduction and valence band
states at a given momentum state of the 2D Dirac material.
In particular, we are concerned with the field stimulated transitions
that occur between the valence and conduction band.
Therefore, as a starting point, we wish to express the
Hamiltonian \eqref{eq:hamk} in the basis of the valence 
and conduction band states \eqref{eq:freepsi}. To this
end let us act \eqref{eq:schrok} by the first unitary transformation
\begin{multline}
    \mathcal{R}_1=\frac{1}{\sqrt{2}}\left(
    \begin{array}{cc}
        1 & 1 \\
        \mathrm{e}^{i\theta(\boldsymbol{k})} &
        -\mathrm{e}^{i\theta(\boldsymbol{k})}\\
    \end{array}
    \right)\\
    =\frac{\mathrm{e}^{i\theta(\boldsymbol{k})/2}}{\sqrt{2}}\bigg\{
    \cos\left[\frac{\theta(\boldsymbol{k})}{2}\right]
    \left(\sigma_x+\sigma_z\right)\\
    +\sin\left[\frac{\theta(\boldsymbol{k})}{2}\right]
    \left(\sigma_y-i\sigma_0\right)
    \bigg\}.
\end{multline}
The Schr\"odinger equation for the transformed
evolution operator
$U=\mathcal{R}_1^\dagger \mathcal{U}(t)\mathcal{R}$
in the conduction and valence band base takes the form
\begin{multline}
    0=\mathcal{R}_1^{\dagger}
    \left[i\hbar\frac{d}{dt}- H(t)\right]\mathcal{U}(t)
    \mathcal{R}_1\\
    =\bigg\{i\hbar\frac{d}{dt}-\left[\gamma(\boldsymbol{k})
      +\lambda(\boldsymbol{k})\cos(\omega t+\phi)\right]\sigma_0\\
       -\left[\epsilon(\boldsymbol{k})
    +\alpha(\boldsymbol{k})\cos(\omega t)
    +\beta(\boldsymbol{k})\cos(\omega t+\phi)
    \right]\sigma_z\\
       -i\left[\mu(\boldsymbol{k})\cos(\omega t)
     -\nu(\boldsymbol{k})\cos(\omega t+\phi)\right]
     \left(\sigma_+-\sigma_-\right)
     \bigg\}U(t),
\end{multline}
where the spin ladder operators are defined by
\begin{eqnarray}
\sigma_+ &=& \frac{\sigma_x+i\sigma_y}{2},\\
\sigma_- &=& \frac{\sigma_x-i\sigma_y}{2},
\end{eqnarray}
the $\boldsymbol{k}$-dependent coefficients as,
\begin{eqnarray}
  \alpha(\boldsymbol{k})
    &=&\hbar\frac{ v_x^2k_x e E_x}{\omega\epsilon(\boldsymbol{k})},\\
  \beta(\boldsymbol{k})
    &=&\hbar\frac{ v_y^2k_y eE_y}{\omega\epsilon(\boldsymbol{k})},\\
  \gamma(\boldsymbol{k})
    &=& \hbar v_t k_y,\\
  \lambda(\boldsymbol{k})
    &=&\frac{v_t e E_y}{\omega},\\
  \mu(\boldsymbol{k}) 
    &=& \hbar\frac{ v_xv_yk_y e E_x}{\omega\epsilon(\boldsymbol{k})},\\
   \nu(\boldsymbol{k}) 
    &=& \hbar\frac{ v_xv_yk_x e E_y}{\omega\epsilon(\boldsymbol{k})},
\end{eqnarray}
and $\epsilon(\boldsymbol{k})$ is given by \eqref{eq:freepsilon}.
The next two transformations are devoted to eliminating
the time-dependent elements in the diagonal of
the Schr\"odinger equation.
First, we remove the terms
proportional to $\sigma_0$ by means of applying
\begin{equation}
    \mathcal{R}_2=\exp\bigg[
      -\frac{i}{\hbar}\Xi(\boldsymbol{k},t)\sigma_0\bigg].
\end{equation}
where $\alpha(\boldsymbol{k})$ and $\beta(\boldsymbol{k})$
have been absorbed in the complex number
\begin{equation}
    \Xi(\boldsymbol{k},t)=\gamma(\boldsymbol{k})t
      +\frac{\lambda(\boldsymbol{k})}{\omega}
       \left[\sin(\omega t+\phi)-\sin(\phi)\right].
\end{equation}
Under this transformation, the Schr\"odinger equation
becomes
\begin{multline}
0=
\mathcal{R}_2^{\dagger}
    \left[i\hbar\frac{d}{dt}- H(t)\right]U(t)
   \mathcal{R}_2\\
    =\bigg\{i\hbar\frac{d}{dt}
       -\big[\epsilon(\boldsymbol{k})
    +\alpha(\boldsymbol{k})\cos(\omega t)
    +\beta(\boldsymbol{k})\cos(\omega t+\phi)
    \big]\sigma_z\\
       -i\big[\mu(\boldsymbol{k})\cos(\omega t)
     -\nu(\boldsymbol{k})\cos(\omega t+\phi)\big]\\
     \times\left(\sigma_+-\sigma_-\right)
     \bigg\}U_2(t),
\end{multline}
where $U_2(t)=\mathcal{R}_2^\dagger
U(t)\mathcal{R}_2$ .
In a similar way, the following transformation
lifts the time-dependence from the entries proportional
to $\sigma_z$. This transformation is given by
the following rotation around the $z$ axis
\begin{equation}
       \mathcal{R}_3=\exp\bigg[
      -\frac{i}{\hbar}\Lambda(\boldsymbol{k},t)\sigma_z\bigg].
\end{equation}
The time-dependent transformation parameter is given by
\begin{multline}
    \Lambda(\boldsymbol{k},t)=\frac{1}{\omega}
    \big[\alpha(\boldsymbol{k})\sin(\omega t)\\
    +\beta(\boldsymbol{k})\sin(\omega t+\phi)
    -\beta(\boldsymbol{k})\sin(\phi)
    \big]\\
    =\frac{\chi(\boldsymbol{k})}{\omega}
    \cos\left(\omega t+\zeta(\boldsymbol{k})\right)
    -\frac{\beta(\boldsymbol{k})}{\omega}\sin\left(\phi\right),
\end{multline}
where
\begin{eqnarray}
\chi(\boldsymbol{k})\mathrm{e}^{i\zeta(\boldsymbol{k})}
  =\frac{\alpha(\boldsymbol{k})+\beta(\boldsymbol{k})\mathrm{e}^{i\phi}}{i}.
\end{eqnarray}
The reduced Schr\"odinger equation  takes the form
\begin{multline}
0=
\mathcal{R}_3^{\dagger}\mathcal{R}_2^{\dagger}
    \left[i\hbar\frac{d}{dt}- H(t)\right]U(t)
    \mathcal{R}_2\mathcal{R}_3\\
    =\bigg[i\hbar\frac{d}{dt}
       -\epsilon(\boldsymbol{k})\sigma_z\\
      +\Gamma(\boldsymbol{k},t)\sigma_+
     +\Gamma^*(\boldsymbol{k},t)\sigma_-
     \bigg]U_3(t),\label{eq:schrofam}
\end{multline}
where the transformed
evolution operator is 
$U_3(t)=\mathcal{R}_3^\dagger\mathcal{R}_2^\dagger
U(t)\mathcal{R}_2\mathcal{R}_3$ and
\begin{multline}
    \Gamma(\boldsymbol{k},t)
    =
    -\frac{i}{2}\left[\mu(\boldsymbol{k})\cos(\omega t)
     +\nu(\boldsymbol{k})\cos(\omega t+\phi)\right]\\
     \times
     \exp\left[-\frac{2i\Lambda(\boldsymbol{k},t)}{\hbar}\right].
\end{multline}

At this stage, the Schr\"odinger equation takes
the more familiar form where the
diagonal elements (proportional to $\sigma_z$) are constant
and the time-dependent ones are
secluded to the off-diagonal entries
(proportional to $\sigma_+$ and $\sigma_-$).
We can, therefore, move on to the rotating wave approximation \cite{snoke2020}.
It consists in setting the system in a rotating frame that
revolves around the $z$ axis at an angular velocity
$\omega$ keeping the constant terms and discarding those
that rapidly oscillate  \cite{snoke2020}.
In the standard analysis of Rabi oscillations,
the radio frequency field may be decomposed into the two
circular components. One rotates in the same direction
as the rotating frame and the other in the
opposite direction\cite{dubbers2013quantum} .
While the component rotating synchronously with the frame becomes
constant, the other  rapidly oscillates at an angular
frequency $2\omega$.
The last one is neglected as it is quickly averaged over time
yielding an approximated constant Hamiltonian.
This approximation is justified as long as the
static field is larger than
the oscillating one\cite{dubbers2013quantum}. 

In our case, however, a fundamental difference surfaces
when we look closely at the parameter $\Gamma(\boldsymbol{k},t)$.
Using the Jacobi-Anger expansion
\begin{equation}
    \exp(i z \cos\varphi)=\sum_{n=-\infty}^{\infty} i^nJ_n(z)\exp(in\varphi),
\end{equation}
we obtain the following Fourier series:
\begin{multline}
    \Gamma(\boldsymbol{k},t)
    =
     \frac{i}{2}
     \exp\left[-\frac{2i\beta(\boldsymbol{k})}{\hbar\omega}\sin\phi\right]\\
     \times\sum_{n=-\infty}^{\infty}\,\, i^n
      \exp\left[i n \zeta(\boldsymbol{k})\right]
     J_n\left[\frac{2\chi(\boldsymbol{k})}{\hbar\omega}\right]
     \\
     \times\bigg\{
     \left[\nu(\boldsymbol{k})\mathrm{e}^{i\phi}
       -\mu(\boldsymbol{k})\right]
     \exp\left[i(n+1)\left(\omega t+\delta\right)\right]\\
     +\left[\nu(\boldsymbol{k})\mathrm{e}^{-i\phi}
       -\mu(\boldsymbol{k})\right]
      \exp\left[i(n-1)\left(\omega t+\delta\right)\right]
     \bigg\},\label{eq:fouriergamma}
\end{multline}
 where in the previous expression we have
restored the initial field phase $\delta$.

In contrast with the classical treatment of Rabi oscillations,
where only one oscillating mode is present,
here, the Hamiltonian is composed of an infinite number of
Fourier modes. This is our first evidence
as to why multi-photon modes are generated in Dirac materials
while in standard Rabi systems only the single-photon
one is present.
We can nevertheless follow the usual prescription of the
rotating wave approximation with a slight modification.
Rotating around the $z$ axis by a phase $-q \omega t$,
where $q\in \mathbb{Z}$, we tune in with each one
of the Fourier modes.
This is accomplished through the rotation operator
\begin{equation}
    \mathcal{R}_4=\exp\bigg(
      -\frac{i}{2}q\omega t\sigma_z\bigg).
\end{equation}
In the rotating frame
the Schr\"odinger equation \eqref{eq:schrofam}
 takes the form
\begin{multline}
0=\mathcal{R}_4^{\dagger}
\mathcal{R}_3^{\dagger}\mathcal{R}_2^{\dagger}\mathcal{R}_1^{\dagger}
    \left[i\hbar\frac{d}{dt}- H(t)\right]U(t)
    \mathcal{R}_1\mathcal{R}_2\mathcal{R}_3\mathcal{R}_4\\
    =\bigg\{i\hbar\frac{d}{dt}
       -\left[\epsilon(\boldsymbol{k})-\frac{q\hbar \omega}{2}\right]\sigma_z\\
      +\Gamma(\boldsymbol{k},t)\mathrm{e}^{iq\omega t}\sigma_+
     +\Gamma^*(\boldsymbol{k},t)\mathrm{e}^{-iq\omega t}\sigma_-
     \bigg\}U_4(t).\label{eq:schrofin}
\end{multline}
where, as before, the transformed evolution operator is
\begin{equation}
    U_4(t)=\mathcal{R}_4^{\dagger}
\mathcal{R}_3^{\dagger}\mathcal{R}_2^{\dagger}
U(t)\mathcal{R}_2\mathcal{R}_3\mathcal{R}_4.\label{eq:u4}
\end{equation}
By the Fourier expansion \eqref{eq:fouriergamma}
we readily identify the synchronous terms
\begin{multline}
    \Gamma_{q}^S(\boldsymbol{k})=
    \frac{i^q}{2}\mathrm{e}^{-iq[\delta+\zeta(\boldsymbol{k})]}
    \exp\left[-\frac{2i\beta(\boldsymbol{k})}{\hbar\omega}\sin\phi\right]\\
    \times
     \bigg\{
      \mathrm{e}^{i\zeta(\boldsymbol{k})}
     J_{q-1}\left[\frac{2\chi(\boldsymbol{k})}{\hbar\omega}\right]
     \left[\nu(\boldsymbol{k})\mathrm{e}^{-i\phi}-\mu(\boldsymbol{k})\right]\\
     -\mathrm{e}^{-i\zeta(\boldsymbol{k})}
     J_{q+1}\left[\frac{2\chi(\boldsymbol{k})}{\hbar\omega}\right]
     \left[\nu(\boldsymbol{k})\mathrm{e}^{i\phi}-\mu(\boldsymbol{k})\right]
     \bigg\}.
     \label{eq:synchronous}
\end{multline}
Neglecting all the components rotating with higher frequencies,
we end up with a static equation,
\begin{multline}
0=\bigg\{i\hbar\frac{d}{dt}
       -\left[\epsilon(\boldsymbol{k})-\frac{q\hbar \omega}{2}\right]\sigma_z\\
      +\Gamma_{q}^{S}(\boldsymbol{k})\sigma_+
     +\Gamma_{q}^{S*}(\boldsymbol{k})\sigma_-
     \bigg\}U_4(t).\label{eq:schrostatic}
\end{multline}
In order for this approximation
to be valid, the static entries [$\epsilon(\boldsymbol{k})$]
must be larger than the oscillating ones
[$\Gamma_{q}^{S}(\boldsymbol{k})$], thus
\begin{equation}
    \epsilon(\boldsymbol{k})
    \gg
    \left\vert \Gamma_{q}^{S}(\boldsymbol{k}) \right\vert.
\end{equation}
The formal solution of \eqref{eq:schrostatic} is
\begin{multline}
    U_4(t)=\exp\bigg\{
    \frac{it}{\hbar}\left[\epsilon(\boldsymbol{k})
    -\frac{q\hbar \omega}{2}\right]\sigma_z\\
    -\frac{it}{\hbar}\left[\Gamma_{q}^{S}(\boldsymbol{k})\sigma_+
     +\Gamma_{q}^{S*}(\boldsymbol{k})\sigma_-\right]
    \bigg\}\\
  =\cos[\Omega_{q}(\boldsymbol{k}) t]\sigma_0
  +\frac{i}{\hbar\Omega_{q}(\boldsymbol{k})}
  \sin[\Omega_{q}(\boldsymbol{k}) t]\\
  \times\big[
  \Delta_{q}(\boldsymbol{k})\sigma_z
       -\Gamma_{q}^{S}(\boldsymbol{k})\sigma_+
       -\Gamma_{q}^{S*}(\boldsymbol{k})\sigma_-
  \big],
\end{multline}
where
\begin{eqnarray}
 \Delta_{q}(\boldsymbol{k}) &=& \epsilon(\boldsymbol{k})
    -\frac{q\hbar\omega}{2},\\
\Omega_{q}(\boldsymbol{k}) &=& \frac{1}{\hbar}\sqrt{\Delta_{q}^2(\boldsymbol{k})
    +\left\vert\Gamma_{q}^S(\boldsymbol{k})\right\vert^2},
\end{eqnarray}
As one of the last steps, we have to work back the solution
for the actual evolution operator from \eqref{eq:u4}
\begin{multline}
    U(t)=\mathcal{R}_2\mathcal{R}_3\mathcal{R}_4
        U_4(t)\mathcal{R}_4^{\dagger}\mathcal{R}_3^{\dagger}
          \mathcal{R}_2^{\dagger}\\
   =\cos[\Omega_{q}(\boldsymbol{k}) t]\sigma_0
    +\frac{i}{\hbar \Omega_{q}(\boldsymbol{k})}
  \sin[\Omega_{q}(\boldsymbol{k}) t]\\
  \times\bigg\{\Delta_{q}(\boldsymbol{k})\sigma_z
       -\exp\left[-\frac{2i\Lambda(\boldsymbol{k},t)}{\hbar}-iq\omega t\right]
      \Gamma_{q}^S(\boldsymbol{k})\sigma_+\\
       -\exp\left[\frac{2i\Lambda(\boldsymbol{k},t)}{\hbar}+iq\omega t\right]
       \Gamma_{q}^{S*}(\boldsymbol{k})\sigma_-
  \bigg\}.
\end{multline}
Finally, the probability of transitioning from the conduction
to the valence band is
\begin{equation}
    P_{CV}(t)=\left\vert\psi_V^\dagger U(t)\psi_C\right\vert^2\\
    =\frac{\left\vert\Gamma_{q}^S(\boldsymbol{k})\right\vert^2}
    {\left[\hbar \Omega_{q}(\boldsymbol{k})\right]^2}
    \sin^2\left[\Omega_{q}^2(\boldsymbol{k})t\right],
    \label{eq:rabiformula}
\end{equation}
where, in the conduction-valence band base we have
\begin{equation}
\psi_{C}=(1,0)^\top \,\, , \,\,\,\,\,\,\,\,\,\,
\psi_{V}=(0,1)^\top.
\end{equation}
Equation \eqref{eq:rabiformula} is
the Rabi equation. From it we can infer
that the detuning parameter is
$\Delta_{q}(\boldsymbol{k})=\epsilon(\boldsymbol{k})-q\hbar\omega/2$.
It is hardly surprising that the maximum transition
probability amplitude is attained when
$\Delta_{q}(\boldsymbol{k})=0$ or equivalently when
$\Omega_{q}(\boldsymbol{k})=\vert\Gamma_{q}^S(\boldsymbol{k})\vert$.
This simply means that, in the $k_x-k_y$ plane, full transitions
between the valence and  conduction bands
may occur between points having the same $\boldsymbol{k}$ values
and
whose energy difference complies with
\begin{equation}
2\epsilon(\boldsymbol{k})=q\hbar\omega.
\label{eq:resonancecond}
\end{equation}
We call this the resonance condition.
One can thus picture electron transitions
as occurring vertically from one ellipse in the valence
band to another one in the conduction band.

The truly significant behaviours
arise from the convoluted structure of the transition time
\begin{equation}
    \tau_{q}(\boldsymbol{k})
    =\frac{\pi\hbar}
      {2\sqrt{\left[\epsilon(\boldsymbol{k})-\frac{q\hbar \omega}{2}\right]^2
      +\left\vert\Gamma_{q}^S(\boldsymbol{k})\right\vert ^2}}.
\end{equation}
Despite the highly symmetrical condition imposed
by the detuning parameter, the transition
time is able to introduce a high degree of anisotropy
as we will show later on.
If the resonance condition \eqref{eq:resonancecond}
is met
\begin{equation}
\tau_{q}(\boldsymbol{k})=\frac{\pi\hbar}
  {\left\vert\Gamma_{q}^S(\boldsymbol{k})\right\vert}.
\end{equation}
This relation deserves special attention since it
establishes a connection between the intensity of the electric field
and multi-photon processes.
When the rotating frame synchronizes with the
$\omega$ ($q=1$) mode, one of the terms of
$\Gamma_{q}^S(\boldsymbol{k})$ is proportional
to the Bessel function of zeroth order
$J_0[2\chi(\boldsymbol{k})/\hbar\omega]$.
This is the only integer order Bessel function
that does not vanish upon being evaluated at zero.
Since $\chi(\boldsymbol{k})$ is proportional
to the magnitude of the electric field,
$\vert \Gamma_{1}^S(\boldsymbol{k})\vert$ ($q=1$)
is the dominant synchronous entry in Eq. \eqref{eq:schrofin}
for low electromagnetic fields.
As the intensity of the electromagnetic field
rises, it activates synchronous terms with larger $q$
that involve more photon interactions.
These are proportional to Bessel functions
of higher order that vanish when evaluated at zero,
and therefore come in to play for large values
of the amplitude of the electric field.
In virtue of the resonance condition \eqref{eq:resonancecond},
a high intensity light pulse might hit
multi-photon resonances ($q>1$) aside from the
single-photon one, that is always triggered regardless of
the field strength.

\section{Numerical Results}\label{sec:results}

The quasienergies of
Hamiltonian \eqref{eq:hamk}
are computed 
through Eq. \eqref{eq:quasieneries}.
By numerically solving the four differential
equations that arise from \eqref{eq:schroUk},
the evolution operator  $\mathcal{U}(t)$
matrix elements are determined
in the time
interval $t\in \left[0,T\right]$.
These, in turn, are used to compute the eigenvalues
$u_{\eta,j}(\boldsymbol{k})$
of $\mathcal{U}(T)$ that enter Eq. \eqref{eq:quasieneries}.
Through these calculations, 
we have analyzed three cases: vanishing,
linearly polarized and circularly polarized intense fields.

Figure \ref{figure1} shows the zero-field quasi-spectrum.
For the sake of simplicity, in this figure and the
ones that follow we have 
removed the tilting by plotting
$\Gamma_{q}^S(\boldsymbol{k})-\hbar v_t k_y$.
Although most readers familiar with
Dirac materials and Floquet theory will
immediately recognize the zero-field quasi-spectrum
in this figure, it is worthwhile
reviewing it in order to be able to make a proper comparison
with the two other cases.
If the field vanishes one can either obtain
the energy or the quasienergy-spectrum given
that, under such conditions, the Hamiltonian is time-independent.
However, when a periodical time-varying electromagnetic field
is present,
the notion of energy is rendered meaningless and
we can only speak of quasienergies.
Even though the energy spectrum in vanisghing fields
is widely known, the quasienergy requires
some further consideration.
It is comprised of two
valence and conduction band cones whose tips touch
in the Dirac point as can be seen
in Fig. \eqref{figure1} (a) . Both cones extend to infinity
in opposite directions in
$\boldsymbol{k}$-space. By definition, the quasienergy is a phase
and, as such, it is only unambiguously defined in
$0\le \mathcal{E}_{\eta,j,m}(\boldsymbol{k})\le \hbar\omega$.
Any quasienergy value lying outside will thus be
echoed in the same range.
As we mentioned earlier, we call this range Floquet zone.
These repetitions are clearly seen in
Fig. \ref{figure1} (b) where the
quasienergy spectrum is plotted as a function
of the momentum components $k_{x,y}$ for zero-field case.
The two bands $j=1$ (transparent green) and
$j=2$ (solid blue) of the first Floquet band ($m=0$)
are shown in Fig. \ref{figure1} (a).
We readily spot the tips of both bands' Dirac cones
touching in $k_x=k_y=0$. As both cones extend and
reach the edges of the Floquet zone, they reemerge
in the opposite side as part of the other band
($m=1 \rightarrow m=2$ or $m=2\rightarrow m=1$).
In this manner, the multiple sections of the
Dirac cone are arranged concentrically around the
tip of the Dirac cone.
Panel (c) shows
the quasi-spectrum density plot.
The Dirac point is located in the center of
the dark elliptical zone ($k_x=k_y=0$) and
the contours of the cylindrical sections that
reach the upper boundary of the Floquet zone
are depicted as bright lines.
Figures \eqref{figure1} (d) and (e) show
cross sections of the quasienergy for $k_y=0$
and $k_x=0$ respectively. In these figures we
clearly see the multiple sections
of the cones switching from one band to the other
as they cross from one
Floquet zone ($m=-1,0,1$) to the next.

The effects of an linearly polarized electromagnetic field
in the quasienergy-spectrum as a function
of the momentum components $k_x$ and $k_y$ can be viewed
in Fig. \ref{figure2}. The polarization
of the electromagnetic field is along the $x$ direction
($E_x=2 \hbar\omega^2/ev_F$ and $Ey=0$).
In panel (a)
we have plotted the quasienergies of the
two bands ($j=1$ and $j=2$). In panel (b) we
can appreciate further details in the
quasienergy density plot of the first band ($j=1$).
We immediately notice
a striking difference with the zero field quasi-spectrum:
the concentric circles are replaced by a grid pattern.
The grid lines are oriented along the direction
perpendicular to the field polarization and
a complex pattern of new Dirac-like points
emerges around the original one.
The quasienergy spectrum cross sections at $k_y=0$
and $k_x=0$ are shown in panels (d) and (e).
While the $k_y=0$ cross section seems
unaltered by the presence of the electromagnetic radiation,
the quasi-spectrum is significantly elongated
in the $k_x=0$ projection.
This can be attributed to a renormalization
of the material parameters, mainly the $v_x$ and
$v_y$ velocities in the vicinity of the original Dirac
point an the newly formed ones.
The deformation of the spectrum may be understood
as the result of electronic dressing \cite{kibis2018electromagnetic}.
It is important to underline that, even though
the spectrum has been considerably deformed,
the original Dirac point is preserved under the
action of a linearly polarized field.
This same behaviour has been observed by us
by a completely different mathematical approach
that relies in the Fourier spectral decomposition
of the wave function\cite{sandoval2020floquet}.

The boundary between the weak ($k>eE/\hbar\omega$)
and strong field ($k<eE/\hbar\omega$)
regimes is indicated
with a solid white line in Figs. \eqref{figure2}
(a), (b) and (c). In Figs. \eqref{figure2} (d) and (e)
the edges of the circle are marked
vertical dotted black lines. It can be seen
that the deformation of the quasienergy-spectrum
with respect to the zero field case
is always restricted to the strong field region.

Figure \ref{figure3} exhibits
the quasienergy-spectrum under
a circularly polarized illumination.
The fundamental difference
with the spectrum
of linearly polarized radiation
is that for circularly polarized light
a Gap opens up in the Dirac point.
Similarly as under linearly polarized light,
the spectrum is strongly distorted
within the strong field region
$k<eE/\hbar\omega$, however, the
distortion has cylindrical symmetry.
In the same manner as for the
linearly polarized field, in Figs. \ref{figure3}
(a), (b) and (c) the white circle
indicates the boundary between
the strong and the weak field regimes.
Likewise, in Figs. \ref{figure3} (d) and (e)
the vertical dotted black lines
mark the edges between both regions.

\section{Discussion}\label{Discussion}
Now we would like to establish a link
between the structure of the
spectrum and the transition probability
under pulsed illumination.
Before proceeding further,
it is instructive to examine
the behaviour of the transition time
$\pi\hbar/\left\vert \Gamma_{1}^S (\boldsymbol{k})\right\vert$
subject to a resonant field ($\hbar\omega=2\epsilon(\boldsymbol{k})$).
A suitable parametrization of the momentum vector
under these conditions is
\begin{equation}
    \boldsymbol{k}=\frac{\omega}{2v_x}\cos\varphi\hat{\boldsymbol{\imath}}+\frac{\omega}{2v_y}\sin\varphi\hat{\boldsymbol{\jmath}},
\end{equation}
where the parameter $\varphi$ is a directional angle
in the $k_x-k_y$-plane.
Figure \ref{figure4} presents plots of the
single-photon mode
transition time $\pi\hbar/\vert \Gamma_{1}^S(\boldsymbol{k})\vert$
as a function of the parameter $\varphi$ for
graphene ($v_t=0$ and $v_x=v_y$) [(a) and (c)]
and borophene-like ($v_t \ne 0$ and $v_x\ne v_y$)
[(b) and (d)] materials submitted
to weak and strong fields.
The rotational symmetry of graphene
provides a reference point to discriminate
if the transition time behaviour originates in the anisotropic
velocities or the field polarization.
In these figures we can appreciate that while
circularly polarized light has no directional effects,
in highly symmetrical materials,
linearly polarized field consistently exhibits variations in the
transition time regardless of the field intensity.
We also infer that small variations of the
transition time are expected for anisotropic materials
under circularly polarized
illumination. Much larger fluctuations are, however, expected
under linearly polarized light.

Having examined the behaviour of the resonant transition time
we now turn to the analysis of the transition probability
of carriers under a pulsed light excitation.
To this end we use the vector potential
of a square pulse 
\begin{multline}
    \boldsymbol{A}(t)=-\frac{E_x}{\omega}
    \cos\left(\omega t+\delta\right)
      \Theta[\tau_{1}(\boldsymbol{k}),t-t_0]
         \hat{\boldsymbol{\imath}}\\
       -\frac{E_y}{\omega}
       \cos\left(\omega t+\delta+\phi\right)
       \Theta[\tau_{1}(\boldsymbol{k}),t-t_0]
       \hat{\boldsymbol{\jmath}},
       \label{eq:vecpotstep}
\end{multline}
where the step function is defined as
\begin{equation}
     \Theta[\tau_{1}(\boldsymbol{k}),t]
    =\begin{cases}
      1, \,\, 0\le t\le \tau_{1}(\boldsymbol{k}),\\
      0, \,\, 0> t>\tau_{1}(\boldsymbol{k}).
    \end{cases}
\end{equation}
Plugging this expression for the vector potential into
the Schr\"odinger equation \eqref{eq:schroUk}
and numerically calculating the evolution operator $\mathcal{U}(t)$
we readily obtain the transition probability
\begin{equation}
    P_{CV}(t)=\left\vert\psi_C^\dagger\, \mathcal{U}(t)\,\psi_V\right\vert^2,
\end{equation}
where the valence and conduction band states are expressed
in the sublattice state basis as
\begin{align}
   & \psi_V=\frac{1}{\sqrt{2}}\left(1
         ,-\mathrm{e}^{i\theta(\boldsymbol{k})}\right)^\top, &
    \psi_C=\frac{1}{\sqrt{2}}\left(1
         ,\mathrm{e}^{i\theta(\boldsymbol{k})}\right)^\top.
\end{align}

The duration of the pulse matches the
single-photon
resonant transition time $\tau_{1,\boldsymbol{k}}$
for a momentum state characterized by $\boldsymbol{k}$.
After the pulse, the wave function of the system will reach a steady state
if it is allowed to evolve for a sufficiently long time
$t>t_0+\tau_{1,\boldsymbol{k}}$.
One would expect that such a pulse will induce a full transition
($P_{CV}=1$) between
the valence $\psi_V$ and the  conduction $\psi_C$ band states
provided that their energy difference matches the resonant condition
$2\epsilon(\boldsymbol{k})=\hbar\omega$.
While this is true in the weak-field regime, the
strong-field regime presents a far more complex behaviour.

Figure \ref{figure5} shows plots of $P_{CV}[t>t_0+\tau_{1}(\boldsymbol{k})]$
as a function of $k_{x}$ and $k_y$ for a linearly polarized 
field ($\phi=0$) in the weak field regime
($E_x e v_F/\hbar \omega^2=E_y e v_F/\hbar \omega^2=0.01$).
We clearly observe that transitions are strictly confined to the
$\boldsymbol{k}$-space region where the resonant condition
$2\epsilon(\boldsymbol{k})=\hbar\omega$ is met. In panel (a)
this region is indicated by the elliptical solid line above
the plot. Furthermore, along this line the probability is not uniform
and is oriented in accord with the electric field direction.
The density plot in panel (b) shows a vanishing probability
along the polarization line. Given that the resonant condition
is perfectly elliptical, this asymmetry is rather attributed to
the anisotropies of the transition time.

Unlike linearly polarized light, circularly polarized light
($\phi=-\pi/2$, $E_x e v_F/\hbar \omega^2=E_y e v_F/\hbar \omega^2=0.01$)
produces a largely even transition probability as can
be seen in Fig. \ref{figure6}. As expected, full transitions
occur only where the resonant condition is fulfilled
as suggested by the consistency between $P_{CV}$
and the solid elliptical line
above the plot in panel (a). Though not visible in panel (b),
smooth variations
of the probability are noticeable close to
$P_{CV}=1$ in panel (a). As we discussed above, these small
changes in $P_{CV}$ are due to the fluctuations
of the transition time as a function of $\boldsymbol{k}$.
These in turn are due to the anisotropic velocities $v_x$ and  $v_y$.

The transition probability as a function of $k_x$ and $k_y$
exhibits
completely different features in the strong field regime.
Figures \ref{figure7} and \ref{figure8} present $P_{CV}(t)$ 
plots alike those in Figs. \ref{figure5} and \ref{figure6},
but in the strong field regime.
A common feature to both polarizations
is that a pulse
of given frequency $\omega$ and duration $\tau_{1,\boldsymbol{k}}$
is capable of exciting states that fall well outside
the single-photon resonant region. This is an
indication that the pulse has also induced
some transitions through multi-photon modes.
Since the pulse duration is tuned in
to given valence and conduction band states,
some of the transitions may not be complete
($P_{CV}<1$) and may take place outside the single
and multi-photon resonance regions.
Though more accentuated in the linear case
(Fig. \ref{figure7}), both plots are consistently anisotropic.
Under linearly polarized light the lack of rotational invariance is due
to the preferred orientation of the polarization direction.
In contrast, the origin of the anisotropy produced in the transition
probability by circularly polarized light is two-fold: the
unmatching velocities, i.e. $v_x\ne v_y$, and the field's initial
phase $\delta$.

\section{Conclusions}\label{sec:conclusions}
We have analyzed the quasienergy-spectrum and the
valence to conduction band
transition probability
of an anisotropic tilted Dirac material subject to an arbitrarily
intense electromagnetic field.
The weak and strong field regimes have been studied
as well as the linear and circular polariztions.
The quasienergy-spectrum in the weak field regime
strongly resembles the energy-spectrum of free carries
regardless of the field polarization.
At the crossover between the weak and the
strong field regimes in  $k = eE/\hbar\omega$,
the structure of the
quasi spectrum changes abruptly.
Mainly two
kinds of deviations with respect to free carriers
can be identified in the strong field regime.
The first type of deviation corresponds to deformation
of the spectra in $\boldsymbol{k}$-space.
Within the strong field region, where $k < eE/\hbar\omega$,
the spectrum stretches as the electronic parameters renormalize
due to electronic-dressing.
Under linearly polarized light this
deformation occurs exclusively in the direction
perpendicular to the field 
polarization.
On the contrary, circularly polarized light stretches the quasienergy
spectrum in both, $k_x$ and $k_y$ directions.
The second type of deviation is associated with the formation
of gaps.
Linearly polarized light produces a complicated pattern of new
gaps and Dirac points around the original one located in $k_x=k_y=0$.
Instead, circularly polarized light opens up a gap in $k_x=k_y=0$.

In the weak-field regime
the light-matter coupling is perturbative and consequently
the free carrier conic quasienergy spectrum
remains almost unaltered.
Hence, in this limit
the electromagnetic field merely induces
transitions between the quantum levels of
free carriers without modifying their spectrum.
Full transitions are therefore
strictly constrained by the single photon resonance condition
$2\epsilon(\boldsymbol{k})=\omega$. For a linearly polarized pulse
of a given duration not all of the transitions that comply with
the resonant condition might take place
due to the directional fluctuation of the transition time.
Put another way, there are transitions that require
very long transition times and therefore do not fully occur.
Circularly polarized pulses  produce very symmetrical
patterns with very smooth variations caused by the small fluctuations
of the transition time and the anisotropic velocities.

Strong electromagnetic fields, on the other hand, profoundly distort
the quasienergy spectrum. Thus, in the strong-field
regime, transitions that escape the single-photon resonance condition
may take place. Linearly polarized pulses generate a pattern
in the transition probability as a function of the $\boldsymbol{k}$
components that is oriented in agreement with the
polarization direction. Quasi-spectra with almost perfect
cylindrical symmetry are obtained under the action
of circularly polarized pulses. The small deviations
from circular symmetry are due to the anisotropic velocities
($v_x\ne v_y$)
and the initial phase $\delta$ of the radiation pulse.

To obtain many of the previous results, a new method to compute the time evolution operator in quantum mechanics was developed.
This method, based on the determination of the monodromy
matrix, has further capacities that were not
exploited in this work. Among other things,
in combination with the density matrix approach
it can be used to compute non-linear electrical
currents to examine light-field-driven currents
in Dirac materials. Another application of this approach
is the study of high harmonic generation.

\section{Acknowledgements}

This work was supported by DCB UAM-A grant numbers
2232214 and 2232215, and UNAM DGAPA PAPIIT IN102620.
 J.C.S.S. and V.G.I.S acknowledge the total support from 
DGAPA-UNAM fellowship. 

\bibliography{biblio.bib}

\begin{thebibliography}{51}%
\makeatletter
\providecommand \@ifxundefined [1]{%
 \@ifx{#1\undefined}
}%
\providecommand \@ifnum [1]{%
 \ifnum #1\expandafter \@firstoftwo
 \else \expandafter \@secondoftwo
 \fi
}%
\providecommand \@ifx [1]{%
 \ifx #1\expandafter \@firstoftwo
 \else \expandafter \@secondoftwo
 \fi
}%
\providecommand \natexlab [1]{#1}%
\providecommand \enquote  [1]{``#1''}%
\providecommand \bibnamefont  [1]{#1}%
\providecommand \bibfnamefont [1]{#1}%
\providecommand \citenamefont [1]{#1}%
\providecommand \href@noop [0]{\@secondoftwo}%
\providecommand \href [0]{\begingroup \@sanitize@url \@href}%
\providecommand \@href[1]{\@@startlink{#1}\@@href}%
\providecommand \@@href[1]{\endgroup#1\@@endlink}%
\providecommand \@sanitize@url [0]{\catcode `\\12\catcode `\$12\catcode
  `\&12\catcode `\#12\catcode `\^12\catcode `\_12\catcode `\%12\relax}%
\providecommand \@@startlink[1]{}%
\providecommand \@@endlink[0]{}%
\providecommand \url  [0]{\begingroup\@sanitize@url \@url }%
\providecommand \@url [1]{\endgroup\@href {#1}{\urlprefix }}%
\providecommand \urlprefix  [0]{URL }%
\providecommand \Eprint [0]{\href }%
\providecommand \doibase [0]{http://dx.doi.org/}%
\providecommand \selectlanguage [0]{\@gobble}%
\providecommand \bibinfo  [0]{\@secondoftwo}%
\providecommand \bibfield  [0]{\@secondoftwo}%
\providecommand \translation [1]{[#1]}%
\providecommand \BibitemOpen [0]{}%
\providecommand \bibitemStop [0]{}%
\providecommand \bibitemNoStop [0]{.\EOS\space}%
\providecommand \EOS [0]{\spacefactor3000\relax}%
\providecommand \BibitemShut  [1]{\csname bibitem#1\endcsname}%
\let\auto@bib@innerbib\@empty
\bibitem [{\citenamefont {Bonaccorso}\ \emph {et~al.}(2010)\citenamefont
  {Bonaccorso}, \citenamefont {Sun}, \citenamefont {Hasan},\ and\ \citenamefont
  {Ferrari}}]{bonaccorso2010graphene}%
  \BibitemOpen
  \bibfield  {author} {\bibinfo {author} {\bibfnamefont {F.}~\bibnamefont
  {Bonaccorso}}, \bibinfo {author} {\bibfnamefont {Z.}~\bibnamefont {Sun}},
  \bibinfo {author} {\bibfnamefont {T.}~\bibnamefont {Hasan}}, \ and\ \bibinfo
  {author} {\bibfnamefont {A.}~\bibnamefont {Ferrari}},\ }\href@noop {}
  {\bibfield  {journal} {\bibinfo  {journal} {Nature photonics}\ }\textbf
  {\bibinfo {volume} {4}},\ \bibinfo {pages} {611} (\bibinfo {year}
  {2010})}\BibitemShut {NoStop}%
\bibitem [{\citenamefont {Bao}\ \emph {et~al.}(2017)\citenamefont {Bao},
  \citenamefont {Hoh},\ and\ \citenamefont {Zhang}}]{bao2017graphene}%
  \BibitemOpen
  \bibfield  {author} {\bibinfo {author} {\bibfnamefont {Q.}~\bibnamefont
  {Bao}}, \bibinfo {author} {\bibfnamefont {H.}~\bibnamefont {Hoh}}, \ and\
  \bibinfo {author} {\bibfnamefont {Y.}~\bibnamefont {Zhang}},\ }\href@noop {}
  {\emph {\bibinfo {title} {Graphene Photonics, Optoelectronics, and
  Plasmonics}}}\ (\bibinfo  {publisher} {CRC Press},\ \bibinfo {year}
  {2017})\BibitemShut {NoStop}%
\bibitem [{\citenamefont {Ponraj}\ \emph {et~al.}(2016)\citenamefont {Ponraj},
  \citenamefont {Xu}, \citenamefont {Dhanabalan}, \citenamefont {Mu},
  \citenamefont {Wang}, \citenamefont {Yuan}, \citenamefont {Li}, \citenamefont
  {Thakur}, \citenamefont {Ashrafi}, \citenamefont {Mccoubrey} \emph
  {et~al.}}]{ponraj2016photonics}%
  \BibitemOpen
  \bibfield  {author} {\bibinfo {author} {\bibfnamefont {J.~S.}\ \bibnamefont
  {Ponraj}}, \bibinfo {author} {\bibfnamefont {Z.-Q.}\ \bibnamefont {Xu}},
  \bibinfo {author} {\bibfnamefont {S.~C.}\ \bibnamefont {Dhanabalan}},
  \bibinfo {author} {\bibfnamefont {H.}~\bibnamefont {Mu}}, \bibinfo {author}
  {\bibfnamefont {Y.}~\bibnamefont {Wang}}, \bibinfo {author} {\bibfnamefont
  {J.}~\bibnamefont {Yuan}}, \bibinfo {author} {\bibfnamefont {P.}~\bibnamefont
  {Li}}, \bibinfo {author} {\bibfnamefont {S.}~\bibnamefont {Thakur}}, \bibinfo
  {author} {\bibfnamefont {M.}~\bibnamefont {Ashrafi}}, \bibinfo {author}
  {\bibfnamefont {K.}~\bibnamefont {Mccoubrey}},  \emph {et~al.},\ }\href@noop
  {} {\bibfield  {journal} {\bibinfo  {journal} {Nanotechnology}\ }\textbf
  {\bibinfo {volume} {27}},\ \bibinfo {pages} {462001} (\bibinfo {year}
  {2016})}\BibitemShut {NoStop}%
\bibitem [{\citenamefont {Salén}\ \emph {et~al.}(2019)\citenamefont {Salén},
  \citenamefont {Basini}, \citenamefont {Bonetti}, \citenamefont {Hebling},
  \citenamefont {Krasilnikov}, \citenamefont {Nikitin}, \citenamefont
  {Shamuilov}, \citenamefont {Tibai}, \citenamefont {Zhaunerchyk},\ and\
  \citenamefont {Goryashko}}]{SALEN20191}%
  \BibitemOpen
  \bibfield  {author} {\bibinfo {author} {\bibfnamefont {P.}~\bibnamefont
  {Salén}}, \bibinfo {author} {\bibfnamefont {M.}~\bibnamefont {Basini}},
  \bibinfo {author} {\bibfnamefont {S.}~\bibnamefont {Bonetti}}, \bibinfo
  {author} {\bibfnamefont {J.}~\bibnamefont {Hebling}}, \bibinfo {author}
  {\bibfnamefont {M.}~\bibnamefont {Krasilnikov}}, \bibinfo {author}
  {\bibfnamefont {A.~Y.}\ \bibnamefont {Nikitin}}, \bibinfo {author}
  {\bibfnamefont {G.}~\bibnamefont {Shamuilov}}, \bibinfo {author}
  {\bibfnamefont {Z.}~\bibnamefont {Tibai}}, \bibinfo {author} {\bibfnamefont
  {V.}~\bibnamefont {Zhaunerchyk}}, \ and\ \bibinfo {author} {\bibfnamefont
  {V.}~\bibnamefont {Goryashko}},\ }\href {\doibase
  https://doi.org/10.1016/j.physrep.2019.09.002} {\bibfield  {journal}
  {\bibinfo  {journal} {Physics Reports}\ }\textbf {\bibinfo {volume}
  {836-837}},\ \bibinfo {pages} {1 } (\bibinfo {year} {2019})},\ \bibinfo
  {note} {matter manipulation with extreme terahertz light: Progress in the
  enabling THz technology}\BibitemShut {NoStop}%
\bibitem [{\citenamefont {Grigorenko}\ \emph {et~al.}(2012)\citenamefont
  {Grigorenko}, \citenamefont {Polini},\ and\ \citenamefont
  {Novoselov}}]{grigorenko2012graphene}%
  \BibitemOpen
  \bibfield  {author} {\bibinfo {author} {\bibfnamefont {A.}~\bibnamefont
  {Grigorenko}}, \bibinfo {author} {\bibfnamefont {M.}~\bibnamefont {Polini}},
  \ and\ \bibinfo {author} {\bibfnamefont {K.}~\bibnamefont {Novoselov}},\
  }\href@noop {} {\bibfield  {journal} {\bibinfo  {journal} {Nature photonics}\
  }\textbf {\bibinfo {volume} {6}},\ \bibinfo {pages} {749} (\bibinfo {year}
  {2012})}\BibitemShut {NoStop}%
\bibitem [{\citenamefont {Fan}\ \emph {et~al.}(2019)\citenamefont {Fan},
  \citenamefont {Shen}, \citenamefont {Zhang}, \citenamefont {Zhao},
  \citenamefont {Wu}, \citenamefont {Fu}, \citenamefont {Wei}, \citenamefont
  {Li},\ and\ \citenamefont {Soukoulis}}]{fan2019graphene}%
  \BibitemOpen
  \bibfield  {author} {\bibinfo {author} {\bibfnamefont {Y.}~\bibnamefont
  {Fan}}, \bibinfo {author} {\bibfnamefont {N.-H.}\ \bibnamefont {Shen}},
  \bibinfo {author} {\bibfnamefont {F.}~\bibnamefont {Zhang}}, \bibinfo
  {author} {\bibfnamefont {Q.}~\bibnamefont {Zhao}}, \bibinfo {author}
  {\bibfnamefont {H.}~\bibnamefont {Wu}}, \bibinfo {author} {\bibfnamefont
  {Q.}~\bibnamefont {Fu}}, \bibinfo {author} {\bibfnamefont {Z.}~\bibnamefont
  {Wei}}, \bibinfo {author} {\bibfnamefont {H.}~\bibnamefont {Li}}, \ and\
  \bibinfo {author} {\bibfnamefont {C.~M.}\ \bibnamefont {Soukoulis}},\
  }\href@noop {} {\bibfield  {journal} {\bibinfo  {journal} {Advanced Optical
  Materials}\ }\textbf {\bibinfo {volume} {7}},\ \bibinfo {pages} {1800537}
  (\bibinfo {year} {2019})}\BibitemShut {NoStop}%
\bibitem [{\citenamefont {Wu}\ \emph {et~al.}(2019)\citenamefont {Wu},
  \citenamefont {Guo}, \citenamefont {Du}, \citenamefont {Xia}, \citenamefont
  {Zeng}, \citenamefont {Tian}, \citenamefont {Shi}, \citenamefont {Tian},
  \citenamefont {Li}, \citenamefont {Tsang},\ and\ \citenamefont
  {Jie}}]{doi:10.1021/acsnano.9b03994f}%
  \BibitemOpen
  \bibfield  {author} {\bibinfo {author} {\bibfnamefont {D.}~\bibnamefont
  {Wu}}, \bibinfo {author} {\bibfnamefont {J.}~\bibnamefont {Guo}}, \bibinfo
  {author} {\bibfnamefont {J.}~\bibnamefont {Du}}, \bibinfo {author}
  {\bibfnamefont {C.}~\bibnamefont {Xia}}, \bibinfo {author} {\bibfnamefont
  {L.}~\bibnamefont {Zeng}}, \bibinfo {author} {\bibfnamefont {Y.}~\bibnamefont
  {Tian}}, \bibinfo {author} {\bibfnamefont {Z.}~\bibnamefont {Shi}}, \bibinfo
  {author} {\bibfnamefont {Y.}~\bibnamefont {Tian}}, \bibinfo {author}
  {\bibfnamefont {X.~J.}\ \bibnamefont {Li}}, \bibinfo {author} {\bibfnamefont
  {Y.~H.}\ \bibnamefont {Tsang}}, \ and\ \bibinfo {author} {\bibfnamefont
  {J.}~\bibnamefont {Jie}},\ }\href {\doibase 10.1021/acsnano.9b03994}
  {\bibfield  {journal} {\bibinfo  {journal} {ACS Nano}\ }\textbf {\bibinfo
  {volume} {13}},\ \bibinfo {pages} {9907} (\bibinfo {year} {2019})},\ \bibinfo
  {note} {pMID: 31361122},\ \Eprint
  {http://arxiv.org/abs/https://doi.org/10.1021/acsnano.9b03994}
  {https://doi.org/10.1021/acsnano.9b03994} \BibitemShut {NoStop}%
\bibitem [{\citenamefont {Scagliotti}\ \emph {et~al.}(2019)\citenamefont
  {Scagliotti}, \citenamefont {Salvato}, \citenamefont {Crescenzi]},
  \citenamefont {Kovalchuk}, \citenamefont {Komissarov}, \citenamefont
  {Prischepa}, \citenamefont {Catone}, \citenamefont {Mario]}, \citenamefont
  {Boscardin}, \citenamefont {Crivellari},\ and\ \citenamefont
  {Castrucci}}]{SCAGLIOTTI2019643}%
  \BibitemOpen
  \bibfield  {author} {\bibinfo {author} {\bibfnamefont {M.}~\bibnamefont
  {Scagliotti}}, \bibinfo {author} {\bibfnamefont {M.}~\bibnamefont {Salvato}},
  \bibinfo {author} {\bibfnamefont {M.~D.}\ \bibnamefont {Crescenzi]}},
  \bibinfo {author} {\bibfnamefont {N.}~\bibnamefont {Kovalchuk}}, \bibinfo
  {author} {\bibfnamefont {I.}~\bibnamefont {Komissarov}}, \bibinfo {author}
  {\bibfnamefont {S.}~\bibnamefont {Prischepa}}, \bibinfo {author}
  {\bibfnamefont {D.}~\bibnamefont {Catone}}, \bibinfo {author} {\bibfnamefont
  {L.~D.}\ \bibnamefont {Mario]}}, \bibinfo {author} {\bibfnamefont
  {M.}~\bibnamefont {Boscardin}}, \bibinfo {author} {\bibfnamefont
  {M.}~\bibnamefont {Crivellari}}, \ and\ \bibinfo {author} {\bibfnamefont
  {P.}~\bibnamefont {Castrucci}},\ }\href {\doibase
  https://doi.org/10.1016/j.carbon.2019.06.051} {\bibfield  {journal} {\bibinfo
   {journal} {Carbon}\ }\textbf {\bibinfo {volume} {152}},\ \bibinfo {pages}
  {643 } (\bibinfo {year} {2019})}\BibitemShut {NoStop}%
\bibitem [{\citenamefont {Liu}\ \emph {et~al.}(2011)\citenamefont {Liu},
  \citenamefont {Yin}, \citenamefont {Ulin-Avila}, \citenamefont {Geng},
  \citenamefont {Zentgraf}, \citenamefont {Ju}, \citenamefont {Wang},\ and\
  \citenamefont {Zhang}}]{liu2011graphene}%
  \BibitemOpen
  \bibfield  {author} {\bibinfo {author} {\bibfnamefont {M.}~\bibnamefont
  {Liu}}, \bibinfo {author} {\bibfnamefont {X.}~\bibnamefont {Yin}}, \bibinfo
  {author} {\bibfnamefont {E.}~\bibnamefont {Ulin-Avila}}, \bibinfo {author}
  {\bibfnamefont {B.}~\bibnamefont {Geng}}, \bibinfo {author} {\bibfnamefont
  {T.}~\bibnamefont {Zentgraf}}, \bibinfo {author} {\bibfnamefont
  {L.}~\bibnamefont {Ju}}, \bibinfo {author} {\bibfnamefont {F.}~\bibnamefont
  {Wang}}, \ and\ \bibinfo {author} {\bibfnamefont {X.}~\bibnamefont {Zhang}},\
  }\href@noop {} {\bibfield  {journal} {\bibinfo  {journal} {Nature}\ }\textbf
  {\bibinfo {volume} {474}},\ \bibinfo {pages} {64} (\bibinfo {year}
  {2011})}\BibitemShut {NoStop}%
\bibitem [{\citenamefont {Sorianello}\ \emph {et~al.}(2018)\citenamefont
  {Sorianello}, \citenamefont {Midrio}, \citenamefont {Contestabile},
  \citenamefont {Asselberghs}, \citenamefont {Van~Campenhout}, \citenamefont
  {Huyghebaert}, \citenamefont {Goykhman}, \citenamefont {Ott}, \citenamefont
  {Ferrari},\ and\ \citenamefont {Romagnoli}}]{sorianello2018graphene}%
  \BibitemOpen
  \bibfield  {author} {\bibinfo {author} {\bibfnamefont {V.}~\bibnamefont
  {Sorianello}}, \bibinfo {author} {\bibfnamefont {M.}~\bibnamefont {Midrio}},
  \bibinfo {author} {\bibfnamefont {G.}~\bibnamefont {Contestabile}}, \bibinfo
  {author} {\bibfnamefont {I.}~\bibnamefont {Asselberghs}}, \bibinfo {author}
  {\bibfnamefont {J.}~\bibnamefont {Van~Campenhout}}, \bibinfo {author}
  {\bibfnamefont {C.}~\bibnamefont {Huyghebaert}}, \bibinfo {author}
  {\bibfnamefont {I.}~\bibnamefont {Goykhman}}, \bibinfo {author}
  {\bibfnamefont {A.}~\bibnamefont {Ott}}, \bibinfo {author} {\bibfnamefont
  {A.}~\bibnamefont {Ferrari}}, \ and\ \bibinfo {author} {\bibfnamefont
  {M.}~\bibnamefont {Romagnoli}},\ }\href@noop {} {\bibfield  {journal}
  {\bibinfo  {journal} {Nature Photonics}\ }\textbf {\bibinfo {volume} {12}},\
  \bibinfo {pages} {40} (\bibinfo {year} {2018})}\BibitemShut {NoStop}%
\bibitem [{\citenamefont {Hao}\ \emph {et~al.}(2019)\citenamefont {Hao},
  \citenamefont {Jiao}, \citenamefont {Peng}, \citenamefont {Zhen},
  \citenamefont {Dagarbek}, \citenamefont {Zou},\ and\ \citenamefont
  {Li}}]{hao2019experimental}%
  \BibitemOpen
  \bibfield  {author} {\bibinfo {author} {\bibfnamefont {R.}~\bibnamefont
  {Hao}}, \bibinfo {author} {\bibfnamefont {J.}~\bibnamefont {Jiao}}, \bibinfo
  {author} {\bibfnamefont {X.}~\bibnamefont {Peng}}, \bibinfo {author}
  {\bibfnamefont {Z.}~\bibnamefont {Zhen}}, \bibinfo {author} {\bibfnamefont
  {R.}~\bibnamefont {Dagarbek}}, \bibinfo {author} {\bibfnamefont
  {Y.}~\bibnamefont {Zou}}, \ and\ \bibinfo {author} {\bibfnamefont
  {E.}~\bibnamefont {Li}},\ }\href@noop {} {\bibfield  {journal} {\bibinfo
  {journal} {Optics letters}\ }\textbf {\bibinfo {volume} {44}},\ \bibinfo
  {pages} {2586} (\bibinfo {year} {2019})}\BibitemShut {NoStop}%
\bibitem [{\citenamefont {Safaei}\ \emph {et~al.}(2019)\citenamefont {Safaei},
  \citenamefont {Chandra}, \citenamefont {Shabbir}, \citenamefont
  {Leuenberger},\ and\ \citenamefont {Chanda}}]{safaei2019dirac}%
  \BibitemOpen
  \bibfield  {author} {\bibinfo {author} {\bibfnamefont {A.}~\bibnamefont
  {Safaei}}, \bibinfo {author} {\bibfnamefont {S.}~\bibnamefont {Chandra}},
  \bibinfo {author} {\bibfnamefont {M.~W.}\ \bibnamefont {Shabbir}}, \bibinfo
  {author} {\bibfnamefont {M.~N.}\ \bibnamefont {Leuenberger}}, \ and\ \bibinfo
  {author} {\bibfnamefont {D.}~\bibnamefont {Chanda}},\ }\href@noop {}
  {\bibfield  {journal} {\bibinfo  {journal} {Nature communications}\ }\textbf
  {\bibinfo {volume} {10}},\ \bibinfo {pages} {1} (\bibinfo {year}
  {2019})}\BibitemShut {NoStop}%
\bibitem [{\citenamefont {Yin}\ \emph {et~al.}(2014)\citenamefont {Yin},
  \citenamefont {Zhu}, \citenamefont {He}, \citenamefont {Cao}, \citenamefont
  {Tan}, \citenamefont {Chen}, \citenamefont {Yan},\ and\ \citenamefont
  {Zhang}}]{yin2014graphene}%
  \BibitemOpen
  \bibfield  {author} {\bibinfo {author} {\bibfnamefont {Z.}~\bibnamefont
  {Yin}}, \bibinfo {author} {\bibfnamefont {J.}~\bibnamefont {Zhu}}, \bibinfo
  {author} {\bibfnamefont {Q.}~\bibnamefont {He}}, \bibinfo {author}
  {\bibfnamefont {X.}~\bibnamefont {Cao}}, \bibinfo {author} {\bibfnamefont
  {C.}~\bibnamefont {Tan}}, \bibinfo {author} {\bibfnamefont {H.}~\bibnamefont
  {Chen}}, \bibinfo {author} {\bibfnamefont {Q.}~\bibnamefont {Yan}}, \ and\
  \bibinfo {author} {\bibfnamefont {H.}~\bibnamefont {Zhang}},\ }\href@noop {}
  {\bibfield  {journal} {\bibinfo  {journal} {Advanced energy materials}\
  }\textbf {\bibinfo {volume} {4}},\ \bibinfo {pages} {1300574} (\bibinfo
  {year} {2014})}\BibitemShut {NoStop}%
\bibitem [{\citenamefont {O’keeffe}\ \emph {et~al.}(2019)\citenamefont
  {O’keeffe}, \citenamefont {Catone}, \citenamefont {Paladini}, \citenamefont
  {Toschi}, \citenamefont {Turchini}, \citenamefont {Avaldi}, \citenamefont
  {Martelli}, \citenamefont {Agresti}, \citenamefont {Pescetelli},
  \citenamefont {Del Rio~Castillo} \emph {et~al.}}]{o2019graphene}%
  \BibitemOpen
  \bibfield  {author} {\bibinfo {author} {\bibfnamefont {P.}~\bibnamefont
  {O’keeffe}}, \bibinfo {author} {\bibfnamefont {D.}~\bibnamefont {Catone}},
  \bibinfo {author} {\bibfnamefont {A.}~\bibnamefont {Paladini}}, \bibinfo
  {author} {\bibfnamefont {F.}~\bibnamefont {Toschi}}, \bibinfo {author}
  {\bibfnamefont {S.}~\bibnamefont {Turchini}}, \bibinfo {author}
  {\bibfnamefont {L.}~\bibnamefont {Avaldi}}, \bibinfo {author} {\bibfnamefont
  {F.}~\bibnamefont {Martelli}}, \bibinfo {author} {\bibfnamefont
  {A.}~\bibnamefont {Agresti}}, \bibinfo {author} {\bibfnamefont
  {S.}~\bibnamefont {Pescetelli}}, \bibinfo {author} {\bibfnamefont
  {A.}~\bibnamefont {Del Rio~Castillo}},  \emph {et~al.},\ }\href@noop {}
  {\bibfield  {journal} {\bibinfo  {journal} {Nano letters}\ }\textbf {\bibinfo
  {volume} {19}},\ \bibinfo {pages} {684} (\bibinfo {year} {2019})}\BibitemShut
  {NoStop}%
\bibitem [{\citenamefont {Castro~Neto}\ \emph {et~al.}(2009)\citenamefont
  {Castro~Neto}, \citenamefont {Guinea}, \citenamefont {Peres}, \citenamefont
  {Novoselov},\ and\ \citenamefont {Geim}}]{neto2009electronic}%
  \BibitemOpen
  \bibfield  {author} {\bibinfo {author} {\bibfnamefont {A.~H.}\ \bibnamefont
  {Castro~Neto}}, \bibinfo {author} {\bibfnamefont {F.}~\bibnamefont {Guinea}},
  \bibinfo {author} {\bibfnamefont {N.~M.~R.}\ \bibnamefont {Peres}}, \bibinfo
  {author} {\bibfnamefont {K.~S.}\ \bibnamefont {Novoselov}}, \ and\ \bibinfo
  {author} {\bibfnamefont {A.~K.}\ \bibnamefont {Geim}},\ }\href {\doibase
  10.1103/RevModPhys.81.109} {\bibfield  {journal} {\bibinfo  {journal} {Rev.
  Mod. Phys.}\ }\textbf {\bibinfo {volume} {81}},\ \bibinfo {pages} {109}
  (\bibinfo {year} {2009})}\BibitemShut {NoStop}%
\bibitem [{\citenamefont {Higuchi}\ \emph {et~al.}(2017)\citenamefont
  {Higuchi}, \citenamefont {Heide}, \citenamefont {Ullmann}, \citenamefont
  {Weber},\ and\ \citenamefont {Hommelhoff}}]{higuchi2017light}%
  \BibitemOpen
  \bibfield  {author} {\bibinfo {author} {\bibfnamefont {T.}~\bibnamefont
  {Higuchi}}, \bibinfo {author} {\bibfnamefont {C.}~\bibnamefont {Heide}},
  \bibinfo {author} {\bibfnamefont {K.}~\bibnamefont {Ullmann}}, \bibinfo
  {author} {\bibfnamefont {H.~B.}\ \bibnamefont {Weber}}, \ and\ \bibinfo
  {author} {\bibfnamefont {P.}~\bibnamefont {Hommelhoff}},\ }\href@noop {}
  {\bibfield  {journal} {\bibinfo  {journal} {Nature}\ }\textbf {\bibinfo
  {volume} {550}},\ \bibinfo {pages} {224} (\bibinfo {year}
  {2017})}\BibitemShut {NoStop}%
\bibitem [{\citenamefont {Heide}\ \emph {et~al.}(2018)\citenamefont {Heide},
  \citenamefont {Higuchi}, \citenamefont {Weber},\ and\ \citenamefont
  {Hommelhoff}}]{heide2018coherent}%
  \BibitemOpen
  \bibfield  {author} {\bibinfo {author} {\bibfnamefont {C.}~\bibnamefont
  {Heide}}, \bibinfo {author} {\bibfnamefont {T.}~\bibnamefont {Higuchi}},
  \bibinfo {author} {\bibfnamefont {H.~B.}\ \bibnamefont {Weber}}, \ and\
  \bibinfo {author} {\bibfnamefont {P.}~\bibnamefont {Hommelhoff}},\ }\href
  {\doibase 10.1103/PhysRevLett.121.207401} {\bibfield  {journal} {\bibinfo
  {journal} {Phys. Rev. Lett.}\ }\textbf {\bibinfo {volume} {121}},\ \bibinfo
  {pages} {207401} (\bibinfo {year} {2018})}\BibitemShut {NoStop}%
\bibitem [{\citenamefont {Orfanos}\ \emph {et~al.}(2019)\citenamefont
  {Orfanos}, \citenamefont {Makos}, \citenamefont {Liontos}, \citenamefont
  {Skantzakis}, \citenamefont {Förg}, \citenamefont {Charalambidis},\ and\
  \citenamefont {Tzallas}}]{doi:10.1063/1.5086773}%
  \BibitemOpen
  \bibfield  {author} {\bibinfo {author} {\bibfnamefont {I.}~\bibnamefont
  {Orfanos}}, \bibinfo {author} {\bibfnamefont {I.}~\bibnamefont {Makos}},
  \bibinfo {author} {\bibfnamefont {I.}~\bibnamefont {Liontos}}, \bibinfo
  {author} {\bibfnamefont {E.}~\bibnamefont {Skantzakis}}, \bibinfo {author}
  {\bibfnamefont {B.}~\bibnamefont {Förg}}, \bibinfo {author} {\bibfnamefont
  {D.}~\bibnamefont {Charalambidis}}, \ and\ \bibinfo {author} {\bibfnamefont
  {P.}~\bibnamefont {Tzallas}},\ }\href {\doibase 10.1063/1.5086773} {\bibfield
   {journal} {\bibinfo  {journal} {APL Photonics}\ }\textbf {\bibinfo {volume}
  {4}},\ \bibinfo {pages} {080901} (\bibinfo {year} {2019})},\ \Eprint
  {http://arxiv.org/abs/https://doi.org/10.1063/1.5086773}
  {https://doi.org/10.1063/1.5086773} \BibitemShut {NoStop}%
\bibitem [{\citenamefont {Krausz}\ and\ \citenamefont
  {Stockman}(2014)}]{krausz2014attosecond}%
  \BibitemOpen
  \bibfield  {author} {\bibinfo {author} {\bibfnamefont {F.}~\bibnamefont
  {Krausz}}\ and\ \bibinfo {author} {\bibfnamefont {M.~I.}\ \bibnamefont
  {Stockman}},\ }\href@noop {} {\bibfield  {journal} {\bibinfo  {journal}
  {Nature Photonics}\ }\textbf {\bibinfo {volume} {8}},\ \bibinfo {pages} {205}
  (\bibinfo {year} {2014})}\BibitemShut {NoStop}%
\bibitem [{\citenamefont {Hentschel}\ \emph {et~al.}(2001)\citenamefont
  {Hentschel}, \citenamefont {Kienberger}, \citenamefont {Spielmann},
  \citenamefont {Reider}, \citenamefont {Milosevic}, \citenamefont {Brabec},
  \citenamefont {Corkum}, \citenamefont {Heinzmann}, \citenamefont {Drescher},\
  and\ \citenamefont {Krausz}}]{hentschel2001attosecond}%
  \BibitemOpen
  \bibfield  {author} {\bibinfo {author} {\bibfnamefont {M.}~\bibnamefont
  {Hentschel}}, \bibinfo {author} {\bibfnamefont {R.}~\bibnamefont
  {Kienberger}}, \bibinfo {author} {\bibfnamefont {C.}~\bibnamefont
  {Spielmann}}, \bibinfo {author} {\bibfnamefont {G.~A.}\ \bibnamefont
  {Reider}}, \bibinfo {author} {\bibfnamefont {N.}~\bibnamefont {Milosevic}},
  \bibinfo {author} {\bibfnamefont {T.}~\bibnamefont {Brabec}}, \bibinfo
  {author} {\bibfnamefont {P.}~\bibnamefont {Corkum}}, \bibinfo {author}
  {\bibfnamefont {U.}~\bibnamefont {Heinzmann}}, \bibinfo {author}
  {\bibfnamefont {M.}~\bibnamefont {Drescher}}, \ and\ \bibinfo {author}
  {\bibfnamefont {F.}~\bibnamefont {Krausz}},\ }\href@noop {} {\bibfield
  {journal} {\bibinfo  {journal} {Nature}\ }\textbf {\bibinfo {volume} {414}},\
  \bibinfo {pages} {509} (\bibinfo {year} {2001})}\BibitemShut {NoStop}%
\bibitem [{\citenamefont {Oka}\ and\ \citenamefont
  {Aoki}(2009)}]{PhysRevB.79.081406}%
  \BibitemOpen
  \bibfield  {author} {\bibinfo {author} {\bibfnamefont {T.}~\bibnamefont
  {Oka}}\ and\ \bibinfo {author} {\bibfnamefont {H.}~\bibnamefont {Aoki}},\
  }\href {\doibase 10.1103/PhysRevB.79.081406} {\bibfield  {journal} {\bibinfo
  {journal} {Phys. Rev. B}\ }\textbf {\bibinfo {volume} {79}},\ \bibinfo
  {pages} {081406} (\bibinfo {year} {2009})}\BibitemShut {NoStop}%
\bibitem [{\citenamefont {McIver}\ \emph {et~al.}(2020)\citenamefont {McIver},
  \citenamefont {Schulte}, \citenamefont {Stein}, \citenamefont {Matsuyama},
  \citenamefont {Jotzu}, \citenamefont {Meier},\ and\ \citenamefont
  {Cavalleri}}]{mciver2020light}%
  \BibitemOpen
  \bibfield  {author} {\bibinfo {author} {\bibfnamefont {J.~W.}\ \bibnamefont
  {McIver}}, \bibinfo {author} {\bibfnamefont {B.}~\bibnamefont {Schulte}},
  \bibinfo {author} {\bibfnamefont {F.-U.}\ \bibnamefont {Stein}}, \bibinfo
  {author} {\bibfnamefont {T.}~\bibnamefont {Matsuyama}}, \bibinfo {author}
  {\bibfnamefont {G.}~\bibnamefont {Jotzu}}, \bibinfo {author} {\bibfnamefont
  {G.}~\bibnamefont {Meier}}, \ and\ \bibinfo {author} {\bibfnamefont
  {A.}~\bibnamefont {Cavalleri}},\ }\href@noop {} {\bibfield  {journal}
  {\bibinfo  {journal} {Nature Physics}\ }\textbf {\bibinfo {volume} {16}},\
  \bibinfo {pages} {38} (\bibinfo {year} {2020})}\BibitemShut {NoStop}%
\bibitem [{\citenamefont {Lopez-Rodriguez}\ and\ \citenamefont
  {Naumis}(2008)}]{lopez2008analytic}%
  \BibitemOpen
  \bibfield  {author} {\bibinfo {author} {\bibfnamefont {F.~J.}\ \bibnamefont
  {Lopez-Rodriguez}}\ and\ \bibinfo {author} {\bibfnamefont {G.~G.}\
  \bibnamefont {Naumis}},\ }\href {\doibase 10.1103/PhysRevB.78.201406}
  {\bibfield  {journal} {\bibinfo  {journal} {Phys. Rev. B}\ }\textbf {\bibinfo
  {volume} {78}},\ \bibinfo {pages} {201406(R)} (\bibinfo {year}
  {2008})}\BibitemShut {NoStop}%
\bibitem [{\citenamefont {L{\'o}pez-Rodr{\'\i}guez}\ and\ \citenamefont
  {Naumis}(2010)}]{lopez2010graphene}%
  \BibitemOpen
  \bibfield  {author} {\bibinfo {author} {\bibfnamefont {F.}~\bibnamefont
  {L{\'o}pez-Rodr{\'\i}guez}}\ and\ \bibinfo {author} {\bibfnamefont
  {G.}~\bibnamefont {Naumis}},\ }\href@noop {} {\bibfield  {journal} {\bibinfo
  {journal} {Philosophical Magazine}\ }\textbf {\bibinfo {volume} {90}},\
  \bibinfo {pages} {2977} (\bibinfo {year} {2010})}\BibitemShut {NoStop}%
\bibitem [{\citenamefont {Kibis}(2010)}]{kibis2010metal}%
  \BibitemOpen
  \bibfield  {author} {\bibinfo {author} {\bibfnamefont {O.~V.}\ \bibnamefont
  {Kibis}},\ }\href {\doibase 10.1103/PhysRevB.81.165433} {\bibfield  {journal}
  {\bibinfo  {journal} {Phys. Rev. B}\ }\textbf {\bibinfo {volume} {81}},\
  \bibinfo {pages} {165433} (\bibinfo {year} {2010})}\BibitemShut {NoStop}%
\bibitem [{\citenamefont {Calvo}\ \emph
  {et~al.}(2011{\natexlab{a}})\citenamefont {Calvo}, \citenamefont {Pastawski},
  \citenamefont {Roche},\ and\ \citenamefont {Torres}}]{Foa2011Tuning}%
  \BibitemOpen
  \bibfield  {author} {\bibinfo {author} {\bibfnamefont {H.~L.}\ \bibnamefont
  {Calvo}}, \bibinfo {author} {\bibfnamefont {H.~M.}\ \bibnamefont
  {Pastawski}}, \bibinfo {author} {\bibfnamefont {S.}~\bibnamefont {Roche}}, \
  and\ \bibinfo {author} {\bibfnamefont {L.~E. F.~F.}\ \bibnamefont {Torres}},\
  }\href {\doibase 10.1063/1.3597412} {\bibfield  {journal} {\bibinfo
  {journal} {Applied Physics Letters}\ }\textbf {\bibinfo {volume} {98}},\
  \bibinfo {pages} {232103} (\bibinfo {year} {2011}{\natexlab{a}})},\ \Eprint
  {http://arxiv.org/abs/https://doi.org/10.1063/1.3597412}
  {https://doi.org/10.1063/1.3597412} \BibitemShut {NoStop}%
\bibitem [{\citenamefont {Sun}\ \emph {et~al.}(2012)\citenamefont {Sun},
  \citenamefont {Aivazian}, \citenamefont {Jones}, \citenamefont {Ross},
  \citenamefont {Yao}, \citenamefont {Cobden},\ and\ \citenamefont
  {Xu}}]{sun2012ultrafast}%
  \BibitemOpen
  \bibfield  {author} {\bibinfo {author} {\bibfnamefont {D.}~\bibnamefont
  {Sun}}, \bibinfo {author} {\bibfnamefont {G.}~\bibnamefont {Aivazian}},
  \bibinfo {author} {\bibfnamefont {A.~M.}\ \bibnamefont {Jones}}, \bibinfo
  {author} {\bibfnamefont {J.~S.}\ \bibnamefont {Ross}}, \bibinfo {author}
  {\bibfnamefont {W.}~\bibnamefont {Yao}}, \bibinfo {author} {\bibfnamefont
  {D.}~\bibnamefont {Cobden}}, \ and\ \bibinfo {author} {\bibfnamefont
  {X.}~\bibnamefont {Xu}},\ }\href@noop {} {\bibfield  {journal} {\bibinfo
  {journal} {Nature nanotechnology}\ }\textbf {\bibinfo {volume} {7}},\
  \bibinfo {pages} {114} (\bibinfo {year} {2012})}\BibitemShut {NoStop}%
\bibitem [{\citenamefont {Kristinsson}\ \emph
  {et~al.}(2016{\natexlab{a}})\citenamefont {Kristinsson}, \citenamefont
  {Kibis}, \citenamefont {Morina},\ and\ \citenamefont {Shelykh}}]{kibis2015}%
  \BibitemOpen
  \bibfield  {author} {\bibinfo {author} {\bibfnamefont {K.}~\bibnamefont
  {Kristinsson}}, \bibinfo {author} {\bibfnamefont {O.~V.}\ \bibnamefont
  {Kibis}}, \bibinfo {author} {\bibfnamefont {S.}~\bibnamefont {Morina}}, \
  and\ \bibinfo {author} {\bibfnamefont {I.~A.}\ \bibnamefont {Shelykh}},\
  }\href {https://doi.org/10.1038/srep20082} {\bibfield  {journal} {\bibinfo
  {journal} {Scientific Reports}\ }\textbf {\bibinfo {volume} {6}},\ \bibinfo
  {pages} {20082 EP } (\bibinfo {year} {2016}{\natexlab{a}})}\BibitemShut
  {NoStop}%
\bibitem [{\citenamefont {Kristinsson}\ \emph
  {et~al.}(2016{\natexlab{b}})\citenamefont {Kristinsson}, \citenamefont
  {Kibis}, \citenamefont {Morina},\ and\ \citenamefont
  {Shelykh}}]{kristinsson2016control}%
  \BibitemOpen
  \bibfield  {author} {\bibinfo {author} {\bibfnamefont {K.}~\bibnamefont
  {Kristinsson}}, \bibinfo {author} {\bibfnamefont {O.~V.}\ \bibnamefont
  {Kibis}}, \bibinfo {author} {\bibfnamefont {S.}~\bibnamefont {Morina}}, \
  and\ \bibinfo {author} {\bibfnamefont {I.~A.}\ \bibnamefont {Shelykh}},\
  }\href@noop {} {\bibfield  {journal} {\bibinfo  {journal} {Scientific
  reports}\ }\textbf {\bibinfo {volume} {6}},\ \bibinfo {pages} {1} (\bibinfo
  {year} {2016}{\natexlab{b}})}\BibitemShut {NoStop}%
\bibitem [{\citenamefont {Kibis}\ \emph {et~al.}(2017)\citenamefont {Kibis},
  \citenamefont {Dini}, \citenamefont {Iorsh},\ and\ \citenamefont
  {Shelykh}}]{kibis2017all}%
  \BibitemOpen
  \bibfield  {author} {\bibinfo {author} {\bibfnamefont {O.}~\bibnamefont
  {Kibis}}, \bibinfo {author} {\bibfnamefont {K.}~\bibnamefont {Dini}},
  \bibinfo {author} {\bibfnamefont {I.}~\bibnamefont {Iorsh}}, \ and\ \bibinfo
  {author} {\bibfnamefont {I.}~\bibnamefont {Shelykh}},\ }\href@noop {}
  {\bibfield  {journal} {\bibinfo  {journal} {Physical Review B}\ }\textbf
  {\bibinfo {volume} {95}},\ \bibinfo {pages} {125401} (\bibinfo {year}
  {2017})}\BibitemShut {NoStop}%
\bibitem [{\citenamefont {Kibis}\ \emph {et~al.}(2018)\citenamefont {Kibis},
  \citenamefont {Dini}, \citenamefont {Iorsh}, \citenamefont {Dragunov},\ and\
  \citenamefont {Shelykh}}]{kibis2018electromagnetic}%
  \BibitemOpen
  \bibfield  {author} {\bibinfo {author} {\bibfnamefont {O.}~\bibnamefont
  {Kibis}}, \bibinfo {author} {\bibfnamefont {K.}~\bibnamefont {Dini}},
  \bibinfo {author} {\bibfnamefont {I.}~\bibnamefont {Iorsh}}, \bibinfo
  {author} {\bibfnamefont {V.}~\bibnamefont {Dragunov}}, \ and\ \bibinfo
  {author} {\bibfnamefont {I.}~\bibnamefont {Shelykh}},\ }\href@noop {}
  {\bibfield  {journal} {\bibinfo  {journal} {Journal of Structural Chemistry}\
  }\textbf {\bibinfo {volume} {59}},\ \bibinfo {pages} {867} (\bibinfo {year}
  {2018})}\BibitemShut {NoStop}%
\bibitem [{\citenamefont {Calvo}\ \emph
  {et~al.}(2011{\natexlab{b}})\citenamefont {Calvo}, \citenamefont {Pastawski},
  \citenamefont {Roche},\ and\ \citenamefont {Torres}}]{doi:10.1063/1.3597412}%
  \BibitemOpen
  \bibfield  {author} {\bibinfo {author} {\bibfnamefont {H.~L.}\ \bibnamefont
  {Calvo}}, \bibinfo {author} {\bibfnamefont {H.~M.}\ \bibnamefont
  {Pastawski}}, \bibinfo {author} {\bibfnamefont {S.}~\bibnamefont {Roche}}, \
  and\ \bibinfo {author} {\bibfnamefont {L.~E. F.~F.}\ \bibnamefont {Torres}},\
  }\href {\doibase 10.1063/1.3597412} {\bibfield  {journal} {\bibinfo
  {journal} {Applied Physics Letters}\ }\textbf {\bibinfo {volume} {98}},\
  \bibinfo {pages} {232103} (\bibinfo {year} {2011}{\natexlab{b}})},\ \Eprint
  {http://arxiv.org/abs/https://doi.org/10.1063/1.3597412}
  {https://doi.org/10.1063/1.3597412} \BibitemShut {NoStop}%
\bibitem [{\citenamefont {Sandoval-Santana}\ \emph {et~al.}(2020)\citenamefont
  {Sandoval-Santana}, \citenamefont {Ibarra-Sierra}, \citenamefont {Kunold},\
  and\ \citenamefont {Naumis}}]{sandoval2020floquet}%
  \BibitemOpen
  \bibfield  {author} {\bibinfo {author} {\bibfnamefont {J.}~\bibnamefont
  {Sandoval-Santana}}, \bibinfo {author} {\bibfnamefont {V.}~\bibnamefont
  {Ibarra-Sierra}}, \bibinfo {author} {\bibfnamefont {A.}~\bibnamefont
  {Kunold}}, \ and\ \bibinfo {author} {\bibfnamefont {G.~G.}\ \bibnamefont
  {Naumis}},\ }\href@noop {} {\bibfield  {journal} {\bibinfo  {journal} {arXiv
  preprint arXiv:2003.12119}\ } (\bibinfo {year} {2020})}\BibitemShut {NoStop}%
\bibitem [{\citenamefont {Champo}\ and\ \citenamefont
  {Naumis}(2019)}]{Champo2019}%
  \BibitemOpen
  \bibfield  {author} {\bibinfo {author} {\bibfnamefont {A.~E.}\ \bibnamefont
  {Champo}}\ and\ \bibinfo {author} {\bibfnamefont {G.~G.}\ \bibnamefont
  {Naumis}},\ }\href {\doibase 10.1103/PhysRevB.99.035415} {\bibfield
  {journal} {\bibinfo  {journal} {Physical Review B}\ }\textbf {\bibinfo
  {volume} {99}},\ \bibinfo {pages} {1} (\bibinfo {year} {2019})}\BibitemShut
  {NoStop}%
\bibitem [{\citenamefont {Ibarra-Sierra}\ \emph {et~al.}(2019)\citenamefont
  {Ibarra-Sierra}, \citenamefont {Sandoval-Santana}, \citenamefont {Kunold},\
  and\ \citenamefont {Naumis}}]{ibarra2019dynamical}%
  \BibitemOpen
  \bibfield  {author} {\bibinfo {author} {\bibfnamefont {V.}~\bibnamefont
  {Ibarra-Sierra}}, \bibinfo {author} {\bibfnamefont {J.}~\bibnamefont
  {Sandoval-Santana}}, \bibinfo {author} {\bibfnamefont {A.}~\bibnamefont
  {Kunold}}, \ and\ \bibinfo {author} {\bibfnamefont {G.~G.}\ \bibnamefont
  {Naumis}},\ }\href@noop {} {\bibfield  {journal} {\bibinfo  {journal}
  {Physical Review B}\ }\textbf {\bibinfo {volume} {100}},\ \bibinfo {pages}
  {125302} (\bibinfo {year} {2019})}\BibitemShut {NoStop}%
\bibitem [{\citenamefont {Goldman}\ and\ \citenamefont
  {Dalibard}(2014)}]{Goldman}%
  \BibitemOpen
  \bibfield  {author} {\bibinfo {author} {\bibfnamefont {N.}~\bibnamefont
  {Goldman}}\ and\ \bibinfo {author} {\bibfnamefont {J.}~\bibnamefont
  {Dalibard}},\ }\href {\doibase 10.1103/PhysRevX.4.031027} {\bibfield
  {journal} {\bibinfo  {journal} {Phys. Rev. X}\ }\textbf {\bibinfo {volume}
  {4}},\ \bibinfo {pages} {031027} (\bibinfo {year} {2014})}\BibitemShut
  {NoStop}%
\bibitem [{\citenamefont {Sandoval-Santana}\ \emph {et~al.}(2019)\citenamefont
  {Sandoval-Santana}, \citenamefont {Ibarra-Sierra}, \citenamefont {Cardoso},
  \citenamefont {Kunold}, \citenamefont {Roman-Taboada},\ and\ \citenamefont
  {Naumis}}]{sandovalannderphysik}%
  \BibitemOpen
  \bibfield  {author} {\bibinfo {author} {\bibfnamefont {J.~C.}\ \bibnamefont
  {Sandoval-Santana}}, \bibinfo {author} {\bibfnamefont {V.~G.}\ \bibnamefont
  {Ibarra-Sierra}}, \bibinfo {author} {\bibfnamefont {J.~L.}\ \bibnamefont
  {Cardoso}}, \bibinfo {author} {\bibfnamefont {A.}~\bibnamefont {Kunold}},
  \bibinfo {author} {\bibfnamefont {P.}~\bibnamefont {Roman-Taboada}}, \ and\
  \bibinfo {author} {\bibfnamefont {G.}~\bibnamefont {Naumis}},\ }\href
  {\doibase 10.1002/andp.201900035} {\bibfield  {journal} {\bibinfo  {journal}
  {Annalen der Physik}\ }\textbf {\bibinfo {volume} {531}},\ \bibinfo {pages}
  {1900035} (\bibinfo {year} {2019})},\ \Eprint
  {http://arxiv.org/abs/https://onlinelibrary.wiley.com/doi/pdf/10.1002/andp.201900035}
  {https://onlinelibrary.wiley.com/doi/pdf/10.1002/andp.201900035} \BibitemShut
  {NoStop}%
\bibitem [{\citenamefont {Kitagawa}\ \emph {et~al.}(2011)\citenamefont
  {Kitagawa}, \citenamefont {Oka}, \citenamefont {Brataas}, \citenamefont
  {Fu},\ and\ \citenamefont {Demler}}]{Calvo}%
  \BibitemOpen
  \bibfield  {author} {\bibinfo {author} {\bibfnamefont {T.}~\bibnamefont
  {Kitagawa}}, \bibinfo {author} {\bibfnamefont {T.}~\bibnamefont {Oka}},
  \bibinfo {author} {\bibfnamefont {A.}~\bibnamefont {Brataas}}, \bibinfo
  {author} {\bibfnamefont {L.}~\bibnamefont {Fu}}, \ and\ \bibinfo {author}
  {\bibfnamefont {E.}~\bibnamefont {Demler}},\ }\href {\doibase
  10.1103/PhysRevB.84.235108} {\bibfield  {journal} {\bibinfo  {journal} {Phys.
  Rev. B}\ }\textbf {\bibinfo {volume} {84}},\ \bibinfo {pages} {235108}
  (\bibinfo {year} {2011})}\BibitemShut {NoStop}%
\bibitem [{\citenamefont {Peng}\ \emph {et~al.}(2016)\citenamefont {Peng},
  \citenamefont {Zhang}, \citenamefont {Shao}, \citenamefont {Xu},
  \citenamefont {Zhang},\ and\ \citenamefont {Zhu}}]{peng2016electronic}%
  \BibitemOpen
  \bibfield  {author} {\bibinfo {author} {\bibfnamefont {B.}~\bibnamefont
  {Peng}}, \bibinfo {author} {\bibfnamefont {H.}~\bibnamefont {Zhang}},
  \bibinfo {author} {\bibfnamefont {H.}~\bibnamefont {Shao}}, \bibinfo {author}
  {\bibfnamefont {Y.}~\bibnamefont {Xu}}, \bibinfo {author} {\bibfnamefont
  {R.}~\bibnamefont {Zhang}}, \ and\ \bibinfo {author} {\bibfnamefont
  {H.}~\bibnamefont {Zhu}},\ }\href@noop {} {\bibfield  {journal} {\bibinfo
  {journal} {Journal of Materials Chemistry C}\ }\textbf {\bibinfo {volume}
  {4}},\ \bibinfo {pages} {3592} (\bibinfo {year} {2016})}\BibitemShut
  {NoStop}%
\bibitem [{\citenamefont {Verma}\ \emph {et~al.}(2017)\citenamefont {Verma},
  \citenamefont {Mawrie},\ and\ \citenamefont {Ghosh}}]{verma2017effect}%
  \BibitemOpen
  \bibfield  {author} {\bibinfo {author} {\bibfnamefont {S.}~\bibnamefont
  {Verma}}, \bibinfo {author} {\bibfnamefont {A.}~\bibnamefont {Mawrie}}, \
  and\ \bibinfo {author} {\bibfnamefont {T.~K.}\ \bibnamefont {Ghosh}},\
  }\href@noop {} {\bibfield  {journal} {\bibinfo  {journal} {Physical Review
  B}\ }\textbf {\bibinfo {volume} {96}},\ \bibinfo {pages} {155418} (\bibinfo
  {year} {2017})}\BibitemShut {NoStop}%
\bibitem [{\citenamefont {Villanova}\ and\ \citenamefont
  {Park}(2016)}]{Villanova2016Spin}%
  \BibitemOpen
  \bibfield  {author} {\bibinfo {author} {\bibfnamefont {J.~W.}\ \bibnamefont
  {Villanova}}\ and\ \bibinfo {author} {\bibfnamefont {K.}~\bibnamefont
  {Park}},\ }\href {\doibase 10.1103/PhysRevB.93.085122} {\bibfield  {journal}
  {\bibinfo  {journal} {Phys. Rev. B}\ }\textbf {\bibinfo {volume} {93}},\
  \bibinfo {pages} {085122} (\bibinfo {year} {2016})}\BibitemShut {NoStop}%
\bibitem [{\citenamefont {Zabolotskiy}\ and\ \citenamefont
  {Lozovik}(2016)}]{Zabolotskiy2016}%
  \BibitemOpen
  \bibfield  {author} {\bibinfo {author} {\bibfnamefont {A.~D.}\ \bibnamefont
  {Zabolotskiy}}\ and\ \bibinfo {author} {\bibfnamefont {Y.~E.}\ \bibnamefont
  {Lozovik}},\ }\href {\doibase 10.1103/PhysRevB.94.165403} {\bibfield
  {journal} {\bibinfo  {journal} {Physical Review B}\ }\textbf {\bibinfo
  {volume} {94}},\ \bibinfo {pages} {1} (\bibinfo {year} {2016})}\BibitemShut
  {NoStop}%
\bibitem [{\citenamefont {Herrera}\ and\ \citenamefont
  {Naumis}(2019)}]{HerreraNaumisKubo2019}%
  \BibitemOpen
  \bibfield  {author} {\bibinfo {author} {\bibfnamefont {S.~A.}\ \bibnamefont
  {Herrera}}\ and\ \bibinfo {author} {\bibfnamefont {G.~G.}\ \bibnamefont
  {Naumis}},\ }\href {\doibase 10.1103/PhysRevB.100.195420} {\bibfield
  {journal} {\bibinfo  {journal} {Phys. Rev. B}\ }\textbf {\bibinfo {volume}
  {100}},\ \bibinfo {pages} {195420} (\bibinfo {year} {2019})}\BibitemShut
  {NoStop}%
\bibitem [{\citenamefont {Sakurai}\ and\ \citenamefont
  {Commins}(1995)}]{sakurai1995modern}%
  \BibitemOpen
  \bibfield  {author} {\bibinfo {author} {\bibfnamefont {J.~J.}\ \bibnamefont
  {Sakurai}}\ and\ \bibinfo {author} {\bibfnamefont {E.~D.}\ \bibnamefont
  {Commins}},\ }\href@noop {} {\enquote {\bibinfo {title} {Modern quantum
  mechanics, revised edition},}\ } (\bibinfo {year} {1995})\BibitemShut
  {NoStop}%
\bibitem [{\citenamefont {Shirley}(1965)}]{Shirley1965Solution}%
  \BibitemOpen
  \bibfield  {author} {\bibinfo {author} {\bibfnamefont {J.~H.}\ \bibnamefont
  {Shirley}},\ }\href {\doibase 10.1103/PhysRev.138.B979} {\bibfield  {journal}
  {\bibinfo  {journal} {Phys. Rev.}\ }\textbf {\bibinfo {volume} {138}},\
  \bibinfo {pages} {B979} (\bibinfo {year} {1965})}\BibitemShut {NoStop}%
\bibitem [{\citenamefont {Verhulst}(2006)}]{verhulst2006nonlinear}%
  \BibitemOpen
  \bibfield  {author} {\bibinfo {author} {\bibfnamefont {F.}~\bibnamefont
  {Verhulst}},\ }\href {https://books.google.com.mx/books?id=oyaFXgcFJ2cC}
  {\emph {\bibinfo {title} {Nonlinear Differential Equations and Dynamical
  Systems}}},\ Universitext\ (\bibinfo  {publisher} {Springer Berlin
  Heidelberg},\ \bibinfo {year} {2006})\BibitemShut {NoStop}%
\bibitem [{\citenamefont {Hale}(2009)}]{hale2009ordinary}%
  \BibitemOpen
  \bibfield  {author} {\bibinfo {author} {\bibfnamefont {J.~K.}\ \bibnamefont
  {Hale}},\ }\href@noop {} {\enquote {\bibinfo {title} {Ordinary differential
  equations. mineola, ny},}\ } (\bibinfo {year} {2009})\BibitemShut {NoStop}%
\bibitem [{\citenamefont {Mojarro}\ \emph {et~al.}(2020)\citenamefont
  {Mojarro}, \citenamefont {Ibarra-Sierra}, \citenamefont {Sandoval-Santana},
  \citenamefont {Carrillo-Bastos},\ and\ \citenamefont {Naumis}}]{mojarro2020}%
  \BibitemOpen
  \bibfield  {author} {\bibinfo {author} {\bibfnamefont {M.~A.}\ \bibnamefont
  {Mojarro}}, \bibinfo {author} {\bibfnamefont {V.~G.}\ \bibnamefont
  {Ibarra-Sierra}}, \bibinfo {author} {\bibfnamefont {J.~C.}\ \bibnamefont
  {Sandoval-Santana}}, \bibinfo {author} {\bibfnamefont {R.}~\bibnamefont
  {Carrillo-Bastos}}, \ and\ \bibinfo {author} {\bibfnamefont {G.~G.}\
  \bibnamefont {Naumis}},\ }\href {\doibase 10.1103/PhysRevB.101.165305}
  {\bibfield  {journal} {\bibinfo  {journal} {Phys. Rev. B}\ }\textbf {\bibinfo
  {volume} {101}},\ \bibinfo {pages} {165305} (\bibinfo {year}
  {2020})}\BibitemShut {NoStop}%
\bibitem [{\citenamefont {Dittrich}(1998)}]{dittrich1998quantum}%
  \BibitemOpen
  \bibfield  {author} {\bibinfo {author} {\bibfnamefont {T.}~\bibnamefont
  {Dittrich}},\ }\href {https://books.google.com.mx/books?id=N6jvAAAAMAAJ}
  {\emph {\bibinfo {title} {Quantum transport and dissipation}}}\ (\bibinfo
  {publisher} {Wiley-VCH},\ \bibinfo {year} {1998})\BibitemShut {NoStop}%
\bibitem [{\citenamefont {Snoke}(2020)}]{snoke2020}%
  \BibitemOpen
  \bibfield  {author} {\bibinfo {author} {\bibfnamefont {D.}~\bibnamefont
  {Snoke}},\ }\href {https://books.google.com.mx/books?id=EnG9DwAAQBAJ} {\emph
  {\bibinfo {title} {Solid State Physics: Essential Concepts}}}\ (\bibinfo
  {publisher} {Cambridge University Press},\ \bibinfo {year}
  {2020})\BibitemShut {NoStop}%
\bibitem [{\citenamefont {Dubbers}\ and\ \citenamefont
  {St{\"o}ckmann}(2013)}]{dubbers2013quantum}%
  \BibitemOpen
  \bibfield  {author} {\bibinfo {author} {\bibfnamefont {D.}~\bibnamefont
  {Dubbers}}\ and\ \bibinfo {author} {\bibfnamefont {H.-J.}\ \bibnamefont
  {St{\"o}ckmann}},\ }\href@noop {} {\emph {\bibinfo {title} {Quantum physics:
  the bottom-up approach: from the simple two-level system to irreducible
  representations}}}\ (\bibinfo  {publisher} {Springer Science \& Business
  Media},\ \bibinfo {year} {2013})\BibitemShut {NoStop}%
\end{thebibliography}%

\end{document}